\documentclass[preprint,preprintnumbers,
amsmath,showkeys,amssymb,aps,floatfix]{revtex4}

\usepackage{graphicx}
\usepackage{dcolumn}
\usepackage{bm}
\usepackage{longtable}
\usepackage{epsfig}
\usepackage{times}

\begin{document}

\title{Nuclear effects in atomic transitions}

\author{Adriana~P\'alffy}\email{Palffy@mpi-hd.mpg.de}

\affiliation{Max-Planck-Institut f\"ur Kernphysik, Saupfercheckweg~1,
69117 Heidelberg, Germany}

\begin{abstract}

Atomic electrons are sensitive to the properties of the nucleus they are bound to, such as 
nuclear mass, charge distribution, spin, magnetization distribution, or even excited level scheme.
These nuclear parameters are reflected in the atomic transition energies.
A very precise determination of atomic spectra  may thus  reveal information about the nucleus, otherwise hardly accessible via nuclear physics experiments. This work reviews theoretical and experimental aspects of 
the nuclear effects that can be identified in atomic structure data. 
An introduction to the theory of isotope shifts and hyperfine splitting of atomic spectra is given, together with an overview of the typical experimental techniques used in high-precision atomic spectroscopy. More exotic effects at the borderline between atomic and nuclear physics, such as parity violation in atomic transitions due to the weak interaction, or  nuclear polarization and nuclear excitation by electron capture, are also addressed.


\end{abstract}

\maketitle
\section{Introduction}

Atomic and nuclear physics historically share a common ancestry. When Niels Bohr proposed in 1913 his first model of the atom, he depicted it as having a small and dense positively charged nucleus, surrounded by the orbiting electrons. The existence of the electron, a discrete unit of negative charge,  had been discovered in the cathode rays experiments performed by J. J. Thomson in 1897.
However, it was the great interest arising from the phenomenon of radioactivity that engaged scientists in the quest for the microscopic world of the atom. The beginning of the XXth century was a very exciting time for physics, with one experimental discovery following the other. Quantum mechanics, the novel theory appropriate to describe the microscopic world of the atom, emerged to a good extent from the need to understand atomic spectra. With time, atomic physics turned to be the
field primarily concerned with the arrangement of electrons around the nucleus and the processes by which these change, i.e., the atomic transitions. Nuclear physics, on the other hand, focused on the atomic nucleus, its building blocks and the interactions between nuclei. The question what  does the nucleus consist of has led over the years to the Standard Model,  the integrated theory of  elementary particles, which is supposed to explain all  processes in Nature ruled by the first three of the four known fundamental forces: the electromagnetic, weak, strong and gravitational forces.

How much information about the nucleus is used in or can be obtained from atomic physics? This work intends to provide an up to date answer to these questions. Traditionally, atomic physics could afford to consider the nucleus as a point-like  charge without loosing in accuracy. With increased precision in atomic physics experiments, the nucleus is going through an odyssey from the simplified picture of a point-like charge to its real size and properties. Different nuclear masses,  charge distributions or spins do have an effect on the electronic levels and transition energies, and the small energy corrections or splittings observed in experiments can give in turn information about the atomic nucleus. Atomic spectroscopy,  at present at an unrivalled level of precision achieved by frequency-comb techniques, can thus be used to determine nuclear parameters which are otherwise hardly accessible via nuclear physics experiments.
The easiest way to determine the nuclear effect related differences in the atomic structure is to compare spectra of different isotopes---nuclei with the same atomic number, i.e., charge, but different mass number---and determine the
small variation of electronic transition energies, the so-called isotope shifts. 
 These shifts can be caused only by nuclear properties other than the  charge.

The correct identification of nuclear effects in atomic transitions is very important in connection with  accurate theoretical atomic structure calculations involving many-body correlations or quantum electrodynamics (QED) corrections, which can be tested in high-precision experiments. A key ingredient for the determination of nuclear effects in atomic transitions and in general for tests of theoretical atomic and nuclear models are highly charged ions, advantageous  due to their simplified electronic structure.  Experiments at facilities with stored or trapped highly charged ions have played in the last decades an important role in the study of nuclear effects and determination of nuclear properties from atomic structure data.

The borderline between atomic and nuclear physics is also inhabited by several nuclear processes that directly involve atomic electrons. The underlying interactions for such processes are either electromagnetic or weak. For instance, the nucleus and the electron can interact through the electromagnetic field and undergo transitions simultaneously.  Such a process is  well-known in nuclear physics, as being sometimes the only decay channel for a nuclear excited state: the internal conversion. A nucleus in an excited state that for some reason cannot decay  radiatively  transfers its energy through the electromagnetic field to one of the atomic electrons which leaves the atom. The inverse process of internal conversion, which might occur in highly charged ions, is called nuclear excitation by electron capture. The weak interaction, on the other hand, can create or annihilate atomic electrons, or can mix the parity of atomic states, leading to parity-violating effects. These processes at the interface between atomic and nuclear physics also belong to the category of {\it nuclear effects in atomic transitions}.

In the following, theoretical and experimental aspects of these effects  are addressed, structured in three parts. The first section deals with isotopic shifts, their origin in the charge volume and mass properties of the nucleus, and some of the established methods to determine them experimentally. The second section addresses the hyperfine structure of atomic spectra, its relation with the determining nuclear parameters, such as the nuclear spin and the nuclear multipole moments, as well as the various experimental methods used for observation. A separate section is dedicated to more exotic aspects such as parity violation and nuclear processes involving atomic electrons. The paper concludes with a discussion of the future directions in probing nuclear effects in atomic structure.

\section{Isotope shifts\label{is}}

By far the most important nuclear property that determines the atomic energy levels is the nuclear charge. 
The energy of the atomic levels can be obtained in the first approximation  by solving the Schr\"odinger equation for the electrons in the central potential of the nuclear charge $+Ze$, where $Z$ is the atomic number. In order to take into account the spin of the electron, the Dirac equation is introduced. However, in the first approximation, both equations  describe the nucleus as a point-like charge of infinite mass. 

Measurements of atomic spectra for different isotopes of the same atomic number $Z$ show slight 
differences---the isotope shifts. This frequency difference of an electronic transition  is usually described separately as due to the finite mass of the nucleus---the mass shift---and to the size of the nuclear charge distribution---the volume of field shift.
 The mass shift is dominating isotope shifts for light atoms, whereas the field shift scales as  $Z^5$ or even up to $Z^6$ and thus becomes the leading contribution in the case of heavy ones. In order  to extract information about the nuclear properties, a proper interpretation of the measured isotope shifts based on theoretical calculations is necessary. 

Since often nuclei with the same $Z$ but different mass number $A$ also have different nuclear spins, spectra of different isotopes may also have  a completely different hyperfine structure. We will focus in this section on the mass and field isotope shifts, which do not give rise to splittings of the energy levels, and address the hyperfine structure in the next section. An introduction to the theory of the mass and field shifts is given in separate subsections, followed by an overview of several experimental methods used for isotope shift measurements. 

\subsection{Mass shift\label{mass_shift}}

Originally  assumed  for the sake of simplicity to be infinitely heavy, the nucleus  obviously  has a finite mass. The approximation of an infinitely heavy nucleus is acceptable for certain cases, but for a precise description  one should consider the effects of nuclear motion or recoil. Nonrelativistically, in the case of hydrogen-like ions, this can be done easily by introducing the reduced mass, (see, for instance, Ref.~\cite{BranJoach}),
\begin{equation}
\mu=\frac{mM}{m+M}\, ,
\end{equation}
where $m$ is the mass of the electron and $M$ the one of the nucleus. The Schr\"odinger equation is then solved in the center-of-mass system, with the reduced mass of the electron with respect to the nucleus replacing the electronic mass $m$. For many-electron systems, apart from the reduced mass, also a mass polarization term enters the Schr\"odinger equation, of the form
\begin{equation}
-\frac{\hbar^2}{M}\sum_{i<j}\nabla_{\vec{r}_i}\cdot \nabla_{\vec{r}_j}\cdot \, ,
\label{smsnonrel}
\end{equation}
where the indices $i$ and $j$ denote the $i$th and $j$th electrons, respectively. 
Hughes and Eckart \cite{HughEck} in 1930 were the ones to separate the mass shift  of atomic spectra into these two parts: the normal mass shift (NMS), due to the introduction of the reduced mass in the kinetic energy term in the  Schr\"odinger equation (also called the one-body part), and the specific mass shift (SMS), which rises from the mass polarization term, or two-body part, introduced in Eq.~(\ref{smsnonrel}). For decades, according to this separation, the NMS and SMS have been treated differently in calculations - the NMS was evaluated exactly, while the SMS was obtained from perturbation theory. This method of evaluation proved to be unreliable even in the non-relativistic case, as pointed out in Ref.~\cite{Palmer}. Also, rigorously speaking, the mass shift theory in the nonrelativistic case is valid only for a point-like nucleus \cite{FrickePRL}.

In the relativistic theory, NMS and SMS both rise naturally from the relativistic nuclear recoil operator, which can be written in the lowest-order relativistic approximation as \cite{ShabaevYF,Palmer},
\begin{equation}
R_{ij}=\frac{1}{2M}\sum_{i,j}\left[\vec{p}_i\cdot\vec{p}_j - \frac{Z\alpha}{r_i}\left(\vec{\alpha}_i+\frac{(\vec{\alpha}_i\cdot\vec{r}_i)\vec{r}_i}{r_i^2}\right)\cdot\vec{p}_j\right]\, ,
\label{nuclear_recoil}
\end{equation}
where $\vec{r}_i$ and $\vec{p}_i$ are the coordinate and the momentum of the $i$th electron, respectively, $\vec{\alpha}_i$ are the vector Dirac matrices, acting on the electron $i$, and $\alpha$ the fine structure constant. The NMS correction is given by the expectation value of the $R_{ij}$ operator for $i=j$, $\langle \sum_{i} R_{ii} \rangle$, while the SMS correction can be obtained from the $i\neq j$ terms, $\langle \sum_{i\neq j} R_{ij} \rangle$, treating the two terms  on equal footage \cite{Palmer}. The full relativistic theory of the nuclear recoil effect can  only be formulated in the framework of QED, see \cite{ShabaevPRA57,ShabaevPR}. A precise optical isotope shift measurement with stored highly charged argon ~\cite{SoriaOrts} ions has confirmed experimentally the necessity to consider the relativistic nuclear recoil effect in  theory. More details about the experimental approach  are given in  Section \ref{exIS}, where the typical experimental methods to determine isotope shifts are outlined.

\subsection{Volume or field shift}

Although often treated as point-like, the nucleus has a finite size.  Consequently, the distribution of the protons in the nuclear matter determine the electrostatic nuclear potential, which will no longer be $1/r$. The Dirac equation should be then solved numerically  using a realistic nuclear potential instead of the usual Coulomb potential generated by a point-like charge $+Ze$. Compared to a point-like nucleus, the extended nuclear  charge distribution leads to a shift of the resonance energies on the order of 200 eV in the case of the $K$-shell electrons in uranium ($Z=92$).

Since different isotopes of the same element with nuclear charge $+Ze$ have different numbers of neutrons, their charge distribution will  not coincide, leading to a small shift in their electronic level energies - the volume or field shift. Coming back to our previous example, in the case of the $K$-shell electrons in uranium, the field shift between $^{235}\mathrm{U}$ and $^{238}\mathrm{U}$ is  less than 1 eV. The effect of the finite nuclear size is strongest for the $s$ electronic orbitals, which have the largest overlap with the nucleus, and among these the field shift for the $1s$ electrons is obviously the largest. The nuclear size effects scale with the nuclear charge as $Z^5$ up to $Z^6$ \cite{Roxana}, such that the field shift value for the K-shell of uranium should be one of the largest measured field shifts. Uranium (Z=92) and californium (Z=98) are among the heaviest nuclei used for atomic physics experiments, both of them long-lived radioactive nuclei (in atomic physics, experiments with stable or long-lived isotopes are preferred due to technical difficulties).

The history of isotope shifts goes back to 1931, when experiments by Sch\"uler and Keyston on the hyperfine structure of thallium  and later on mercury \cite{SchKeyTa,SchKeyHg} led to the discovery of a structure which was not due to the nuclear spin  but to a displacement of the atomic levels in different isotopes. The authors  pointed out that their observations have to be explained by some differences in the nuclear fields of the isotopes. 
Ever since, several models for the nuclear charge distribution have been used in the literature. One of the obvious choices is to consider the nucleus as a homogeneously charged sphere of radius R. In this model, the electrostatic potential $V(r)$ due to the nucleus is then given by \cite{BranJoach}
\begin{equation}
V(r)=\left\{\begin{array}{cc} \frac{Ze^2}{2R}\left(\frac{r^2}{R^2}-3\right) & r\le R\, , \\
-\frac{Ze^2}{r} & r\ge R\, .
 \end{array} \right.
\end{equation}
The nuclear charge distribution is in this case constant, $\rho(r)=3Ze/(4\pi R^3)$. Using the nuclear charge distribution, radial moments $\langle r^n \rangle$ can be defined (see, for instance, Ref.~\cite{FrickeADNDT60}), of which the most important  is the  root mean square charge (RMS) radius of the nucleus,
\begin{equation}
\langle r^2\rangle= \frac{1}{Ze}\int d^3r \rho(r) r^2\, .
\end{equation}
It is actually the RMS radius $\langle r^2\rangle^{1/2}$ that the nuclear size effect depends on, rather than the charge distribution itself. Indeed, isotope shift measurements are (for nowadays precision) basically insensitive to the nuclear charge distribution, as long as the same RMS radius is reproduced \cite{Otten,FrickeADNDT60,ShabaevJPB26}. That is why most experiments aiming at the determination of the nuclear charge distribution  extract the RMS radius out of the measured values with the help of various models. For the homogeneous sphere nucleus model, the RMS radius is given by $\langle r^2\rangle=3R^2/5$.

\begin{figure}
\begin{center}
\includegraphics[width=0.50\textwidth]{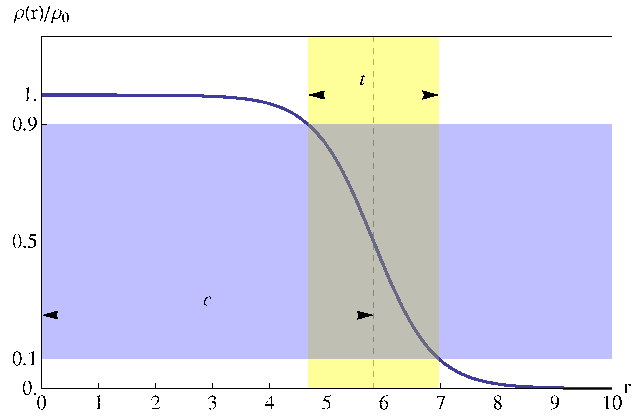}
\end{center}
\caption{A two-parameter Fermi nuclear charge distribution determined by  the surface thickness $t$ and the half-density radius $c$.} 
\label{rho_nucleus}
\end{figure}

A more physical description of the nuclear charge distribution is given by a two-parameter Fermi-type  distribution \cite{JohnSoff}
\begin{equation}
\rho(r)=\frac{\rho_0}{1+e^{(r-c)/a}}
\, ,
\end{equation}
with the surface thickness 
\begin{equation}
t=4a \mathrm{ln}3
\, .
\end{equation}
The quantity $c$ is the half-density radius which gives the
distance to the point at which the nuclear charge density
is one-half of its maximum value.  The distance in the surface
layer over which the nuclear density decreases from 90$\%$ to 10$\%$ of its maximum is called surface thickness and this is denoted by $t$ in the above equation. A plot of the Fermi distribution highlighting the two parameters is presented in Figure~\ref{rho_nucleus}. The Fermi distribution with $t = 0$ reduces to a
uniform distribution with radius $R = c$.  A typically used value for all nuclei is  $t=2.30$ fm. The constant $\rho_0$ can be obtained by writing the total nuclear charge $Ze$ as the integral over the Fermi-like charge distribution. 
Alternatively, the Fermi distribution parameters $c$ and $\rho_0$ can be expressed with the help of the RMS radius \cite{ShabaevJPB26,VYerokhin},
\begin{eqnarray}
c^2&=& \frac{5}{3}\langle r^2\rangle -\frac{7}{3}a^2\pi^2\, , \nonumber \\
\rho_0&=&\frac{3}{4\pi c^3}\left(1+\frac{\pi^2a^2}{c^2}\right)^{-1}\, .
\end{eqnarray}
A commonly used shortcut to the nuclear charge distribution problem is that in most calculations the thickness parameter is taken as $t=2.30$ fm, and the RMS radius is obtained from the empirical formula \cite{JohnSoff}
\begin{equation}
\langle r^2\rangle^{1/2}=(0.836 A^{1/3} + 0.570)\ \ \mathrm{fm}\, ,
\end{equation}
which is a weighted fit to the measured RMS radii for nuclei with $A>9$. Its accuracy was found to be better than 0.05 fm and is in most cases sufficient for the present experimental accuracy. 

The two models for the nuclear charge distribution described so far assume a spherical nucleus.  It is however well known  that most of the isotopes have a deformed nuclear ground state. To account for the nuclear deformation, a deformed Fermi distribution can be introduced \cite{ZumbroPRL53}:
\begin{equation}
\rho(r,\theta,\varphi)=\frac{\rho_0}{1+e^{(r-c)/a}}
\, ,
\end{equation}
where this time we have
\begin{equation}
c=c_0\left(1+\sum_{l=1}^{\infty}\sum_{m=-l}^{l}\beta_{lm}Y_{lm}(\theta,\varphi)\right)
\, .
\end{equation}
In the equation above, $\beta_{lm}$ are the nuclear multipole deformation parameters and $Y_{lm}$ the spherical harmonics. 
In most cases,  axial symmetry is assumed, and considering only quadrupole and hexadecapole nuclear deformation, this expression reduces to
\begin{equation}
c=c_0[1+\beta_{20}Y_{20}(\theta,\varphi)+\beta_{40}Y_{40}(\theta,\varphi)]
\, .
\end{equation}
This deformed nuclear charge distribution has been taken into account to calculate field shifts and derive approximate analytical formulas for nuclear-size corrections to binding energies of uranium and neodymium isotopes \cite{Kozhedub}. The authors of Ref.~\cite{Kozhedub} conclude that, if a precise value of the RMS radius is available from experiments, for an accurate  calculation of the finite nuclear size effect on the electronic binding energies the nuclear deformation should be taken into account. The value of the finite-nuclear-size correction obtained from their analytical approximative method for H-like uranium $^{238}\mathrm{U}^{91+}$ is 198.39 eV when  taking  into account the nuclear deformation and 198.61 eV when using a spherical nucleus with the same RMS radius. The theory thus predicts a 220~meV energy shift between the spherical and deformed nucleus models. Whether it is possible to discern between nuclear models on the basis of such a small difference relying on experimental results is a different matter. 
In the following we address this practical issue and present some of the experimental methods used to determine the isotope shifts.

\subsection{Experimental determination of  isotope shifts\label{exIS}}

Isotope shifts are typically very small, and can only be determined in high-precision  experiments. The better the experimental precision, the more one can learn about the nucleus and about the theoretical formalism that we use to obtain electronic transition energies. Usually atomic transition energies do not directly provide  information about the nucleus (other than the atomic number $Z$). Other parameters  have to be extracted and interpreted by first taking into account all possible contributions to the electronic transition energy, whether from electron correlation, first- and second-order QED effects, nuclear recoil, nuclear finite size or nuclear polarization. Measured isotope shifts are evaluated in theory as the sum of the mass and field shift, plus QED and nuclear polarization contributions. Among these often the nuclear polarization contribution (which is the subject of Section~\ref{np}) is the one with the largest theoretical error bars. As a bottom line, if one wants to extract out of the measured isotope shifts the most interesting nuclear property, the difference in the RMS radii of the different isotopes, one has to make sure that (i)  the field shift is obtained out of the isotope shift by taking the correct theoretical values for all other contributions and (ii) a model for the field shift calculation delivers the RMS radius values. Thus, a reliable interpretation of the data is just as important as the measurement of the isotope shift  itself.

At the present level of accuracy, atomic physics measurements provide important information about the nucleus in the form of the already introduced RMS radius.  The absolute values of the RMS nuclear radius can be obtained by two experimental methods: {\it electron scattering} and {\it muonic atom x-rays}. In both cases the theoretical analysis of the results is crucial for obtaining the nuclear parameter.

Direct isotope shift measurements only provide the difference between RMS radii of two isotopes, $\delta\langle r^2\rangle$. Usually either the $K_{\alpha}$ x-ray isotope shifts, or optical isotope shifts are used to determine the variation of the RMS radii. Optical isotope shift measurements are performed by {\it laser spectroscopy}, a highly-sensitive method involving laser excitation of the transition of interest and subsequent detection of the decay. The difficulty here consists of interpreting the measured isotope shifts, especially due to the fact that the method is limited to low-charge states, i.e., many-electron systems. Atomic many-body effects make the theoretical description difficult and bring large uncertainties due to SMS and QED contributions in the calculation of the electronic density at the nucleus. {\it X-ray spectroscopy} can be used for the $K_{\alpha}$ isotope shifts in highly charged ions. Less electrons make the theoretical interpretation easier, with the disadvantage that the experimental precision (one needs meV precision for x-ray transition energies) is limited. Alternatively, {\it dielectronic recombination measurements} on highly charged ions for a number of nuclei can achieve nowadays the necessary precision and provide complementary data to isotope shifts and RMS radii obtained via other methods. A brief description of the five methods, with special focus on the new and less traditional dielectronic recombination  measurements will be given. Apart of dielectronic recombination isotope shift measurements, which are relatively new, the RMS radii obtained from the other four methods have been combined in an attempt to elude the model-dependencies in the evaluation of the data in several compilations, from which the most recent is Ref.~\cite{Angeli2004}. For an insight in the combined theoretical analysis of the four methods, see also \cite{FrickeADNDT60}.

\subsubsection{ Elastic electron scattering\label{escat}}
Elastic electron scattering refers to electron scattering in the Coulomb field of nuclei. Electrons are accelerated up to several hundreds MeV in a linear accelerator and then focused on a nuclear target. This energy is required in order to achieve a small wavelength for the electron despite its small mass. The electron beam should also be very intense, because of the very small scattering cross section, and have a very good energy resolution (unfortunately, these two criteria are not easily achieved simultaneously, leading to the use of the so-called energy-loss spectrometer systems \cite{DonSick}). What is measured is the differential cross section  $d\sigma(E, \theta)/d\Omega$ for the elastic scattering of an electron of energy $E$ at the angle $\theta$. The angular differential scattering cross section is then compared with the relativistic formula of Mott for elastic scattering of electrons  (see, for instance, a nice introduction of the scattering cross sections and form factors in Ref.~\cite{Hof56}). The Mott formula (in relativistic units for which $4\pi\varepsilon_0=1$),
\begin{equation}
\frac{d\sigma}{d\Omega}=\frac{Z^2e^4\mathrm{cos}^2\theta/2}{4E^2\mathrm{sin}^4\theta/2}\, ,
\end{equation}
is the relativistic equivalent of the well-known Rutherford formula for scattering of charged point-like particles on a charged point-like center. The finite size of the nucleus is introduced via  a so-called form factor. For a spin-zero nucleus, in the first Born approximation, the scattering cross section can be factorized in $(d\sigma/d\Omega)_{\mathrm{Mott}}$ for a point-like nucleus, and the  form factor $F(q)$, a structure function which depends on the momentum transfer in the scattering, $q=(2E/\hbar c)\ \mathrm{sin}(\theta/2)$, provided that $m_ec^2\ll E$,
\begin{equation}
\left. \frac{d\sigma}{d\Omega}\right|_{\mathrm{exp}}=\left.\frac{d\sigma}{d\Omega}\right|_{\mathrm{Mott}}|F(q)|^2 \, .
\end{equation}
The form factor is related to the nuclear charge distribution $\rho_n(r)$ through a Fourier-Bessel transformation \cite{FrickeADNDT60}
\begin{equation}
F(q)=\int d^3r \rho_n(r)j_0(qr), \hspace{1cm} 0<q<\infty\, ,
\end{equation}
where $j_0(qr)=(\mathrm{sin}\,qr)/(qr)$ represents the spherical Bessel function of zero order. This Fourier-Bessel analysis has been introduced by Dreher et al.~\cite{Dreher} and provides a model-independent method to obtain information about the nuclear charge distribution in elastic electron scattering experiments and in particular to extract the RMS radii. The first Born approximation is only suitable for scattering on light nuclei, while for medium and heavy nuclei a careful analysis of the experimental cross sections should be done taking into account the phase shift \cite{Hof56}. A number of RMS radii compilations for stable isotopes from electron scattering experiments  and sometimes in conjunction with other methods are available in Refs.~\cite{Hof56,deVries,FrickeADNDT60,Angeli2004}.

In the last half a century, following the pioneering experiments by Hofstadter {\it et al.} \cite{Hof56}  at Stanford, USA, electron-nucleus scattering has proved to be an excellent tool for studying the nuclei. Among the advantages of the method we have the generating electromagnetic interaction, which is on the one hand well understood, and on the other hand weak enough for the electrons neither to perturb the nucleus, nor to get absorbed or undergo  multiple scattering.
 Also, this is the only electromagnetic method which can provide information about the radial dependence of the nuclear charge density $\rho_n(r)$. 
It should be noted, however, that the accuracy of the method is limited,  with typical accuracies of 0.2$\%$ for RMS radii. Also, typical experiments work only with stable nuclear targets. With the development of Radioactive Ion Beam  facilities,  properties of exotic short-lived nuclei can become object of study in the laboratory. Experimental facilities designed to perform elastic electron scattering off exotic nuclei are at present under construction at RIKEN in Japan and GSI in Germany \cite{Suda,Wakasugi,Simon,FAIR}.

\subsubsection{ Muonic atoms } 
Muonic atoms were actually the first to provide accurate measurements of nuclear size \cite{FitRain53}, at a time when the nuclear radius was still believed to be $1.4A^{1/3}$. The muon is a heavy unstable  lepton with negative charge $-e$, and  mass which is about 207 times heavier than that of the electron. A muon can form a bound system with a nucleus, just like the electron. However, because of its much heavier mass, the muon comes much closer to the nucleus, acting as a probe---in the $1s$ state, for instance, the muon spends a significant fraction of time inside the nucleus. The ionization potential and the spectral lines corresponding to transitions between bound states are also much higher than the ones  encountered in  electronic atoms. For nuclei with large $Z$, the ionization potential is on the order of several MeV, and the transitions between muonic bound states range between $\gamma$ rays and X rays.

When the muon is stopped in matter, it first scatters from atom to atom just as a free electron, until it is captured into a high atomic orbit (a Rydberg state) of a particular atom. From this excited capture state the muon will cascade down through the electron cloud, ejecting Auger electrons and emitting radiation, until it arrives to a region inside the innermost electron orbit. From there,  with no further electrons in sight, the muon decays mostly radiatively. The energies of these transitions provide information about the size and shape of the nucleus as well as tests of our understanding of QED. The capture and cascade processes up to the muonic $1s$ state take place in $10^{-12}-10^{-9}$~s, while the lifetime of the muon is of 2.2$\times 10^{-6}$~s. The muon spends thus most of its lifetime in the $1s$ state, probing the nucleus. 

A comparison of the theoretical and experimental values for the  energies and relative rates for various radiative transitions provide information about the nuclear size and the RMS radii. The muonic wavefunctions can be usually obtained with good accuracy, since the muonic atom is basically a  hydrogen-like system---there  is always only one muon bound to a nucleus, and the presence of the atomic electrons far away from the nucleus can be safely neglected. However, unlike in electronic atoms, the muon orbit is comparable to the nuclear size, and the transition energies of the same order of magnitude as the nuclear excitation energies, such that the nuclear polarization effects are considerably large. 
Also, even though the nucleus is still considerably heavier than the muon, the effect of nuclear motion (nuclear recoil, described in subsection~\ref{mass_shift}) should be taken into account \cite{FrickePRL}. 
It is beyond the purpose of this review to present the theory of muonic atoms and all the contributions to the transitions energies.  The interested reader is referred to the  review of Borie and Rinker \cite{BorRink84}.

The procedure for deducing the nuclear charge radii from the measured transition energies assuming a Fermi charge distribution has been described in detail in Ref.~\cite{Engfer74}, which also gives a compilation of RMS radii, isotope and isomer shifts and magnetic hyperfine constants from muonic atoms. A more recent compilation with more precise data (also using a combined analysis of all electromagnetic methods to determine nuclear charge parameters) can be found in Ref.~\cite{FrickeADNDT60}. It should be noted that the main accuracy for the RMS radii deduced from transition energies in heavy  muonic atoms is not given by the experimental errors, but by the uncertainty of the calculated nuclear polarization contribution.

Just as in the case of electron scattering, RMS radii from  muonic atom experiments have been determined for  all stable isotopes already many years ago. Radioactive muonic atoms  nowadays open the possibility to address exotic nuclei far from the valley of stability, whose properties are yet to be investigated. New, intense muon sources,  together with techniques related to cold films of radioactive atoms as host material, merging of ion and muon beams, or trapping of exotic isotopes and muons in Penning traps, will make a number of interesting muonic atoms experiments  possible (see Ref.~\cite{Jungmann} on muon physics possibilities at  muon-neutrino factories and references therein).

\subsubsection{ Optical isotope shifts }
 Optical isotope shifts  are mostly obtained from laser spectroscopy on atoms or ions. Laser spectroscopy involves transitions with optical frequencies, which means that (with the exception of very light nuclei, discussed separately) this time {\it outer atomic electrons} get to probe the nuclear structure. The high resolution obtained by using laser-atomic beam spectroscopy and the high sensitivity of optical detection techniques allow precision measurements of the optical isotope shifts and hyperfine structure of stable as well as of rare and radioactive isotopes. Laser spectroscopy can also provide an alternative isotope separation method \cite{Demtroeder}. The advantage of this experimental method is its precision and the fact that it  allows the determination of isotope shifts also for unstable nuclei. The drawback is that the data is more difficult to interpret due to the presence of many electrons. In order to extract the most interesting nuclear parameter, the RMS radius, one needs to understand and discern the mass and field shift contributions out of the total isotope shift. This becomes very difficult for many-electron atoms  with large mass shift, for which the theory does not offer reliable  SMS predictions. Usually this problem is solved by performing a combined analysis of isotope shift results from different types of experiments, and extracting scaling factors by hand by means of King plots (two-dimensional diagrams that represent isotope shifts obtained from two different methods on the $x$ and $y$ axis \cite{King} for purposes of calibrating the mass  and field shifts in optical spectra). Unfortunately, the combined analysis is not possible for radioactive isotopes, for which no isotope shifts were determined from electron scattering or muon atom experiments. 

The typical optical isotope shift measurement involves a tunable laser which  excites the  optical transition of interest in atoms or ions. The transition energy is the object of a very precise determination by monitoring the excitation as a function of the laser wavelength. Usually  either  fluorescence or  resonant ionization and subsequent ion detection (so-called Resonance Ionization Mass Spectrometry, RIMS) in the case of laser spectroscopy of atoms are used to detect the induced excitation. The typical problem which affects the accuracy is the Doppler effect. The atoms (usually in a beam) will have a velocity spread which shifts the resonance energy for the laser photons and thus broadens the fluorescence line. Doppler-free measurements are achieved by two-photon laser spectroscopy, using one collinear and one anticollinear laser beams \cite{Demtroeder}. Indeed, if the two absorbed photons  with equal frequency $\omega_1=\omega_2=\omega$, travel in opposite directions (their wave vectors are $\vec{k}_1=-\vec{k}_2$), all particles, independent of their velocities, absorb at the same sum frequency $\omega_1+\omega_2=2\omega$.

\begin{figure}
\begin{center}
\includegraphics[width=0.50\textwidth]{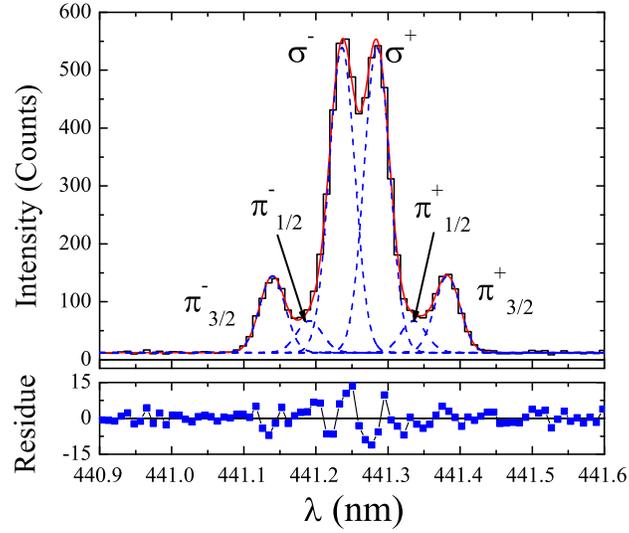}
\end{center}
\caption{A typical spectral line obtained from the
$1s^22s^22p\ ^2P_{1/2} -\ ^2P_{3/2}$ transition in B-like $^{40}\mathrm{Ar}^{13+}$. The six
dashed curves represent a fit to the Zeeman components. The obtained experimental wavelength for the $M1$ transition is $\lambda=441.2559(1)$~nm \cite{Draganic}. Reprinted with permission from R. Soria~Orts {\it et al.}, Phys. Rev. Lett. 97, 103002, 2006. Copyright (2006) by the Americal Physical Society.}
\label{ArRosario}
\end{figure}

\begin{figure}
\begin{center}
\includegraphics[width=0.50\textwidth]{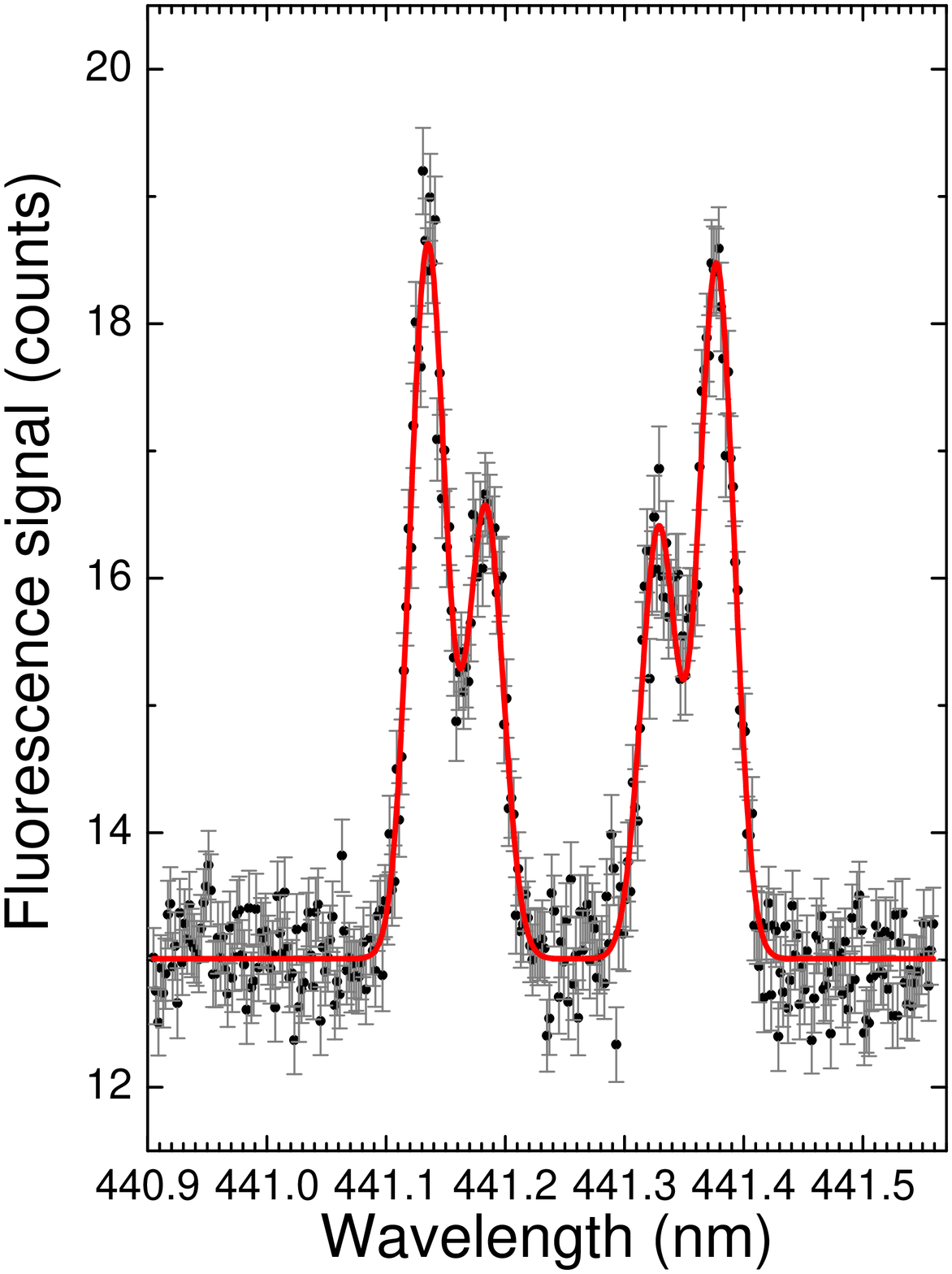}
\end{center}
\caption{The $\pi^{\pm}_{3/2}$ Zeeman components of the 
$1s^22s^22p\ ^2P_{1/2} -\ ^2P_{3/2}$ transition in B-like $^{40}\mathrm{Ar}^{13+}$ obtained from laser spectroscopy at the Heidelberg EBIT. }
\label{ArLaser}
\end{figure}

One unusual measurement of optical isotope shifts that does not involve laser spectroscopy has been already mentioned in Section~\ref{mass_shift}. The experiment was performed  on B-like and Be-like argon ions and confirmed the necessity of considering the relativistic nuclear recoil for the mass shift theory. The isotope shifts of $^{36}\mathrm{Ar}$ versus $^{40}\mathrm{Ar}$ were determined by measuring  the $1s^22s^22p\ ^2P_{1/2}-\ ^2P_{3/2}$ transition in $\mathrm{Ar}^{13+}$ and the $1s^22s2p\ ^3P_{1}-\ ^3P_{2}$ transition in $\mathrm{Ar}^{14+}$ for the two isotopes  with sub-ppm accuracy \cite{SoriaOrts}. These optical frequency transitions both have magnetic dipole ($M1$) multipolarity, such that the excited states, populated by collisions with the electron beam, are long-lived (or metastable) states. The experiment was performed at the Electron Beam Ion Trap (EBIT)  in Heidelberg, Germany.  The recorded spectrum is shown in Figure~\ref{ArRosario}. A dedicated theoretical calculation showed that the NMS and SMS and their relativistic corrections are on the same order of magnitude. Furthermore, the theoretical isotope shifts including the relativistic corrections showed an excellent agreement with the experimental data, confirming the necessity to consider the nuclear recoil relativistically. Recently, the wavelength of the $1s^22s^22p\ ^2P_{1/2}-\ ^2P_{3/2}$ transition in $\mathrm{Ar}^{13+}$ has been confirmed by the same group in Heidelberg, this time using laser spectroscopy. For a comparison, the determined fluorescence spectrum for the $\pi^{\pm}_{3/2}$ Zeeman components is shown in Figure~\ref{ArLaser}.

A special case of optical isotope shifts are the very light nuclei ($Z<10$), for which the field shift represents only a very small fraction of the total isotope shift (for Li, for instance, this fraction is $10^{-4}$). This means that RMS radii can be extracted from optical isotope shift measurements only if the accuracy of the experiment and the accuracy of the calculated mass shifts are on the order of $10^{-5}$ or better. Until recently, the SMS could not be calculated with sufficient accuracy for atomic systems with more than two electrons. For this reason, RMS radii determination for light atoms was restricted to electron scattering or muonic atoms experiments, which only work for stable isotopes. Laser spectroscopy experiments could provide RMS radii only for  hydrogen and helium isotopes, for which the SMS was calculated with the necessary accuracy. The proton radius, for instance, has been determined from the $1s-2s$ transition frequency in hydrogen with $10^{-14}$ accuracy \cite{Haensch} in the group of T. H\"ansch in Munich, Germany.

\begin{figure}
\begin{center}
\includegraphics[width=0.30\textwidth]{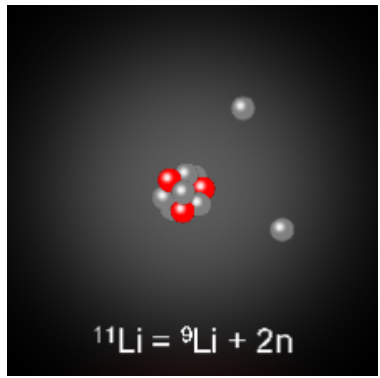}
\end{center}
\caption{Artist's view of the halo $^{11}\mathrm{Li}$ nucleus. The two additional neutrons form the halo around the $^{9}\mathrm{Li}$ core. } 
\label{halo}
\end{figure}

Sufficiently accurate calculations of mass shifts for systems with up to three electrons became reality recently \cite{Drake1,Drake2,Drake3,Pachucki1,Pachucki2}. These calculations have enabled also the study of {\it halo nuclei}, a special variety of light nuclei that has received great attention since their discovery in 1985. In the $^{11}\mathrm{Li}$ isotope it was found that the matter radius  is much larger than that of its neighbours, suggesting a large deformation and/or a long tail in the matter distribution \cite{Tanihata85}.  Later,  such isotopes became known as {\it halo nuclei}, which have diffuse outer neutron distributions at the limit of stability. In particular, $^{11}\mathrm{Li}$ is a three-body system with a $^{9}\mathrm{Li}$ core and two halo neutrons, as depicted in Figure~\ref{halo}. Nuclear forces are not strong enough to bind one neutron to the $^{9}\mathrm{Li}$ nucleus, nor are they able to bind two neutrons together in a dineutron. Yet the 
$^{9}\mathrm{Li}$ and the two neutrons together can form a bound nucleus, $^{11}\mathrm{Li}$, with lifetime $\tau=8.4$~ms.  

A  feature of the halo nuclei is that the matter radius (given by the two halo neutrons) and the charge radius (of the three protons concentrated in the $^{9}\mathrm{Li}$ core) do not coincide. Matter radii can be deduced from nuclear reactions. High-precision isotope shift measurements deliver the RMS charge radius. These two quantities are most important for understanding details of the nuclear structure and halo-core interactions. The RMS charge radius of $^{8}\mathrm{Li}$, $^{9}\mathrm{Li}$ and $^{11}\mathrm{Li}$ have been obtained by high-resolution laser spectroscopy on an atomic beam using two-photon Doppler-free excitation and resonance-ionization detection \cite{Li8-9,Li11}. Other laser spectroscopy experiments for the determination of the charge radii of other halo nuclei such as $^{8}\mathrm{He}$ \cite{He8}, $^{11}\mathrm{Be}$ \cite{Be10} or $^{17}\mathrm{Ne}$ \cite{Ne17} in ion beams or traps have been also performed. Ab initio   descriptions of the structure of light nuclei
based on realistic two- and three-nucleon interactions between individual nucleons have been developed during the past years, with work still in progress.

\subsubsection{Isotope shifts from x-ray transitions}
Isotope shifts from x-ray transitions are provided by x-ray spectroscopy experiments with atoms or highly charged ions. The $K_{\alpha}$ x-ray transitions are the most obvious candidates. The very first such experiment in 1965 determined the isotope shift in $K_{\alpha}$ fluorescence of $^{233}\mathrm{U}$ and $^{238}\mathrm{U}$ neutral atoms which were irradiated  by broadband x-rays \cite{KalfaU}. While from the theoretical point of view isotope shifts of atomic $K\alpha$ x-ray transitions are much easier to handle, the experimental precision of x-ray spectroscopy is considerably lower than the one in the optical regime, where lasers are available. With the operation of the x-ray free electron laser LCLS at SLAC in Stanford, USA, providing photons with up to 8 keV energy, high-precision experiments in this energy regime will hopefully become possible in the near future.

Compared to experiments with neutral atoms, precision measurements of x-ray transitions in highly charged ions are even more advantageous due to the substantially simplified theory to extract the nuclear charge parameters from the isotope shifts. X-ray spectroscopy in highly charged ions of different uranium isotopes has been performed at the SuperEBIT at the Lawrence Livermore National Laboratory in the USA \cite{SuperEBIT-U}. The ion beam of the EBIT was used to ionize, excite and radially trap the ions, and the characteristic x-rays of the $2s_{1/2}-2p_{3/2}$ transition of nearly 4.5 keV for Li-like, Be-like, B-like, C-like and O-like $^{233}\mathrm{U}$ and $^{238}\mathrm{U}$ ions were detected with a curved-crystal spectrometer with an energy resolution of 1.1 eV. This resolution of 100 ppm in highly charged ions was already good enough to improve  the accuracy of the RMS radii for uranium by about one order or magnitude. The obtained difference in the RMS radius  $\delta\langle r^2 \rangle^{233,238}=-0.457\pm 0.043$~fm$^2$ lies in between earlier measurements based on different techniques.  

\subsubsection{Dielectronic recombination experiments}
\begin{figure}
\begin{center}
\includegraphics[width=0.7\textwidth]{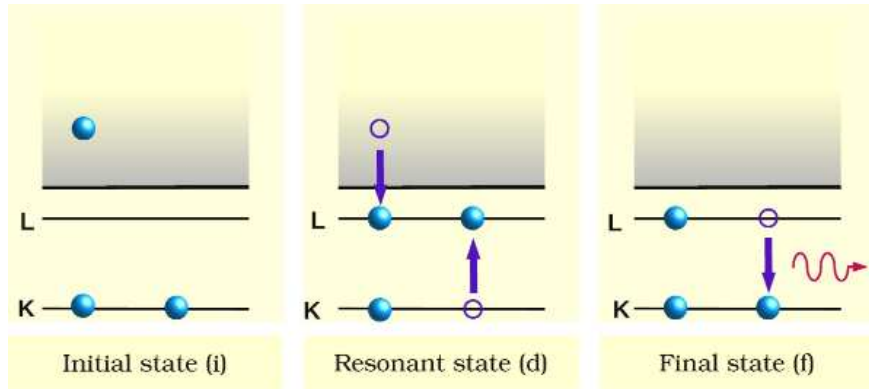}
\end{center}
\caption{Dielectronic capture of a free electron into the $L$ shell of a He-like atom with the simultaneous excitation of one of the core $K$-shell electrons. The resonant recombination is followed by the radiative decay of the excited electron to the $K$ shell. Such a DR process is called $KLL$-DR, denoting the involved atomic shells. } 
\label{DR}
\end{figure}

Recently, another type of atomic physics experiments involving electron recombination in highly charged ions have been used to obtain high-precision isotope shifts. This method involves a resonant channel of photorecombination, namely, {\it  dielectronic recombination}. When a free electron recombines into an ion, it can release its excess energy and angular momentum through a photon. This is the direct process of radiative recombination (RR), possible at all incoming electron energies. However, there also exists a resonant channel: free electrons with matching kinetic energy can recombine resonantly and excite a bound electron, as shown schematically in Figure~\ref{DR}. The formed excited electronic state can decay either by an Auger process (which is  the reverse  of the dielectronic capture), or by emitting one or more photons. The dielectronic capture followed by the emission of photon(s) from the bound electron(s) is known as {\it dielectronic recombination} (DR). DR is of great importance as the predominant electron recombination channel in solar and other dense astrophysical plasmas. Its study in highly charged ions provides important information about transition energies, isotope shifts, hyperfine splitting, in one word, electronic structure. By studying the electron recombination rate as a function of the free electron energy one obtains the spectra of DR resonances, whose positions give the atomic transition energies.

\begin{figure}
\begin{center}
\includegraphics[width=0.70\textwidth]{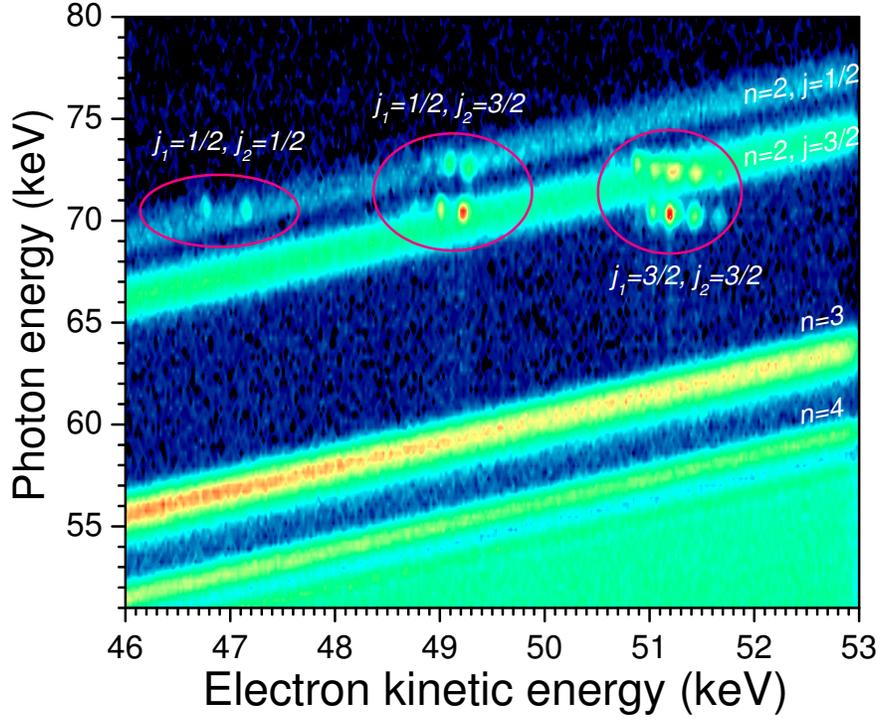}
\end{center}
\caption{Experimental data for RR and $KL_1L_2$-DR in  mercury ions with charge states from He-like to Be-like. The emitted photon energy is plotted as a function of the energy of the recombining electron. On top of the RR lines one can observe the DR resonances. The principal  and total angular momentum quantum numbers $n$ and $j$ of the atomic shell into which the electron recombines radiatively are given on each RR line. The DR resonances are indexed by $j_1$ and $j_2$, the total angular momentum quantum numbers of the excitation level for the bound electron and the recombination state of the free electron, respectively. } 
\label{JoseDR}
\end{figure}

DR experiments involve highly charged ions, either in a storage ring like the Experimental Storage Ring (ESR) at the GSI in Darmstadt, Germany, or in an EBIT. Depending on the experiment, the photons following the recombination or the recombined ions  are detected. In any case, both RR and DR channels are observed, with the resonance structure of DR on top of the flat RR dependence on the free electron energy. An example is presented in  Figure~\ref{JoseDR}, which shows the  photon energy distribution following recombination of mercury ions ranging from He-like to Be-like in an EBIT DR experiment. The electron energy was scanned across the region of the $KLL$ resonances.
 From their positions,   the atomic transition energies can be determined. In a storage ring recombination setup, the free electrons for DR are provided by a co-propagating  electron target or by the electron cooler used to cool the ion beam. Particularly at very low ion-electron collision energies, storage ring experiments  have enabled  precise  determination of Lamb shifts in several highly charged ions \cite{Brandau2003}. In order to determine very precise isotope shifts, DR of an electron with very small kinetic energy, preferably less than 10 eV, is required. For some nuclear charges $Z$, this is possible by allowing the capture into a Rydberg state to match a core $n=2$ intrashell transition. Recombination into Rydberg states also has the advantage that the loosely bound Rydberg electron itself is unaffected by the small variations of nuclear potential due to different charge distributions and its contribution to the isotope shift can be safely neglected.

\begin{figure}
\begin{center}
\includegraphics[width=0.60\textwidth]{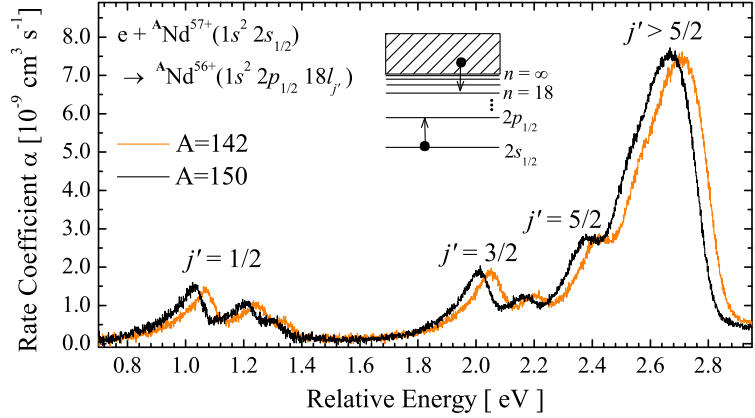}
\end{center}
\caption{Dielectronic recombination of the Li-like
neodymium isotopes $^{142}\mathrm{Nd}^{57+}$ (orange or gray line) and
 $^{150}\mathrm{Nd}^{57+}$ (black line) in the energy range of the $1s^22p_{1/2}18l_{j'}$
resonance groups. The $j'$ labels indicate the individual fine structure components of the $n=18$ Rydberg electron. The DR recombination scheme is presented in the inset.  Reprinted  with permission from  C. Brandau {\it et al.}, Phys. Rev. Lett. 100, 073201, 2008. Copyright (2008) by the American Physical Society.} 
\label{NdDR}
\end{figure}

Recently, isotope shift measurements in a DR experiment and subsequent RMS radii determination have been achieved,  opening a new direction for isotope shift experiments.  Isotope shifts have been measured  using the transitions $e^-+\mathrm{Nd}^{57+} (1s^22s_{1/2})\rightarrow  \mathrm{Nd}^{56+} (1s^22p\ nl_j)^{**}$ of two Nd isotopes $^{142}\mathrm{Nd}$ and $^{150}\mathrm{Nd}$ \cite{NdPRL}. Here $^{**}$ denotes the doubly excited state created in the dielectronic capture. The resonant electron capture occurs in a Rydberg state with principal quantum numbers $n\ge 18$ for the $2s\rightarrow 2p_{1/2}$ excitation and $n\ge 8$ for the $2s\rightarrow 2p_{3/2}$ excitation. Figure~\ref{NdDR} shows the dielectronic recombination rates for the two isotopes as a function of the recombination energy, with visible shifts of the resonance positions. The interpretation of these shifts in terms of RMS radii is in this simple three-electron system very reliable and could be done within state-of-the-art structure calculations that take into account relativistic and QED contributions assuming a two-parameter Fermi charge distribution for the nucleus. Out of the measured isotope shifts for the $2s-2p_{1/2}$ ($\delta E^{142-150}$=40.2(3)(6) meV) and $2s-2p_{3/2}$ ($\delta E^{142-150}$=42.3(12)(20) meV) transitions, with statistical and systematical error bars given in the first and second bracket, respectively, the RMS radii difference was determined, $\delta \langle r^2\rangle^{142-150}=-1.36(1)(3)$~fm$^2$. This value is slightly  larger than the average experimental value of $-1.291(6)$~fm$^2$ from the combined analysis in Ref.~\cite{Angeli2004}, but still in good agreement with the
upper end of the individual data that build the basis of the evaluation.

\section{Hyperfine splitting}
Hyperfine splitting  (HFS) of atomic levels has been first observed by Michelson in 1881, but it was first in 1924 that Pauli suggested  the observed effects  might be due to magnetic interactions between  electron and  nucleus,  possibly possessing a total angular momentum or ``nuclear spin''. Of course, at the time the concept of ``spin'' was not really established in quantum mechanics, but analogously to the fine structure that was attributed to the electron orbital angular momentum, the hyperfine features could be explained by assuming the existence of a nuclear magnetic moment.

The HFS effects  are due to the interaction of the nuclear multipole moments with the electromagnetic field created by the electrons at the nucleus. The most important of these moments are the magnetic dipole moment, associated with the nuclear spin itself, and the electric quadrupole moment, which gives a measure of the deviation from a spherical charge distribution in the nucleus, i.e., of the nuclear deformation. For an atom or ion with nuclear spin $\vec{I}$ and electronic total angular momentum  $\vec{J}$, the value of the energy level splitting is given by (see, for instance, Ref.~\cite{BranJoach})
\begin{equation}
\Delta E=\frac{A}{2}K+\frac{B}{4}\frac{\frac{3}{2}K(K+1)-2I(I+1)J(J+1)}{I(2I-1)J(2J-1)}\, ,
\label{Ehfs}
\end{equation}
where $K$ is given by
\begin{equation}
K=F(F+1)-I(I+1)-J(J+1)\, ,
\end{equation}
and $F$ is the quantum number of the total angular momentum $\vec{F}=\vec{I}+\vec{J}$. The first term in Eq.~(\ref{Ehfs}) is the magnetic dipole HFS, which is responsible for the splitting of one fine structure atomic energy level into a HFS multiplet labeled by $F$. The number of hyperfine structure components corresponding to a fine structure energy level is the smaller of the two numbers $(2J+1)$ and $(2I+1)$. The hyperfine separation, i.e., the energy difference between two neighbouring hyperfine levels is then given by $AF$, with the quantity $A$ in Eq.~(\ref{Ehfs})  given by
\begin{equation}
A=\frac{\mu_0}{4\pi}4g_I\mu_B\mu_N\frac{1}{J(J+1)(2l+1)}\frac{Z^3}{a^3_{\mu}n^3}\, ,
\end{equation}
where $\mu_0$ is the magnetic constant or vacuum permeability, $g_I$ the nuclear $g$ factor, $\mu_B$ and $\mu_N$ the Bohr and nuclear magneton, respectively, $l$ the orbital quantum number, $n$ the principal quantum number and $a_{\mu}$ the reduced Bohr radius, $a_{\mu}=a_0(m/\mu)$ obtained by using the reduced mass $\mu$ of the electron with respect with the nucleus. From the number of lines and the size of the splitting the nuclear spin and nuclear magnetic dipole moment can be extracted.

The second term in Eq.~(\ref{Ehfs}) represents the electric quadrupole correction. Because the dependence on the quantum number $F$ is different from that of the magnetic dipole correction, the hyperfine separation will no longer be simply proportional to $F$. The quadrupole coupling constant $B$ is proportional to the electric quadrupole moment $Q$ and the average gradient of the electric field produced by an electron at the nucleus,
\begin{equation}
B=Q\langle \frac{\partial^2V_e}{\partial z^2}\rangle\, .
\end{equation}
The electric quadrupole contribution provides information about the spectroscopic quadrupole moments and nuclear deformation parameters. 

\begin{figure}
\begin{center}
\includegraphics[width=0.45\textwidth]{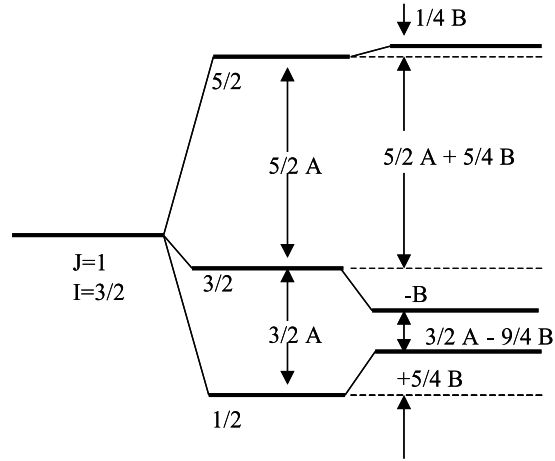}
\end{center}
\caption{Illustration of the hyperfine splitting scheme for $I=3/2$ and $J=1$ considering the example of $^{201}\mathrm{Hg}$. The original fine structure atomic level (left) is split into a magnetic HFS (middle) corrected by the quadrupole HFS (right). Compare with Eq.~(\ref{Ehfs}). } 
\label{Hg_hfs}
\end{figure}

An illustrative example of the HFS scheme for $I=3/2$ and $J=1$ is presented in Figure~\ref{Hg_hfs}. Since the hyperfine energy corrections are independent of the quantum number $M_F$ corresponding to the projection of the total angular momentum $F$ on the quantization axis, the hyperfine levels show a $(2F+1)$-fold  degeneracy which can be lifted in a magnetic field. Note that transitions between different HFS levels corresponding to the same  fine structure level are forbidden in the dipole approximation. They occur as magnetic dipole ($M1$) or electric quadrupole transitions ($E2$), and can be measured by radio-frequency methods. Due to the very good precision of such measurements, hyperfine transitions are used as time standards - the SI fundamental unit of time, the second, is defined in terms of the frequency of the transition between the levels with $F=4$, $M_F=0$ and $F=3$, $M_F=0$ of the ground state of $^{133}\mathrm{Cs}$. 

\subsection{Isotope shifts in the hyperfine structure}

Isotope shifts were first observed  in the HFS, and already in 1931 it was pointed out that the observed displacements were greater than expected from the mass correction of the Rydberg constant \cite{SchKeyTa}. It was suggested that these shifts are due to deviations of the nuclear electric field from the Coulomb law. This represented the starting point for the whole theory of the field shift, outlined in Section \ref{is}. For HFS, the discussion  should be  extended to the nuclear magnetization distribution and the effects of the extended nuclear charge distributions on magnetic interactions.

In order to introduce the corrections to the HFS expression in Eq.~(\ref{Ehfs}), let us consider the simplest case of hydrogen-like ions, with one $1s$ electron and a nuclear spin $I$. The relativistic {\it magnetic} HFS for the sublevel with largest $F$ can be written  as  \cite{ShabaevJPB27}
\begin{equation}
\Delta E=\frac{\alpha^4Z^3}{n^3}\frac{\mu_I}{I}\frac{m_e}{m_p}\frac{(I+J)m_ec^2}{J(J+1)(2l+1)}[A(\alpha Z)(1-\delta)(1-\varepsilon)+\kappa_{\mathrm{rad}}]\, ,
\end{equation}
where $\alpha$ is the fine structure constant, $\mu_I$ the nuclear magnetic moment, $A(\alpha Z)$ a relativistic correction factor and $\kappa_{\mathrm{rad}}$ QED radiative corrections. The product in front of the square bracket is the $AK/2$ term from the non-relativistic expression in Eq.~(\ref{Ehfs}), evaluated for $F=I+J$.
The remaining corrections $\delta$ and $\varepsilon$ stand for the nuclear charge distribution and magnetization distribution corrections, respectively.

In 1932, G. Breit and J. Rosenthal \cite{BreitRosenth} studied the effect of the deviations of the nuclear field from the Coulomb law on the interaction of the electron with the nuclear spin. In this calculation, the nuclear magnetization was assumed to be concentrated at the center of the nucleus. The ratio between the HFS interaction assuming a realistic nuclear charge distribution and the one with a point-like nucleus is known as the Breit-Rosenthal correction $\delta$, sometimes  simply called the nuclear charge distribution correction for HFS. The difference arises from the change of the electronic wave functions when assuming a nuclear charge distribution instead of a point-like charge, just as in the case of the field shift.

In 1950, A. Bohr and V. Weisskopf \cite{BohrWeisskopf}  studied the influence of nuclear structure on the HFS of heavy elements. The issue started from the so-called ``hyperfine anomaly'', an unexplained isotopic dependence in the ratio between HFS and nuclear $g$ factors. For the first time they took into account the possibility that the nuclear dipole moment is not point-like, but should be represented by some distribution of magnetism over the nuclear volume. Indeed, since the total nuclear spin results from the individual spins of neutrons and protons and their angular momenta, the distribution of magnetic dipole density over the nuclear volume may vary greatly from nucleus to nucleus depending on the relative spin and orbital contributions  to the total nuclear moment.  The ratio between the HFS interaction energy assuming the extended nuclear magnetization and the one for point-like nuclear dipole moments is known as the Bohr-Weisskopf correction $\varepsilon$. For heavy H-like systems  the Breit-Rosenthal and Bohr-Weisskopf corrections are much larger than the QED corrections. For the case of H-like $^{165}\mathrm{Ho}$, for instance, vacuum polarization and self-energy contributions (included in $\kappa_{\mathrm{rad}}$) account for 0.5$\%$, the nuclear magnetization distribution (Bohr-Weisskopf) for 2$\%$ and the Breit-Rosenthal effect for 8$\%$ of the total energy splitting. 

Accurate HFS measurements in H-like systems provide a way to determine nuclear magnetization distribution radii. By using previously determined nuclear magnetic moments, and applying appropriate corrections for the nuclear charge distribution and radiative effects, the experimentally determined  HFS  can be interpreted in terms of nuclear magnetization radii. The aim is to distinguish between models of the magnetic structure of the nucleus. This is another opportunity to learn more about the nucleus via high-precision atomic physics experiments assisted by an adequate theoretical interpretation. Since often several effects contribute to the atomic transition energies, increased experimental and theoretical precision refines the search for other features at the borderline between atomic and nuclear physics. For instance, the uncertainty in the neutron distribution inside the nucleus affects the interpretation of parity non-conservation experiments, and knowledge about the nuclear magnetization distribution is vital for the interpretation of experiments looking for nuclear anapole moments or a proton electric dipole moment.

\subsection{Highlights of hyperfine splitting experiments}
The year 2007 marked the 40$^{\mathrm{th}}$ anniversary of the redefinition of the International System of Units (SI) second based on a hyperfine transition in cesium. Until 1967, the second had been always defined based on astronomical time scales. The hyperfine transition $F=4$, $M_F=0$ and $F=3$, $M_F=0$ of the ground state of $^{133}\mathrm{Cs}$  has a defined frequency of 9192.631770 GHz, commensurate with the accuracy in measuring this transition frequency achieved in microwave experiments.

The original idea of {\it microwave atomic beam magnetic resonance experiments} (or, as they were  called at the time, {\it radiofrequency experiments}) is the following \cite{Rabi1,Rabi2}: a beam of atoms propagates through two inhomogeneous magnetic fields  oriented such that the deflections they cause are equal but opposite. The net deflection is then zero if and only if the magnetic moment of the atom in the direction of each  field is the same. If the orientation of the atomic magnetic moment changes  in the space between the two  magnets, the atoms will not reach the detector placed at the zero deflection position, and the latter will register an intensity drop.
Between these inhomogeneous fields a region of  homogeneous variable magnetic field is placed and perpendicular to this, a weak radiofrequency oscillating field is present. If the oscillating frequency corresponds to the energy difference between  two hyperfine levels of interest, detectable transitions  changing the orientation of the atomic magnetic moment will occur.  Thus, the beam intensity dependence as a function of the applied frequency provides the atomic radiofrequency spectrum. A further improvement  by Ramsey \cite{Ramsey} utilizes  two separated oscillatory fields basically building up an interferometer for the atomic beams. Nowadays, the oscillating field is applied in a microwave interrogation cavity which allows narrowing of the resonance line, reduces the sensitivity to microwave power fluctuations and magnetic fields by orders or magnitude, and it eliminates the Doppler effect. The Ramsey method yields the until recently unrivalled  accuracy of the cesium beam primary frequency standards, which achieved with NIST-7  a fractional uncertainty of $5\times 10^{-15}$ \cite{Metrologia}. 

Microwave experiments were the first ones to drive  transitions between hyperfine atomic levels. Apart of cesium and other alkali atoms suitable because of their single valence electron, in particular the 1.4 GHz splitting frequency in hydrogen has been intensively studied, eventually providing the hydrogen maser with a relative frequency accuracy of $7\times 10^{-13}$ \cite{Hmaser}. This  illustrates the potential for extending ground state HFS investigation to H-like ions, due to the simplicity of the theoretical modeling for one-electron systems. However, the microwave measurement of the $1s$ splitting in $^3\mathrm{He}^+$ \cite{HeHFS} has not been followed by further steps towards higher-$Z$ ions, mainly due to difficulties in producing and storing highly charged ions. Obviously, due to the energy splitting dependence on $Z^3$, the HFS transitions in highly charged ions are also no longer accessible to microwave spectroscopy. At around $Z\sim 65$, the transition between the two $F$ levels of the $1s$ ground state already scales up into the infrared, the visible and UV region of the electromagnetic spectrum, depending on the particular value of the magnetic moment of the nucleus. High-precision laser spectroscopy then becomes the appropriate tool to address such transitions. 

Similar to the case of isotope shift measurements,  HFS have been determined also  in muonic atoms or  in x-ray atomic spectra. 
Ground state HFS has been observed in x-ray spectra recorded after nuclear electron capture from the $K$-shell of cesium atoms \cite{XeXray}. The precision of this measurement was  not sufficient to draw any conclusions about the nuclear magnetization. For a long time, the main tool of studying the nuclear magnetization distribution has been by measuring $\gamma$ rays emitted in muonic atoms \cite{Buettgenbach}. The HFS of the ground state is in muonic atoms on the order of a few keV, which unfortunately have to be extracted from $\gamma$-ray transitions corresponding to the nuclear or muonic transitions, with energies of several MeV. The ground state HFS usually cannot be resolved directly, but it has to be determined from the broadening of the line using calibration lines of isotopes with no nuclear spin \cite{Buettgenbach}. The HFS of the $1s$ level can be determined with several percent accuracy, and since the  effect of the nuclear magnetization distribution reduces the separation  by  50~$\%$, a value accurate to 10~$\%$ can be obtained for the Bohr-Weisskopf correction. The ground state HFS have been measured in muonic atoms to a precision of a few percent in $^{209}\mathrm{Bi}$  \cite{Bi209M} and $^{203,205}\mathrm{Tl}$ \cite{TlMuonic}. Unfortunately, the uncertainties do not allow to distinguish between different nuclear models.

For electronic atoms, the advantage of determining HFS in H-like ions is obvious. The HFS is large, the $1s$ electron is close to the nucleus and feels the nuclear magnetization distribution strongly, and the theoretical interpretation of the measured spectra is reliable. Furthermore, nuclear polarization effects are smaller than in muonic atoms. As soon as H-like heavy ions could be produced, the two types of highly charged ions facilities, the storage ring and the EBIT, aimed at performing  HFS experiments. Out of this competition, HFS for the ground state of $^{165}\mathrm{Ho}^{66+}$ \cite{Ho165}, $^{185,187}\mathrm{Re}^{74+}$ \cite{Re185187}, $^{203,205}\mathrm{Tl}^{80+}$ \cite{Tl203205}, $^{207}\mathrm{Pb}^{81+}$ \cite{Pb207} and $^{209}\mathrm{Bi}^{82+}$ \cite{Bi209} were determined. The experiments at the ESR at GSI Darmstadt, which determined the HSF of $^{209}\mathrm{Bi}^{82+}$ (visible) and $^{207}\mathrm{Pb}^{81+}$ (infrared) were performed by precision laser spectroscopy on the electron-cooled beam of H-like ions in a collinear antiparallel setup \cite{Bi209,Pb207}. Laser fluorescence was detected as the laser frequency was tuned in and out of resonance with the Doppler-shifted transition, with total accuracy  limited by ion velocity uncertainties. Note that a certain a priori knowledge of the hyperfine transition is required, since otherwise the scan in the laser frequency is prohibitively time consuming.

\begin{figure}
\begin{center}
\includegraphics[width=0.55\textwidth]{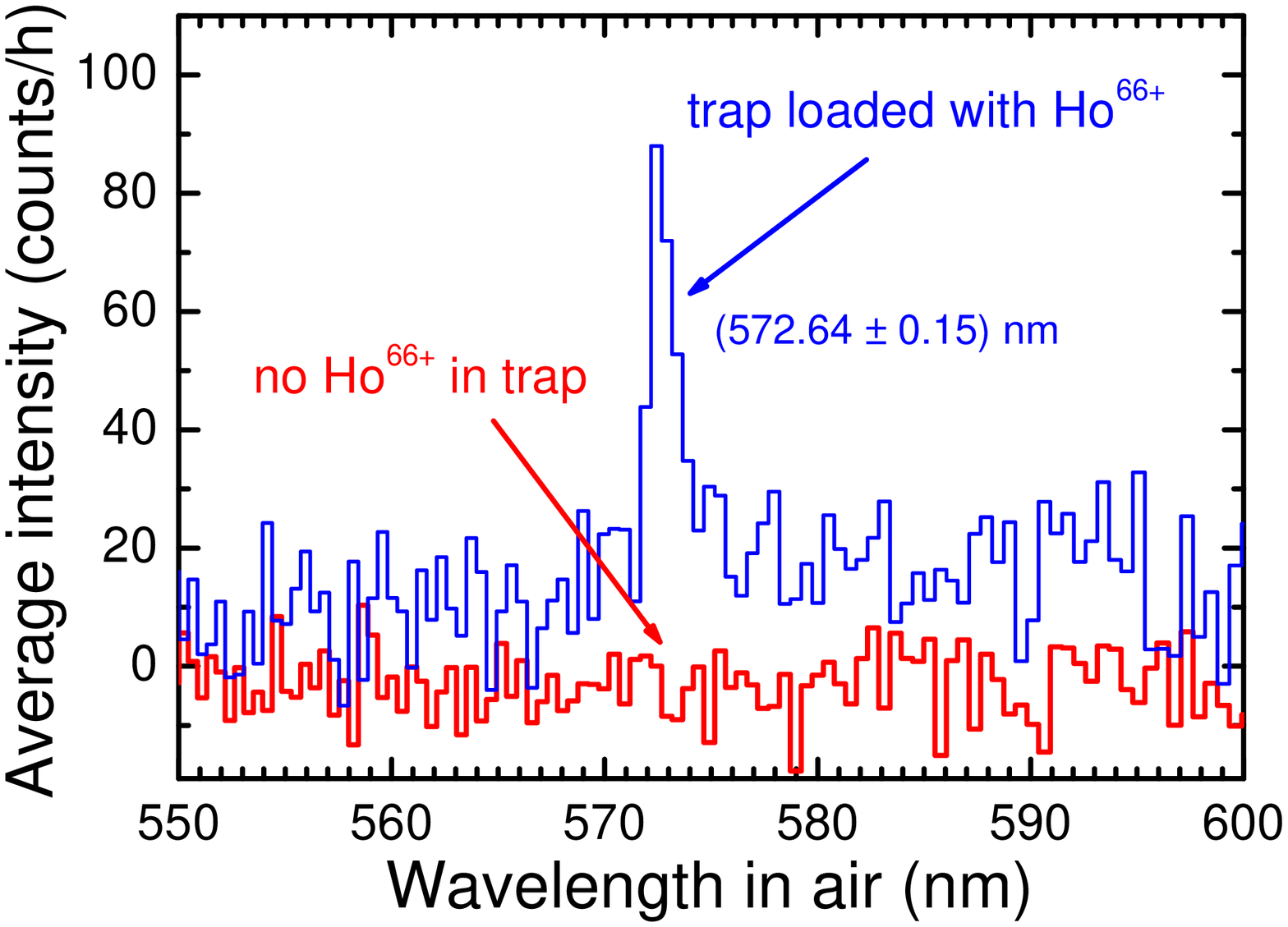}
\end{center}
\caption{Experimental determination of the {$F=4$} to {$F=3$} hyperfine transition in the ground state of hydrogenlike $^{165}\mathrm{Ho}^{66+}$. The signal is compared to the background radiation in the trap when no holmium ions are stored. } 
\label{Holmium}
\end{figure}

At the EBIT there is no such impediment because  laser spectroscopy is not needed. The hyperfine transition of the $1s$ ground state level is determined using ``passive'' emission spectroscopy, since the transitions are excited by electron collisions with the electron beam itself. To match the precision obtained by laser spectroscopy on highly charged ions at the GSI, a  sensitive spectrometer with   good resolving power was used.
In Figure~\ref{Holmium} the measured signal in the direct observation of the spontaneous emission of the hyperfine transitions {$F=4$} to {$F=3$} in the ground state of hydrogenlike $^{165}\mathrm{Ho}^{66+}$ at the Livermore EBIT \cite{Ho165} is shown. The HFS in $^{165}\mathrm{Ho}^{66+}$, $^{203,205}\mathrm{Tl}^{80+}$ and  $^{185,187}\mathrm{Re}^{74+}$ were determined 
at the Livermore EBIT with accuracies comparable or better than those   of the laser-fluorescence measurements of $^{209}\mathrm{Bi}^{82+}$ and $^{207}\mathrm{Pb}^{81+}$ at the GSI. The nuclear magnetization radii deduced from EBIT experiments for the rhenium and thallium isotopes can be found in Refs.~\cite{Re185187,Tl203205}.

Last but not least, DR can also be used to determine HFS of highly charged ions. The precision required by the HFS experiments is usually not matched by the typical accuracies of DR electronic transition energies. However, as  already discussed in Section \ref{exIS},  precision can be enhanced using resonances that lie very close to the ionization threshold. With relative velocities between  ion and  recombining electron very close to zero, the accuracy of the free electron energy is substantially increased. While there is no possibility to obtain HFS for H-like ions in DR, which always involves  two electrons, experiments with other isoelectronic sequences offer the possibility to obtain precise transition energies and resolve HFS  for different ion charges where QED, nuclear magnetization distribution and nuclear charge distribution contribute differently.

Determination of the HFS of $4s_{1/2}$ and 
$4p_{1/2}$ in $^{207}\mathrm{Pb}^{53+}$ \cite{EvaPRL95} was possible by comparing DR resonances of $^{207}\mathrm{Pb}^{53+}$ and $^{208}\mathrm{Pb}^{53+}$, since the latter has zero nuclear spin and does not present HFS. The experiments were performed at the heavy-ion storage ring CRYRING in Stockholm, Sweden. The comparison between the observed resonances in the two lead isotopes was performed by taking into account all isotope dependent contributions using relativistic many-body perturbation theory and a multi-configuration Dirac-Fock method \cite{EvaPRL95}. A careful evaluation extracted the HFS with an accuracy on the order of 10$\%$ of the HFS constant. 

DR close to the recombination threshold can actually be used as a very accurate tool to obtain electronic transitions energies. Recently, DR Rydberg resonances below 0.07 eV were determined for $^{45}\mathrm{Sc}^{18+}$ with absolute accuracies below 0.0002 eV, a factor of 10 smaller than previous storage ring studies \cite{Lestinsky}. This remarkable precision (4.6 ppm) in the measured value of the $2s_{1/2}-2p_{3/2}$ transition energy enabled features of the HFS of the $2s$ state to be resolved. In conjunction with accurate atomic structure calculations, this experiment was sensitive to QED screening corrections.  

Finally, DR has been also used to determine the hyperfine-induced $2s2p\ ^3P_0 \rightarrow 2s^2\ ^1S_0$ transition rate in divalent beryllium-like Ti ions \cite{StefanSchippers} at the heavy-ion storage ring TSR in Heidelberg. Hyperfine-induced transitions are forbidden electronic transitions rendered possible by the magnetic interaction with the nuclear spin. For instance, it is well known that one-photon transitions between two states for which the electronic wave functions both have a total angular momentum equal to zero   are forbidden by selection rules of any electric or magnetic multipole order. These transitions can occur, with very low probabilities, via two-photon emission. However, if the nucleus has a spin $I$ different from zero, the hyperfine interaction mixes states with different angular momenta and thus  one-photon transitions are no longer forbidden. Such hyperfine induced transitions are interesting for obtaining ultra precise optical frequency standards and for cold-atom studies, or in astrophysics, where they can be used for diagnostics of low-density plasmas. 

The rate of the hyperfine-induced $2s2p\ ^3P_0 \rightarrow 2s^2\ ^1S_0$ in $^{47}\mathrm{Ti}^{18+}$ was determined by comparing DR spectra of two Ti isotopes, $^{47}\mathrm{Ti}^{18+}$ and $^{48}\mathrm{Ti}^{18+}$, with nuclear spins $I=5/2$ and  $I=0$, respectively. The ions were produced mainly in the ground state $2s^2\ ^1S_0$, with about $5\%$ in the metastable  $2s2p\ ^3P_0$ state. The electron-ion collision energy in the storage ring was tuned to a value where DR occurs only for ions in the excited  metastable $2s2p\ ^3P_0$ state, and the rate of recombined ions $\mathrm{Ti}^{17+}$ was monitored as a function of the storage time. The ions were stored in the ring for about  200~s. During this time the two-photon decay of the metastable state, with an estimated lifetime of approximately 4 days \cite{Ti_metastable}, did not play any role. 

 While for the $^{48}\mathrm{Ti}$ ions the $0 \rightarrow 0$ transition is forbidden, for the $^{47}\mathrm{Ti}$ ions the lifetime of the metastable state is substantially reduced and the DR resonance essentially disappears.   Out of the recombination rates as a function of the ion storage time the hyperfine-induced decay rate of 1.8~s was determined -- about 40$\%$ lower than the theoretical value of 2.8~s. This determination is almost one order of magnitude more precise than the previous experimental value for the same $2s2p\ ^3P_0 \rightarrow 2s^2\ ^1S_0$ transition in $\mathrm{N}^{3+}$ that was obtained from astrophysical observations and modeling \cite{Brage}. Hyperfine quenching of forbidden transitions has also been proposed as a method to probe the nuclear state. In collisions of $^{238}\mathrm{U}$, which has  ground state nuclear spin equal to zero, a fast decay of the $2s2p\ ^3P_0$ state  could provide evidence of the nucleus being in an excited state \cite{Labzowsky}. HFS experiments on excited nuclear states would be  particularly interesting for identification of certain long-lived nuclear states, as pointed out later in Section~\ref{Conclusions}.

\section{Exotic nuclear properties: nuclear excited states and the weak interaction }
So far, we have addressed changes in atomic transitions that are due to the nuclear size, mass and spin. However, 
these are not the only nuclear properties that can play a role in the life of an atom. 
More exotic features such as the nuclear excited states that determine nuclear polarization corrections for the atomic transitions, or of which excitation and decay mechanisms can directly involve atomic electrons, may have an important contribution to atomic transitions. Going to the weak interaction between nuclear constituents, the coupling of nucleons to the atomic electrons gives rise to atomic parity violation effects.  

\subsection{Atomic parity violation}

In the absence of any external field to define a specific orientation in space, atoms do not have any preference whether to interact with left  or right handed circularly polarized photons. Atomic states have a definite parity, and photon absorption or emission is equally probable for both left handed or right handed circularly polarized photons. 

This holds, however, only if the weak interaction between the atomic electrons and the nuclear constituents is neglected. The weak interaction can mix atomic states of different parity, producing observable effects: left  and right handed  photons are absorbed and emitted slightly differently, i.e., the atoms show optical activity,  so-called circular dichroism. Parity is no longer conserved---one refers to parity violation or parity non-conservation in atoms. Such results are in conflict with QED, and support the theory of unification of the electromagnetic and weak interactions.

In quantum mechanics the parity operator $P$  transforms  each point of space $\vec{r}$  into its opposite $-\vec{r}$. The operator $P$ thus performs a space inversion and transforms a quantum state into its mirror image. A parity quantum number is associated to $P$, namely, $p=\pm 1$, given for atomic states by $p=(-1)^{l}$, where $l$ is the orbital quantum number. The parity quantum number is a multiplicative quantum number, for instance, for a system of $N$ independent electrons, it will be the product of the $N$ individual quantum numbers $p_j= (-1)^{l_j}$. For electromagnetic interactions, parity is conserved, meaning that in any transition the products of the parity quantum numbers of the initial and final sub-systems are identical. 

For a long time it was believed that parity conservation is a universal law, since there were no experiments to prove the contrary. In the early 1950s, a puzzling particle physics experiment was the first to question the universality of parity conservation. The K meson was found to decay sometimes in two pions and sometimes in three pions, and due to the $p=-1$ parity of the individual pions, this implied that the two observed final states had different parity! Lee and Yang \cite{LeeYang}  stepped into the unknown and suggested that parity is not conserved in the decay - since they noted that at the time there was no experimental evidence for parity conservation in transitions induced by the weak interaction. One year later, in 1957,  Wu  confirmed experimentally that parity is violated in the nuclear $\beta$ decay \cite{Wu}. 

Shortly afterwards, the idea to search for parity violation in atomic transitions was considered \cite{Zeldovich}.   However, a first estimate of the left-right asymmetry in atomic transitions was so exceedingly small ($10^{-15}$) that its observation  seemed completely hopeless.  The rotation of the polarization plane of visible light propagating through optically inactive matter was estimated to be only $10^{-13}$ rad/m. Luckily, this estimate proved to be by far too pessimistic, and atomic parity violation was experimentally observed some decades later.

Parity violation in the weak interaction was first observed in  nuclear $\beta$ decay, where always  a change in the electric charge and consequently in the identity of the decaying particles occurs.  It was therefore difficult to conceive that the weak interaction and correspondingly its parity violation feature could have any influence in neutral stable atoms, where the identity of the interacting particles is obviously preserved. Zel'dovich ~\cite{Zeldovich}, when giving the first discussion on the possible effects in atomic physics of weak electron-nucleon interaction, referred to so-called neutral currents. 

\begin{figure}
\begin{center}
\includegraphics[width=0.2\textwidth]{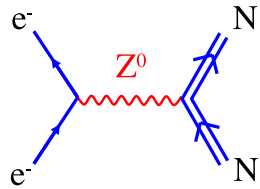}
\end{center}
\caption{ Electroweak interaction between the electron and the nucleus, mediated by the neutral gauge boson $Z^0$.}
\label{APV}
\end{figure}

It took many years for the theoretical understanding of the weak interaction to become mathematically satisfactory---by the late 1960s, Glashow \cite{Glashow}, Weinberg \cite{Weinberg} and Salam and Ward \cite{SalamWard} had suggested independently that the weak and the electromagnetic interactions can be understood as different manifestations of a unified underlying {\it electroweak interaction}. The family of gauge bosons, the particles that mediate the weak interaction, was known to contain the $W^+$ and $W^-$ bosons, both carrying a unit of electric charge. The  $W^{\pm}$ gauge bosons mediate the weak interaction in $\beta$ decay. With the theoretical unification of the weak and electromagnetic interactions, an important prediction was the existence of a new  gauge boson, called $Z^0$, mediating  weak neutral current interactions.
The $Z^0$ boson carries no electric charge, so one way it couples to the fundamental particles such as electrons and quarks is like a heavy photon, as depicted in the Feynman diagram in Figure~\ref{APV}. However, an important aspect of the $Z^0$ boson is that it has chiral behaviour---it changes its coupling to constituents when reflected in a mirror--- similar to its kins the $W^{\pm}$ bosons. The coupling of $Z^0$ to the electron is proportional to the electron helicity $h_e$, a quantity odd under space reflection, defined as the scalar product between the electron spin and velocity, $h_e=\vec{\sigma}_e\cdot \vec{v}_e$. 
The short-range electron-nucleus $Z^0$ exchange potential thus 
 leads to the parity violation effects in atoms. 

The Coulomb interaction and $Z^0$ boson exchange electron-nuclear potentials have certain similarities. In the vicinity of the nucleus (assumed to be pointlike), the Coulomb potential can be expressed as 
\begin{equation}
V_{C}(r_e)=\frac{Ze^2}{r_e}\, ,
\label{VC}
\end{equation}
and the parity-violating potential can be written in the non-relativistic limit as \cite{Bouchiats_review}
\begin{equation}
V_{pv}(r_e)=\frac{1}{2}\mathcal{Q}_Wg^2_{Z^0}\frac{\mathrm{exp}(-M_{Z^0}c\ r_e/\hbar)}{r_e}(\vec{\sigma}_e\cdot \vec{v}_e)/c + \mathrm{h.c.}\, ,
\label{Vpv}
\end{equation}
where $g_{Z^0}$ is a coupling constant whose size, as a consequence of the electroweak unification, is of the order of $e$. The strengths of the Coulomb and electroweak interactions between the electron and the nucleus are given by the $Ze^2$ and $\mathcal{Q}_Wg^2_{Z^0}$ factors, respectively. The quantity $\mathcal{Q}_W$ is therefore known as the {\it  weak charge} of the nucleus---in analogy to the nuclear electric charge. Due to the very small Compton wavelength $\hbar/(M_{Z^0}c)$ of the $Z^0$ boson, it is legitimate to take the limit of an infinite $Z^0$ mass, which then allows replacing the Yukawa potential in Eq.~(\ref{Vpv}) by a Dirac distribution,
\begin{equation}
\lim_{M_{Z^0}\to \infty} \left(\frac{M_{Z^0}c}{\hbar}\right)^2 \frac{\mathrm{exp}(-M_{Z^0}c\ r_e/\hbar)}{r_e}=\delta^3(\vec{r}_e)\, .
\end{equation}
Usually the $Z^0$ boson exchange potential is written with the help of the Fermi constant, $G_F=4\times 10^{-14}\mathrm{Ryd} \times a_0^3$, where $\mathrm{Ryd}$ is the atomic Rydberg constant and $a_0$ the Bohr radius. The parity-violating potential can then be written as
\begin{equation}
V_{pv}(r_e)=\frac{\mathcal{Q}_WG_F}{4\sqrt{2}}[\delta^3(\vec{r}_e)(\vec{\sigma}_e\cdot \vec{v}_e)/c + \mathrm{h.c.}]\, .
\label{VpvF}
\end{equation}
The Coulomb potential in Eq.~(\ref{VC}) is even under space reflection, while the $Z^0$ boson exchange potential in Eq.~(\ref{VpvF}) is odd, giving rise to left-right asymmetry in atomic transitions. 

The basic principle of most parity violation experiments is to compare the transition rate between two states $A$ and $B$ with the one between their mirrored states $\tilde{A}$ and $\tilde{B}$. The left-right asymmetry is defined then as the difference between these rates divided by their sum. With an even amplitude coming from the electromagnetic interaction and the weak amplitude that has an odd contribution, we have the transition rates $P_{L/R}=|A_{em}\pm A_W^{odd}|^2$, which leads to the left-right asymmetry
\begin{equation}
A_{LR}=\frac{P_L-P_R}{P_L+P_R}=2Re(A^{odd}_W/A_{em})\, .
\end{equation}
While at first  $A_{LR}$ was estimated at $10^{-15}$,   which seemed to make observation impossible, Marie-Anne and Claude Bouchiat~\cite{Bouchiats1974} predicted that the electroweak effects grow slightly faster than the cube of the atomic number $Z$. Thus, heavy atoms are most appropriate to look for atomic parity violation, e.g.,  for cesium ($Z$=55) the parity violation effects are increased by a factor of $10^6$. The cesium atom still represents the best compromise between  high $Z$ and  simple atomic structure  allowing for precise atomic structure calculations. Another way to enhance the effect is the study of forbidden atomic transitions such as the $6s_{1/2}-7s_{1/2}$ transition in Cs atoms. This transition is a magnetic dipole, strongly suppressed because of the different principal quantum numbers of the two states. Parity mixing due to the weak interaction induces a small  electric dipole ($E$1) transition amplitude, which has been observed experimentally. The left-right asymmetry in this case can reach  $10^{-4}$. 

Unfortunately, due to the very small oscillator strength of this $6s_{1/2}-7s_{1/2}$ transition, there were a number of practical problems to see any signal. The Bouchiats \cite{Bouchiats1975} suggested a clever technique for enhancing the signal relying on the interference of a Stark-induced mixing of opposite parity states in an atom and the mixing caused by weak neutral currents.
The Stark-induced amplitude works as a lever arm to magnify the neutral current amplitude.  Following this proposal, a number of groups endeavored to
search for weak neutral currents in atomic parity violation experiments, using either the Stark-mixing idea or by
studying the rotation of plane-polarized light as it passes through a gas of atoms. The reader is referred to the comprehensive review in Ref.~\cite{Bouchiats_review}.

The most precise atomic  parity-non-conservation experiment (experimental accuracy of less than 0.4$\%$) \cite{Wood_Science} examined the mixing of
$S$ and $P$ states in atomic cesium. Specifically following the original proposal in Ref.~\cite{Bouchiats1975}, it compared the mixing due to the parity-violating neutral weak current interaction
to the $S$-$P$ mixing caused by an applied electric field.  The measurements were interpreted in
terms of the nuclear weak charge $\mathcal{Q}_W$, quantifying the strength of the electroweak coupling between atomic electrons
and quarks of the nucleus. The main goal is to compare this experimental result with the lowest order predictions of the Standard Model for the weak charge, where $\mathcal{Q}_W$ is the sum of all the weak charges of the atomic nucleus constituents, the $u$ and $d$ quarks,
\begin{equation}
\mathcal{Q}_W=(2Z+N)\mathcal{Q}_W(u)+(Z+2N)\mathcal{Q}_W(d)\, .
\end{equation}
The weak charge happens to lie close to the neutron number $N$,
\begin{equation}
\mathcal{Q}_W=-N-Z(4 \sin^2\theta_W-1)\simeq-N\, ,
\end{equation}
where the weak mixing angle $\theta_W$ is a free parameter of the theory, determined experimentally to be $\sin^2\theta_W=0.23$.

In order to obtain the experimental value of $\mathcal{Q}_W$, atomic structure calculations for the amount of
Stark mixing and the relevant parity-violating electronic matrix elements are needed.
The  uncertainties in the determination of $\mathcal{Q}_W$ are actually  dominated by the uncertainties in these two calculated
quantities. In Ref.~\cite{BennettPRL} the evaluation of the experimental data in Ref. \cite{Wood_Science} was improved by also determining experimentally the Stark mixing and incorporating further experimental results into the
evaluation of the uncertainty in the calculation of the relevant parity-violating  matrix elements. The most
accurate to-date determination of this coupling strength by
combining these two measurements  using  high-precision
calculations for the cesium atom yield a weak charge of $\mathcal{Q}_W=-73.16(29)_{\mathrm{exp}}(20)_{\mathrm{theor}}$ \cite{Porsev}, which is in excellent agreement with the predictions of the Standard Model.

Atomic parity-non-conservation measurements are uniquely sensitive to a variety of new physics, such as
the existence of additional $Z$ bosons,  because they probe a different set of
model-independent electron-quark coupling constants than
those measured by high-energy experiments, in a different energy regime \cite{Langacker}.   Given the huge effort
made in testing the Standard Model in high-energy collider experiments, the results of the
atomic parity violation measurements represent a significant triumph for table-top physics. Improved precision of the atomic experiments, possibly to be achieved by measurements for several atoms along an isotopic chain, as a way to circumvent the atomic theory uncertainties \cite{Musolf}, is expected to keep  parity-violation experiments in atoms interesting and useful for the continuing quest for new physics beyond the Standard Model.

\subsection{Nuclear polarization effects\label{np}}
With increased precision of atomic physics experiments and QED as a powerful tool to calculate radiative corrections up to high orders in $\alpha$, one would be tempted to think that it is only a matter of time until excellent agreement between theory and experiment will be achieved. This is however not the case, since there are also other corrections that have to be accounted for. The nucleus itself is one of the main sources of undesirable interferences. At a certain level of accuracy, it is no longer enough to consider the nucleus as source of the classical external Coulomb field modeled by a classical charge density distribution. One should also take into account the interaction between the radiation field and the nucleus due to intrinsic nuclear dynamics. Due to the exchange of virtual photons, the nucleus can undergo virtual transitions  to  excited states, giving rise to  energy shifts of electronic  states. These contributions are called nuclear polarization  effects. The nuclear polarization contribution is always present  in atomic transition energies and represents a natural limitation for any high-precision tests of QED. As already mentioned in Section~\ref{exIS}, nuclear polarization is often the largest uncertainty factor when it comes to identifying the various contributions to the measured isotope shifts. Nonetheless, one can imagine reaching a level of experimental and theoretical accuracy which allows testing the influence of internal nuclear structure as well as higher-order QED effects. Until then, the nuclear polarization corrections have to be calculated theoretically up to a certain level of accuracy and then used in the complicated procedure of extracting from the experimental data the atomic or nuclear quantity of interest.

\begin{figure}
\begin{center}
\includegraphics[width=0.45\textwidth]{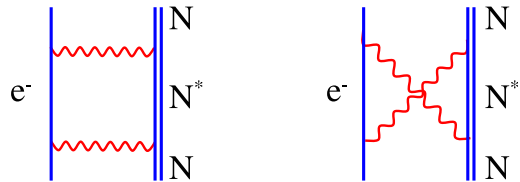}
\end{center}
\caption{ The ladder and cross diagrams contributing to  nuclear polarization in the lowest order. Here $e^{-}$ denotes the electron and $N$ and $N^*$ the ground state and excited nucleus, respectively.}
\label{NPdiagrams}
\end{figure}

Some general remarks are due here. The nuclear polarization correction arises from virtual transitions  as  shown in the Feynman diagrams in Figure~\ref{NPdiagrams}. Since  in the  nuclear polarization energy shift expression the energy difference $E_a-E_n$ between the atomic and nuclear virtual transitions appears in the denominator \cite{PlunienPRA43}, the closer the two energies are, the larger the nuclear polarization correction. This explains why 
the first systems for which nuclear polarization effects have been  studied are the muonic atoms, for which the energy corrections can be as large as  several keV. In atomic spectra, on the other hand,  nuclear polarization corrections  have been disregarded  for a long time as being too small to be experimentally observed. The muon transition energies are of the order of MeV, on the same scale as  typical nuclear transitions. This does not hold for atomic transitions, where the transition energies are usually below or around 100 keV, thus much smaller than the nuclear excitation energies. Exceptionally, there are a number of low-lying rotational and vibrational bands, mostly in even-even nuclei, which can also reach values below 100 keV.  Some very strong  nuclear transitions in the MeV range, called  giant dipole resonances, also play an important role for nuclear polarization corrections, not because of their match in energy, but because of their large resonance strength. Unfortunately, since  nuclei have a complex inner structure and their excited level schemes present more exceptions than rules, for any particular nucleus under consideration an adequate choice of the relevant excitations and nuclear models to describe them is necessary. One usually considers only those having strong excitation strengths, that are expected to contribute considerably.

Collective nuclear models are the simplest alternative, describing collective nuclear rotational and vibrational states as well as giant dipole resonances. These phenomenological collective models of the nucleus have as underlying physical picture the classical charged liquid drop, disregarding  the interior structure  (i.e. the existence of the individual nucleons) in favor of the picture of a homogeneous fluid-like nuclear matter. The model  is applicable only if the size of the nucleon can be neglected with respect to the size of the nucleus as a whole, as it happens in the case of heavy nuclei. Furthermore, collective models are only appropriate to describe even nuclei. For low-lying single-particle excitations of a valence proton in odd nuclei, nuclear polarization calculations rely on the Weisskopf approximation \cite{PlunienPRA51} and require a different approach to calculate the effective electron-nucleus interaction.

A review of the theory of nuclear polarization in muonic atoms can be found in Ref.~\cite{BorRink84}. A relativistic, field-theory approach to nuclear polarization calculations for electronic heavy atoms has been put forward by Mohr, Plunien and Soff \cite{MPSoff}. With the introduction of an effective photon propagator with nuclear polarization insertions, the nuclear polarization corrections are treated on equal footage as  radiative corrections to the electron energy, and appear in an effective self-energy. It is beyond the scope of this work to present the full theoretical treatment of nuclear polarization corrections to atomic transition energies. Instead, the reader is referred to the detailed description of the effective photon propagator technique in Ref.~\cite{PlunienPRA43}. For very light atoms such as deuterium, the theory of nuclear polarization has been developed in Refs.~\cite{PachuckiPRA48,PachuckiPRA49}.

Because of the unreliable nuclear part, any calculation of nuclear polarization shifts is inherently phenomenological and depends on the parameters of the nuclear model used. On average a typical error of $25\%$ is assumed, most of it coming from the uncertainty in the nuclear parameters, e.g., the nuclear transition energies and strengths. The choice of relevant nuclear states is  another source of uncertainty, since  among the neglected ones  some important contributor could have been missed.  

The nuclear polarization corrections are usually small, on the order of 200 meV for the $K$ shell electron in $^{238}_{92}\mathrm{U}$ (compared to the 200~eV shift due to the extended nuclear size).  An evaluation of the nuclear polarization energy shifts for the $K$ shell of $^{238}_{92}\mathrm{U}$ and $^{208}_{82}\mathrm{Pb}$ \cite{PlunienPRA43}  shows that  for $^{208}_{82}\mathrm{Pb}$, with the first excited nuclear state already as high as 2614 keV, the nuclear polarization shift is three orders of magnitude smaller than in $^{238}_{92}\mathrm{U}$ \cite{Yamanaka}, which has a low-lying rotational band with a number of excited nuclear levels below 600 keV, the first   at 44 keV. Nuclear polarization corrections have been calculated mostly for  systems for which the interpretation of  experimental data required high accuracy in theoretical calculations and inclusion off all contributions in the atomic transition energy shifts. Among these are the nuclear polarization correction for the recently studied halo nucleus $^{11}\mathrm{Li}$ \cite{NPLi11}, or for various helium isotopes \cite{NPHe}, the nuclear polarization correction to the bound-electron g-factor in heavy H-like ions \cite{PlunienGFactor}, or for QED tests  in heavy, highly-charged ions \cite{BiXray,QEDNP}.

Unfortunately, small as the nuclear polarization corrections are, they remain a theoretical challenge and a  source of uncertainty for atomic structure calculations, being the major accuracy limitation for the determination of important nuclear parameters such as the RMS radii from atomic transition energies.

\subsection{Nuclear excitation mechanisms involving atomic electrons}
In some cases,  excited nuclear states can  directly drive atomic transitions.
It is well known that for low-lying nuclear excited states, a frequent and sometimes preferred decay channel is internal conversion  (IC). Instead of emitting a $\gamma$ photon, the nucleus transfers  its excess energy and angular momentum   to a bound  electron, which leaves the atom. IC and $\gamma$ decay are usually competing nuclear decay channels. Their ratio defines the internal conversion coefficient, $\alpha=A_{IC}/A_{\gamma}$, where $A_{IC}$ and $A_{\gamma}$ are the IC and $\gamma$ decay rates, respectively. IC is predominant for small transition energies ($\lesssim 100$~keV) and is the primary decay channel for $0^+\rightarrow 0^+$ forbidden transitions whose energy does not allow decay via pair production.

\begin{figure}
\begin{center}
\includegraphics[width=0.3\textwidth]{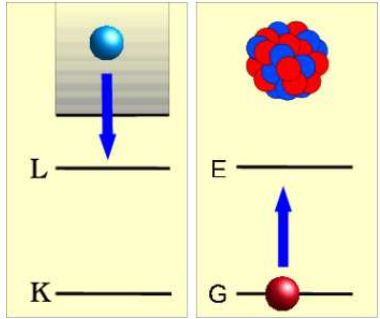}
\end{center}
\caption{The inverse process of internal conversion: NEEC. If the atomic and nuclear transition energies match, an electron can recombine into an ion (left panel) with the simultaneous excitation of the nucleus (right panel) from the ground state $G$ to the excited state $E$.} 
\label{neec}
\end{figure}

The inverse process of IC can occur only when ions and free electrons come across, i.e., in a process of electron recombination. The two most important channels of electron recombination have been discussed when introducing DR: if the electron capture into the atomic bound state occurs with the emission of a photon, one speaks of radiative recombination. When the capture occurs in the presence of a bound electron, at the resonance energy that permits recombination with the simultaneous excitation of the bound electron in the ion, one refers to dielectronic recombination. The inverse process of IC, {\it nuclear excitation by electron capture} (NEEC), is the  nuclear physics analogue of DR, in which the place of the bound electron is taken by the nucleus. If the electronic and nuclear energy levels match, the recombination can take place with the simultaneous excitation of the nucleus, as shown schematically in Figure \ref{neec}.
The nuclear excited state decays then radiatively or can undergo internal conversion.  In contrast to DR, NEEC can also occur in bare ions, as the presence of a bound electron is not required.

IC and NEEC also have  more exotic siblings, the {\it bound} internal conversion (BIC) and its inverse process nuclear excitation by electron transition (NEET). Bound IC is a resonant nuclear decay channel which may occur if the nuclear excitation energy is not enough to ionize the atomic electron, but can induce a transition between two bound electronic states.  Nuclear excitation by electron transition is  a simultaneous excitation of the nucleus  during an atomic decay transition when the energies of the two transitions match. There are only very few atoms in which such a match of the atomic and nuclear transition energies exist, see for instance the list of candidates for NEET in \cite{NEET_list}. 
Although it is difficult to find  systems that fulfill the  resonance condition, BIC and NEET have already been observed experimentally in the year 2000 \cite{Carreyre,Kishimoto}. IC and NEEC together with BIC and NEET couple atomic and nuclear transitions and convert electronic orbital energy into nuclear energy and vice versa. They offer therefore the possibility to explore the spectral properties of heavy
nuclei, such as  nuclear transition energies and strengths, through atomic physics experiments.  NEEC and NEET are also expected to allow the study of atomic vacancy effects on nuclear level population and lifetimes.

That the atomic configuration can influence the nucleus has been long known. Maybe the most spectacular case arises when a bound $\beta$ decay channel is opened in  highly charged ions. In $\beta$ decay the nucleus of an atom emits an electron into the continuum. If the nuclear decay lacks the energy to emit the  electron into the continuum, the $\beta$ decay channel is closed. This can change dramatically if instead of an  atom  we consider a highly charged ion. The $\beta$ decay electron is then emitted in a bound state, requiring less energy for the nuclear transition. The opening of the new bound $\beta$ decay channel in highly charged ions considerably  influences the lifetime of unstable levels in nuclei \cite{Takahashi1,Takahashi2}.  As an example, the lifetime of the   ground state $^{187}_{75}\mathrm{Re}$ decreases by more than nine orders of magnitude  from 42~Gyr for the neutral atom to 32.9 yr for bare ions as a consequence of new bound $\beta$ decay branches to the ground and excited states of the $^{187}_{76}\mathrm{Os}$ daughter \cite{Bosch2,Bosch1}. The case  of $^{187}_{75}\mathrm{Re}$ is particularly interesting in astrophysical context, since  it severely affects the accuracy of the $^{187}\mathrm{Re}$-$^{187}\mathrm{Os}$ cosmochronometer \cite{Bosch1}. The motivation in investigating the behavior of nuclei in highly charged ions is thus related to  nuclear astrophysics, studies of nuclear decay properties and tests of the  relativistic description of atomic inner-shell processes. For highly charged ions, also NEEC is expected to play an important role for the nuclear lifetimes and the  excited-state  population in high-density astrophysical plasma environments.

Historically, the NEEC recombination mechanism has been proposed for the first time  for laser-assisted plasma environments in 1976 in Ref. \cite{Goldanskii}. While this original proposal was followed by a number of theoretical studies of the process, in plasmas \cite{Goldanskii,Harston}, solid targets \cite{Cue,Kimball1,Kimball2} or for highly charged ions  in a relativistic regime \cite{PalffyPRA73}, NEEC has not been observed experimentally yet. The theory of NEEC in highly charged ions involves accurate calculations of the atomic structure together with a phenomenological description of the nucleus. The free electron and the nucleus interact via the Coulomb field (mostly responsible when describing electric multipole transitions of the nucleus) or via a current-current interaction (necessary to picture magnetic multipole nuclear transitions). Both the electron-nucleus interaction and the phenomenological description of the nuclear dynamics via collective models for NEEC are very similar to nuclear polarization calculations \cite{PalffyPRA73}. The cross section for the two-step process of  NEEC followed by the $\gamma$ decay of the nucleus as a function of the recombined electron energy $E$ can be separated as
\begin{equation}
\sigma(E) = \frac{2\pi^2}{p^2}
\frac{A_{\gamma} Y_n}{\Gamma_n} L_n(E-E_0) \,,
\label{CS_NEEC}
\end{equation}
where $Y_n$ is the NEEC rate and $A_{\gamma}$ the $\gamma$ decay rate of the excited nucleus. Furthermore, $p$ is the continuum electron momentum, $\Gamma_n$ the width of the nuclear excited state and $ L_n(E-E_0)$ a Lorenzian profile
\begin{equation}
L_n(E-E_0) = \frac{\Gamma_n / 2\pi}{(E-E_0)^2 + \frac{1}{4} \Gamma_n^2}
\end{equation}
centered on the resonance energy $E_0$ and with the width given by the width of the nuclear resonance. Due to the very narrow natural width of the nuclear excites states $\Gamma_n$, on the order of $10^{-5}$ to $10^{-8}$~eV, the Lorenzian is 
almost a $\delta$ peak, and the NEEC cross section is non-vanishing only very close to the resonance. Since  an electron energy resolution on the sub-meV scale is not available in present experiments, one should convolute the cross section in Eq.~(\ref{CS_NEEC}) with a realistic electron energy distribution to simulate experimental data. 

\begin{figure}
\begin{center}
\includegraphics[width=0.60\textwidth]{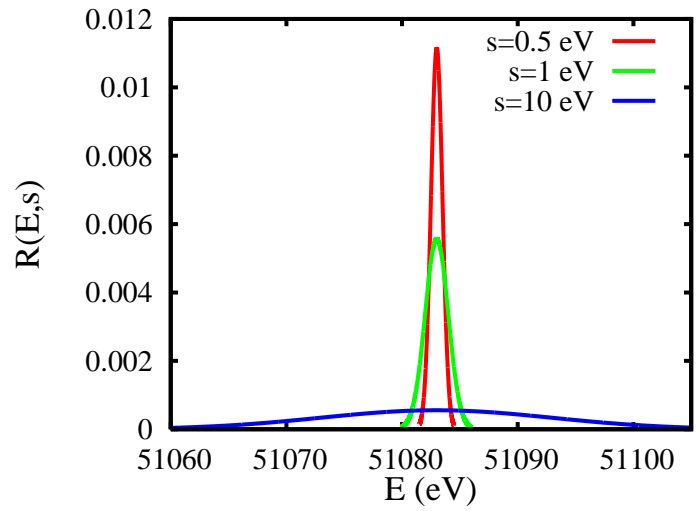}
\end{center}
\caption{The ratio $R(E,s)$ in Eq.~(\ref{r_s}) for recombination into
bare rhenium as a function of the energy of the continuum electron for
three different experimental electron energy width parameters $s$ \cite{interference}. See
text for further explanations.} 
\label{sigback}
\end{figure}

A major difficulty in observing  NEEC experimentally arises from the  background of atomic photorecombination processes, in particular from the non-resonant channel of RR.
Due to identical initial and final states of NEEC followed by $\gamma$ decay of the nucleus and RR, and to the dominance of the latter,  it is
practically impossible to distinguish between the two 
processes \cite{interference}. The RR photon yields for typical experimental conditions exceed the ones of the $\gamma$ photons following NEEC by orders of magnitude, resulting in a signal-to-background ratio of less than $10^{-3}$ even for the most promising cases \cite{interference}.
 The convoluted signal to background ratio 
\begin{equation}
R(E,s)=\frac{\tilde{\sigma}_{\rm
NEEC}(E,s)}{\tilde{\sigma}_{\rm RR}(E,s)} 
\label{r_s}
\end{equation}
for the $M1$ nuclear transition from the ground state to the first excited state at $E_n$=125.358~keV  
with the capture of a free electron into the $K$ shell of a bare  $^{185}_{75}\mathrm{Re}$ ion as a function of the free electron energy can be seen in Figure~\ref{sigback}.  The energy
distribution of the incoming electrons is assumed to be described by a
Gaussian function with the width~$s$. The RR cross section has
a practically constant value on the energy interval of~$s$.
Figure ~\ref{sigback} shows the signal to background ratio for 
  three different experimental width parameters $s=0.5$~eV,
$1$~eV and $10$~eV. While for a width parameter $s=0.5$~eV the
contributions of the NEEC and interference terms can be clearly
discerned from the RR background, for presently more realistic widths on
the order of  eVs or tens of eV the values of the ratio $R(E,s)$
are too small to be observed experimentally.

There are at present two ideas how to tackle the problem of observing experimentally NEEC, or, to be more specific, how to separate the NEEC signal from the RR background. One of them relies on a specific storage ring experimental setup. For NEEC to occur, atomic vacancies are needed, i.e., ions. Usually, the higher the ion charge, the larger is the NEEC cross section, such that experiments with highly charged ions are desirable. Highly charged ions can be produced in accelerator facilities and then transferred and stored in storage rings, where they cycle at almost the speed of light. For recombination experiments, the ion beam is driven through an electron target. In practice,
the different time scales on which NEEC and RR occur can be
useful for eliminating the background in a storage ring experiment where the ions are cycling in the ring at nearly speed of light. While RR is practically instantaneous and its photon signals last typically for $10^{-14}$~s,  NEEC involves intermediate excited nuclear states which decay after $10^{-10}$~s or even later following the recombination.  While the RR photons
will be emitted instantaneously in the region of the electron
target, the radiative decay of the nucleus will occur
later, after the ions have already traveled a certain distance in
the storage ring. Considering typical experimental parameters of the GSI ESR, 
with the ion velocity being 71$\%$ of the speed of light, typical nuclear excited state lifetimes of
10-100~ns correspond to 2-20~m distance between the prompt and delayed photon emission. The electron cooler is used as a target for free electrons and can be tuned to match the resonance condition for NEEC. 
Ions recombined in the cooler are separated from the primary beam in the first bending magnet of the storage ring and can be counted with an efficiency close to unity. By choosing isotopes with convenient excited state lifetimes on
the order of 10-100 ns, the spatial separation can be
determined such that a direct observation of the $\gamma$ photons following
NEEC could be performed with almost complete suppression  of the RR
background \cite{PalffyPLB}.

\begin{figure}
\begin{center}
\includegraphics[width=0.45\textwidth]{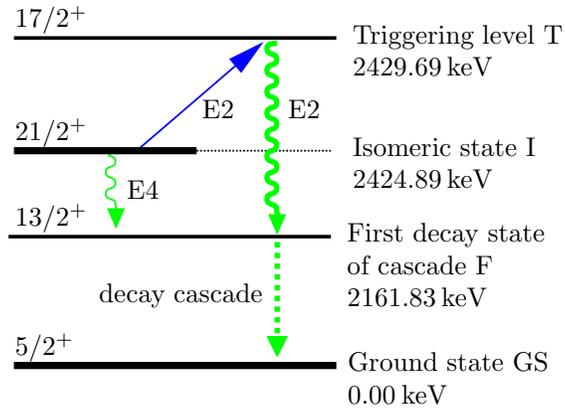}
\end{center}
\caption{Partial level scheme of $^{93}_{42}\mathrm{Mo}$. The isomeric state ($I$) can be excited (for instance via NEEC) to the triggering level ($T$) which subsequently decays back to $I$ or to a level $F$, initiating a cascade via different intermediate states (dashed line) to the ground state ($GS$) \cite{PalffyJMO}. The direct $I\to F$ decay is a strongly hindered $E4$ transition, while $I\to T$  and $T\to F$ are $E2$ transitions.} 
\label{isomerscheme}
\end{figure}

Another approach aims at separating the signal and background photons in energy. Typically NEEC experiments and theoretical studies envisaged excitation of the nucleus from its ground state to the first excited state. However, one can also consider NEEC from a long-lived excited state, for instance an isomeric state. Metastable nuclear states (nuclear isomers) are an interesting topic in nuclear physics, since it is often not clear why they exist, and which isomer excitation and decay mechanisms are prevailing in different environments \cite{walker,nat_phys}. In an advantageous configuration of the nuclear excited levels (see Figure~\ref{isomerscheme}), if NEEC occurs from the isomeric state to an upper {\it triggering} level, the subsequent nuclear decay is possible also to a state below the isomer. Such a process is called {\it isomer triggering} or {\it depletion}, since it allows the depopulation of the isomeric state and thus a controlled release of the energy stored in the metastable state. The advantage 
when considering isomer triggering via NEEC is that the energies of the excitation from the isomeric state (equal to the RR background photons) and of 
the photons emitted in the decay of the triggering level (which are the signature that NEEC occurred) are different, as shown in Figure~\ref{isomerscheme}, 
and thus the experimental background due to RR is much reduced.

The search for practical methods to trigger isomeric states has been the subject 
of a number of sometimes controversial investigations in the last decades. 
Major motivation for this was the challenge of understanding the formation and the role 
of nuclear isomers in the creation of the elements in the 
universe. However, the interest has been not least generated by 
a number of fascinating potential applications related to the controlled release of nuclear energy on demand, such as nuclear batteries that would operate without fission or fusion.
A comparison between different triggering mechanisms, i.e., different ways to reach the triggering level from the metastable state, has surprisingly showed that NEEC is the most efficient way 
of releasing the isomer energy on demand when low-lying triggering levels are used \cite{PalffyPRL}. These are preferable for obtaining high energy gain and facilitating the excitation to the triggering level. 
An experimental verification of NEEC and isomer triggering via NEEC at the borderline 
of atomic and nuclear physics would be therefore very desirable and will be hopefully  rendered possible by upcoming
ion storage ring facilities and ion beam traps 
commencing operation in the near future.


\section{Conclusions\label{Conclusions}}
In atoms,  nucleus and  electrons are bound together and get to step on each other's toes. 
The electrons are sensitive to the nuclear mass, charge distribution, magnetization distribution,  spin, and finally also to the nuclear excited levels that can be (virtually) populated. In the search to better understand  the underlying physics, 
the last decades have witnessed a substantial increase of precision in atomic physics experiments, supported by  more and more accurate theoretical calculations of electronic energies.  A  summary of the discussed nuclear properties and their effects are given in Table \ref{summary}. It is usually a challenging  task to extract nuclear parameters out of atomic physics experimental findings, since most of the time several  effects play a role, and each contribution has to be identified accordingly. Successful approaches often involve putting together theory and data obtained with different complementary experimental techniques, many of which have been outlined here.

Strong efforts  towards improving both experimental precision and theoretical accuracy in atomic and nuclear  physics are needed. With the commissioning of new accelerator facilities such as the Facility for Antiproton and Ion Research (FAIR) \cite{FAIR} at the GSI, Germany, and the Heavy Ion Research Facility in Lanzhou (HIRFL) \cite{HIRFL}, China, as well as the increasing number of  EBIT facilities around the world, many new challenging high-precision experiments with highly charged ions are in sight. A very exciting perspective is opened by the Radioactive Ion Beam facilities, such as the one of the Michigan State University, USA, TRIUMF in Canada or in RIKEN, Japan, which offer the possibility to investigate the atomic and nuclear properties  of nuclides far from the valley of stability in ionic states. Nuclei not yet available for experiments could be produced in quantities that allow  determination of their properties. Halo nuclei have already raised a question mark on whether our nuclear structure theory at present is applicable only to  the limited number of nuclei along the stability line. Just as in the case of halo nuclei, atomic physics experiments with  unstable nuclei might bring new insights in the structure of the nucleus. 

Challenging experiments such as the observation of NEEC or the determination of the nuclear lifetime and properties of the $^{229\mathrm{m}}\mathrm{Th}$ isomer will probably  benefit of the increasing high precision in atomic physics experiments. The $^{229\mathrm{m}}\mathrm{Th}$ isomer lies very close to the ground state, the latest determination of the excitation energy being 7.6~eV \cite{Th229m}. For the decay of the isomeric state, IC is orders of magnitude more probable than $\gamma$ decay, and one cannot hope to determine the transition energy or the isomer lifetime by detecting the  low-energy photon. Instead, one can probe and identify the nuclear state by determining the HFS of the electronic levels, taking advantage of the different spins and magnetic moments of the ground and isomeric states. The determination of the nuclear excited state HFS is demanding but promising for this case, with a number of groups around the world engaged in this project. 

Finally, high-precision atomic physics experiments with neutral atoms in parity-violation measurements on different atoms aim at probing the electroweak interaction and testing the Standard Model in search of new physics. A key ingredient is the reliable and accurate theoretical input. Whether the
atomic theory errors can be reduced to the level of the experimental
uncertainty remains a challenge. An experimental
strategy for circumventing the atomic theory uncertainties is to
measure parity-violating observables for different atoms along an isotopic
chain. Standard Model predictions for ratios of such observables
are largely independent of atomic theory. The next decades may show  how productive the studies at the borderline between atomic and nuclear physics still are and what can atomic transitions reveal about the nucleus, its structure and its properties.

\section*{Acknowledgments}
I would like to thank Klaus Blaum, Stefan Schippers, Zolt\'an Harman and Christoph H. Keitel for many fruitful discussions at the borderline between atomic and nuclear physics. I am deeply indebted to Jos\'e Crespo~L{\'o}pez-Urrutia for carefully reading the manuscript and for technical support with the experimental data figures.

This article contains figures which have been reprinted with permission from the American Physical Society. 
Readers may view, browse, and/or download material for temporary copying purposes only, provided these uses are for 
non-commercial personal purposes. Except as provided by law, this material may not be further reproduced, distributed, transmitted, modified, adapted, performed, displayed, published, or sold in whole or part, without prior written permission from the American Physical Society.

\section*{Short biography}

Adriana P\'alffy has studied physics in Bucharest, Romania, and received her PhD in theoretical physics at the Justus Liebig University in Giessen, Germany. Since 2006 she has been a research fellow at the Max Planck Institute for Nuclear Physics in Heidelberg. Among her research interests, covering the interface between atomic and nuclear physics and quantum optics, are nuclear excitation mechanisms involving atomic electrons, strong laser-nucleus interactions or coherence effects in nuclei.

\bibliography{palffyarxiv}

\begin{thebibliography}{120}
\expandafter\ifx\csname natexlab\endcsname\relax\def\natexlab#1{#1}\fi
\expandafter\ifx\csname bibnamefont\endcsname\relax
  \def\bibnamefont#1{#1}\fi
\expandafter\ifx\csname bibfnamefont\endcsname\relax
  \def\bibfnamefont#1{#1}\fi
\expandafter\ifx\csname citenamefont\endcsname\relax
  \def\citenamefont#1{#1}\fi
\expandafter\ifx\csname url\endcsname\relax
  \def\url#1{\texttt{#1}}\fi
\expandafter\ifx\csname urlprefix\endcsname\relax\def\urlprefix{URL }\fi
\providecommand{\bibinfo}[2]{#2}
\providecommand{\eprint}[2][]{\url{#2}}

\bibitem[{\citenamefont{Bransden and Joachain}(2003)}]{BranJoach}
\bibinfo{author}{\bibfnamefont{B.}~\bibnamefont{Bransden}} \bibnamefont{and}
  \bibinfo{author}{\bibfnamefont{C.}~\bibnamefont{Joachain}},
  \emph{\bibinfo{title}{Physics of Atoms and Molecules}}
  (\bibinfo{publisher}{Pearson Education}, \bibinfo{address}{Harlow, England},
  \bibinfo{year}{2003}), \bibinfo{edition}{2nd} ed.

\bibitem[{\citenamefont{Hughes and Eckart}(1930)}]{HughEck}
\bibinfo{author}{\bibfnamefont{D.}~\bibnamefont{Hughes}} \bibnamefont{and}
  \bibinfo{author}{\bibfnamefont{C.}~\bibnamefont{Eckart}},
  \bibinfo{journal}{Phys. Rev.} \textbf{\bibinfo{volume}{36}},
  \bibinfo{pages}{694} (\bibinfo{year}{1930}).

\bibitem[{\citenamefont{Palmer}(1987)}]{Palmer}
\bibinfo{author}{\bibfnamefont{C.}~\bibnamefont{Palmer}}, \bibinfo{journal}{J.
  Phys. B: At. Mol. Phys.} \textbf{\bibinfo{volume}{20}}, \bibinfo{pages}{5987}
  (\bibinfo{year}{1987}).

\bibitem[{\citenamefont{Fricke}(1973)}]{FrickePRL}
\bibinfo{author}{\bibfnamefont{B.}~\bibnamefont{Fricke}},
  \bibinfo{journal}{Phys. Rev. Lett.} \textbf{\bibinfo{volume}{30}},
  \bibinfo{pages}{119} (\bibinfo{year}{1973}).

\bibitem[{\citenamefont{Shabaev}(1988)}]{ShabaevYF}
\bibinfo{author}{\bibfnamefont{V.~M.} \bibnamefont{Shabaev}},
  \bibinfo{journal}{Yad. Fiz.} \textbf{\bibinfo{volume}{47}},
  \bibinfo{pages}{107} (\bibinfo{year}{1988}).

\bibitem[{\citenamefont{Shabaev}(1998)}]{ShabaevPRA57}
\bibinfo{author}{\bibfnamefont{V.~M.} \bibnamefont{Shabaev}},
  \bibinfo{journal}{Phys. Rev. A} \textbf{\bibinfo{volume}{57}},
  \bibinfo{pages}{59} (\bibinfo{year}{1998}).

\bibitem[{\citenamefont{Shabaev}(2002)}]{ShabaevPR}
\bibinfo{author}{\bibfnamefont{V.~M.} \bibnamefont{Shabaev}},
  \bibinfo{journal}{Phys. Rep.} \textbf{\bibinfo{volume}{356}},
  \bibinfo{pages}{119} (\bibinfo{year}{2002}).

\bibitem[{\citenamefont{Soria~Orts et~al.}(2006)\citenamefont{Soria~Orts,
  Harman, Crespo~L{\'o}pez-Urrutia, Artemyev, Bruhns,
  Mart{\'i}nez~Gonz{\'a}lez, Jentschura, Keitel, Lapierre, Mironov
  et~al.}}]{SoriaOrts}
\bibinfo{author}{\bibfnamefont{R.}~\bibnamefont{Soria~Orts}},
  \bibinfo{author}{\bibfnamefont{Z.}~\bibnamefont{Harman}},
  \bibinfo{author}{\bibfnamefont{J.~R.}
  \bibnamefont{Crespo~L{\'o}pez-Urrutia}},
  \bibinfo{author}{\bibfnamefont{A.~N.} \bibnamefont{Artemyev}},
  \bibinfo{author}{\bibfnamefont{H.}~\bibnamefont{Bruhns}},
  \bibinfo{author}{\bibfnamefont{A.~J.}
  \bibnamefont{Mart{\'i}nez~Gonz{\'a}lez}},
  \bibinfo{author}{\bibfnamefont{U.~D.} \bibnamefont{Jentschura}},
  \bibinfo{author}{\bibfnamefont{C.~H.} \bibnamefont{Keitel}},
  \bibinfo{author}{\bibfnamefont{A.}~\bibnamefont{Lapierre}},
  \bibinfo{author}{\bibfnamefont{V.}~\bibnamefont{Mironov}},
  \bibnamefont{et~al.}, \bibinfo{journal}{Phys. Rev. Lett.}
  \textbf{\bibinfo{volume}{97}}, \bibinfo{pages}{103002:1}
  (\bibinfo{year}{2006}).

\bibitem[{\citenamefont{{\c{S}}chiopu et~al.}(2004)\citenamefont{{\c{S}}chiopu,
  Harman, Scheid, and Gr{\"u}n}}]{Roxana}
\bibinfo{author}{\bibfnamefont{R.}~\bibnamefont{{\c{S}}chiopu}},
  \bibinfo{author}{\bibfnamefont{Z.}~\bibnamefont{Harman}},
  \bibinfo{author}{\bibfnamefont{W.}~\bibnamefont{Scheid}}, \bibnamefont{and}
  \bibinfo{author}{\bibfnamefont{N.}~\bibnamefont{Gr{\"u}n}},
  \bibinfo{journal}{Eur. Phys. J. D} \textbf{\bibinfo{volume}{31}},
  \bibinfo{pages}{21} (\bibinfo{year}{2004}).

\bibitem[{\citenamefont{Sch{\"u}ler and
  Keyston}(1931{\natexlab{a}})}]{SchKeyTa}
\bibinfo{author}{\bibfnamefont{H.}~\bibnamefont{Sch{\"u}ler}} \bibnamefont{and}
  \bibinfo{author}{\bibfnamefont{J.~E.} \bibnamefont{Keyston}},
  \bibinfo{journal}{Z. Phys.} \textbf{\bibinfo{volume}{70}}, \bibinfo{pages}{1}
  (\bibinfo{year}{1931}{\natexlab{a}}).

\bibitem[{\citenamefont{Sch{\"u}ler and
  Keyston}(1931{\natexlab{b}})}]{SchKeyHg}
\bibinfo{author}{\bibfnamefont{H.}~\bibnamefont{Sch{\"u}ler}} \bibnamefont{and}
  \bibinfo{author}{\bibfnamefont{J.~E.} \bibnamefont{Keyston}},
  \bibinfo{journal}{Z. Phys.} \textbf{\bibinfo{volume}{72}},
  \bibinfo{pages}{423} (\bibinfo{year}{1931}{\natexlab{b}}).

\bibitem[{\citenamefont{Fricke et~al.}(1995)\citenamefont{Fricke, Bernhardt,
  Heilig, Schaller, Schellenberg, Shera, and de~Jager}}]{FrickeADNDT60}
\bibinfo{author}{\bibfnamefont{G.}~\bibnamefont{Fricke}},
  \bibinfo{author}{\bibfnamefont{C.}~\bibnamefont{Bernhardt}},
  \bibinfo{author}{\bibfnamefont{K.}~\bibnamefont{Heilig}},
  \bibinfo{author}{\bibfnamefont{L.~A.} \bibnamefont{Schaller}},
  \bibinfo{author}{\bibfnamefont{L.}~\bibnamefont{Schellenberg}},
  \bibinfo{author}{\bibfnamefont{E.~B.} \bibnamefont{Shera}}, \bibnamefont{and}
  \bibinfo{author}{\bibfnamefont{C.~W.} \bibnamefont{de~Jager}},
  \bibinfo{journal}{At. Dat. Nucl. Dat. Tabl.} \textbf{\bibinfo{volume}{60}},
  \bibinfo{pages}{177} (\bibinfo{year}{1995}).

\bibitem[{\citenamefont{Otten}(1989)}]{Otten}
\bibinfo{author}{\bibfnamefont{E.~W.} \bibnamefont{Otten}},
  \emph{\bibinfo{title}{Nuclei far from stability}}, vol.~\bibinfo{volume}{8}
  of \emph{\bibinfo{series}{Treatise of Heavy-Ion Science}}
  (\bibinfo{publisher}{Plenum}, \bibinfo{address}{New York},
  \bibinfo{year}{1989}).

\bibitem[{\citenamefont{Shabaev}(1993)}]{ShabaevJPB26}
\bibinfo{author}{\bibfnamefont{V.~M.} \bibnamefont{Shabaev}},
  \bibinfo{journal}{J. Phys. B: At. Mol. Phys.} \textbf{\bibinfo{volume}{26}},
  \bibinfo{pages}{1103} (\bibinfo{year}{1993}).

\bibitem[{\citenamefont{Johnson and Soff}(1985)}]{JohnSoff}
\bibinfo{author}{\bibfnamefont{W.~R.} \bibnamefont{Johnson}} \bibnamefont{and}
  \bibinfo{author}{\bibfnamefont{G.}~\bibnamefont{Soff}},
  \bibinfo{journal}{At.\ Dat.\ Nucl.\ Dat.\ Tabl.}
  \textbf{\bibinfo{volume}{33}}, \bibinfo{pages}{405} (\bibinfo{year}{1985}).

\bibitem[{\citenamefont{Yerokhin et~al.}(1999)\citenamefont{Yerokhin, Artemyev,
  Beier, Plunien, Shabaev, and Soff}}]{VYerokhin}
\bibinfo{author}{\bibfnamefont{V.~A.} \bibnamefont{Yerokhin}},
  \bibinfo{author}{\bibfnamefont{A.~N.} \bibnamefont{Artemyev}},
  \bibinfo{author}{\bibfnamefont{T.}~\bibnamefont{Beier}},
  \bibinfo{author}{\bibfnamefont{G.}~\bibnamefont{Plunien}},
  \bibinfo{author}{\bibfnamefont{V.~M.} \bibnamefont{Shabaev}},
  \bibnamefont{and} \bibinfo{author}{\bibfnamefont{G.}~\bibnamefont{Soff}},
  \bibinfo{journal}{Phys. Rev. A} \textbf{\bibinfo{volume}{60}},
  \bibinfo{pages}{3522} (\bibinfo{year}{1999}).

\bibitem[{\citenamefont{Zumbro et~al.}(1984)\citenamefont{Zumbro, Shera,
  Tanaka, Jr., Naumann, Hoehn, Reuter, and Steffen}}]{ZumbroPRL53}
\bibinfo{author}{\bibfnamefont{J.~D.} \bibnamefont{Zumbro}},
  \bibinfo{author}{\bibfnamefont{E.~B.} \bibnamefont{Shera}},
  \bibinfo{author}{\bibfnamefont{Y.}~\bibnamefont{Tanaka}},
  \bibinfo{author}{\bibfnamefont{C.~E.~B.} \bibnamefont{Jr.}},
  \bibinfo{author}{\bibfnamefont{R.~A.} \bibnamefont{Naumann}},
  \bibinfo{author}{\bibfnamefont{M.~V.} \bibnamefont{Hoehn}},
  \bibinfo{author}{\bibfnamefont{W.}~\bibnamefont{Reuter}}, \bibnamefont{and}
  \bibinfo{author}{\bibfnamefont{R.~M.} \bibnamefont{Steffen}},
  \bibinfo{journal}{Phys. Rev. Lett.} \textbf{\bibinfo{volume}{53}},
  \bibinfo{pages}{1888} (\bibinfo{year}{1984}).

\bibitem[{\citenamefont{Kozhedub et~al.}(2008)\citenamefont{Kozhedub, Andreev,
  Shabaev, Tupitsyn, Brandau, Kozhuharov, Plunien, and
  St{\"o}hlker}}]{Kozhedub}
\bibinfo{author}{\bibfnamefont{Y.~S.} \bibnamefont{Kozhedub}},
  \bibinfo{author}{\bibfnamefont{O.~V.} \bibnamefont{Andreev}},
  \bibinfo{author}{\bibfnamefont{V.~M.} \bibnamefont{Shabaev}},
  \bibinfo{author}{\bibfnamefont{I.~I.} \bibnamefont{Tupitsyn}},
  \bibinfo{author}{\bibfnamefont{C.}~\bibnamefont{Brandau}},
  \bibinfo{author}{\bibfnamefont{C.}~\bibnamefont{Kozhuharov}},
  \bibinfo{author}{\bibfnamefont{G.}~\bibnamefont{Plunien}}, \bibnamefont{and}
  \bibinfo{author}{\bibfnamefont{T.}~\bibnamefont{St{\"o}hlker}},
  \bibinfo{journal}{Phys. Rev. A} \textbf{\bibinfo{volume}{77}},
  \bibinfo{pages}{032501:1} (\bibinfo{year}{2008}).

\bibitem[{\citenamefont{Angeli}(2004)}]{Angeli2004}
\bibinfo{author}{\bibfnamefont{I.}~\bibnamefont{Angeli}}, \bibinfo{journal}{At.
  Data. Nucl. Data. Tabl.} \textbf{\bibinfo{volume}{87}}, \bibinfo{pages}{185}
  (\bibinfo{year}{2004}).

\bibitem[{\citenamefont{Donnelly and Sick}(1984)}]{DonSick}
\bibinfo{author}{\bibfnamefont{T.~W.} \bibnamefont{Donnelly}} \bibnamefont{and}
  \bibinfo{author}{\bibfnamefont{I.}~\bibnamefont{Sick}},
  \bibinfo{journal}{Rev. Mod. Phys.} \textbf{\bibinfo{volume}{56}},
  \bibinfo{pages}{461} (\bibinfo{year}{1984}).

\bibitem[{\citenamefont{Hofstadter}(1956)}]{Hof56}
\bibinfo{author}{\bibfnamefont{R.}~\bibnamefont{Hofstadter}},
  \bibinfo{journal}{Rev.~Mod.~Phys.} \textbf{\bibinfo{volume}{28}},
  \bibinfo{pages}{214} (\bibinfo{year}{1956}).

\bibitem[{\citenamefont{Dreher et~al.}(1974)\citenamefont{Dreher, Friedrich,
  Merle, Rothaas, and L{\"u}hrs}}]{Dreher}
\bibinfo{author}{\bibfnamefont{B.}~\bibnamefont{Dreher}},
  \bibinfo{author}{\bibfnamefont{J.}~\bibnamefont{Friedrich}},
  \bibinfo{author}{\bibfnamefont{K.}~\bibnamefont{Merle}},
  \bibinfo{author}{\bibfnamefont{H.}~\bibnamefont{Rothaas}}, \bibnamefont{and}
  \bibinfo{author}{\bibfnamefont{G.}~\bibnamefont{L{\"u}hrs}},
  \bibinfo{journal}{Nucl. Phys.} \textbf{\bibinfo{volume}{A 235}},
  \bibinfo{pages}{219} (\bibinfo{year}{1974}).

\bibitem[{\citenamefont{Vries et~al.}(1987)\citenamefont{Vries, Jager, and
  Vries}}]{deVries}
\bibinfo{author}{\bibfnamefont{H.~D.} \bibnamefont{Vries}},
  \bibinfo{author}{\bibfnamefont{C.~W.~D.} \bibnamefont{Jager}},
  \bibnamefont{and} \bibinfo{author}{\bibfnamefont{C.~D.} \bibnamefont{Vries}},
  \bibinfo{journal}{Atomic Data and Nuclear Data Tables}
  \textbf{\bibinfo{volume}{36}}, \bibinfo{pages}{495 } (\bibinfo{year}{1987}).

\bibitem[{\citenamefont{Suda and Wakasugi}(2005)}]{Suda}
\bibinfo{author}{\bibfnamefont{T.}~\bibnamefont{Suda}} \bibnamefont{and}
  \bibinfo{author}{\bibfnamefont{M.}~\bibnamefont{Wakasugi}},
  \bibinfo{journal}{Prog. Part. Nucl. Phys.} \textbf{\bibinfo{volume}{55}},
  \bibinfo{pages}{417} (\bibinfo{year}{2005}).

\bibitem[{\citenamefont{Wakasugi et~al.}(2008)\citenamefont{Wakasugi, Emoto,
  Furukawa, Ishii, Ito, Koseki, Kurita, Kuwajima, Masuda, Morikawa
  et~al.}}]{Wakasugi}
\bibinfo{author}{\bibfnamefont{M.}~\bibnamefont{Wakasugi}},
  \bibinfo{author}{\bibfnamefont{T.}~\bibnamefont{Emoto}},
  \bibinfo{author}{\bibfnamefont{Y.}~\bibnamefont{Furukawa}},
  \bibinfo{author}{\bibfnamefont{K.}~\bibnamefont{Ishii}},
  \bibinfo{author}{\bibfnamefont{S.}~\bibnamefont{Ito}},
  \bibinfo{author}{\bibfnamefont{T.}~\bibnamefont{Koseki}},
  \bibinfo{author}{\bibfnamefont{K.}~\bibnamefont{Kurita}},
  \bibinfo{author}{\bibfnamefont{A.}~\bibnamefont{Kuwajima}},
  \bibinfo{author}{\bibfnamefont{T.}~\bibnamefont{Masuda}},
  \bibinfo{author}{\bibfnamefont{A.}~\bibnamefont{Morikawa}},
  \bibnamefont{et~al.}, \bibinfo{journal}{Phys. Rev. Lett.}
  \textbf{\bibinfo{volume}{100}}, \bibinfo{pages}{164801:1}
  (\bibinfo{year}{2008}).

\bibitem[{\citenamefont{Simon}(2007)}]{Simon}
\bibinfo{author}{\bibfnamefont{H.}~\bibnamefont{Simon}},
  \bibinfo{journal}{Nucl. Phys.} \textbf{\bibinfo{volume}{A787}},
  \bibinfo{pages}{102} (\bibinfo{year}{2007}).

\bibitem[{\citenamefont{Simon}(2005)}]{FAIR}
\bibinfo{author}{\bibfnamefont{S.~H.} \bibnamefont{Simon}},
  \bibinfo{type}{Report}, \bibinfo{institution}{GSI},
  \bibinfo{address}{Darmstadt, Germany} (\bibinfo{year}{2005}),
  \bibinfo{note}{available at http://www.gsi.de/GSI-Future/cdr/}.

\bibitem[{\citenamefont{Fitch and Rainwater}(1953)}]{FitRain53}
\bibinfo{author}{\bibfnamefont{V.~L.} \bibnamefont{Fitch}} \bibnamefont{and}
  \bibinfo{author}{\bibfnamefont{J.}~\bibnamefont{Rainwater}},
  \bibinfo{journal}{Phys. Rev.} \textbf{\bibinfo{volume}{92}},
  \bibinfo{pages}{789} (\bibinfo{year}{1953}).

\bibitem[{\citenamefont{Borie and Rinker}(1982)}]{BorRink84}
\bibinfo{author}{\bibfnamefont{E.}~\bibnamefont{Borie}} \bibnamefont{and}
  \bibinfo{author}{\bibfnamefont{G.~A.} \bibnamefont{Rinker}},
  \bibinfo{journal}{Rev. Mod. Phys.} \textbf{\bibinfo{volume}{54}},
  \bibinfo{pages}{67} (\bibinfo{year}{1982}).

\bibitem[{\citenamefont{Engfer et~al.}(1974)\citenamefont{Engfer, Schneuwly,
  Vuilleumier, Walter, and Zehndercheck}}]{Engfer74}
\bibinfo{author}{\bibfnamefont{R.}~\bibnamefont{Engfer}},
  \bibinfo{author}{\bibfnamefont{H.}~\bibnamefont{Schneuwly}},
  \bibinfo{author}{\bibfnamefont{J.}~\bibnamefont{Vuilleumier}},
  \bibinfo{author}{\bibfnamefont{H.}~\bibnamefont{Walter}}, \bibnamefont{and}
  \bibinfo{author}{\bibfnamefont{A.}~\bibnamefont{Zehndercheck}},
  \bibinfo{journal}{At.\ Dat.\ Nucl.\ Dat.\ Tabl.}
  \textbf{\bibinfo{volume}{14}}, \bibinfo{pages}{509} (\bibinfo{year}{1974}).

\bibitem[{\citenamefont{Jungmann}(2001)}]{Jungmann}
\bibinfo{author}{\bibfnamefont{K.}~\bibnamefont{Jungmann}},
  \bibinfo{journal}{Hyp. Interact.} \textbf{\bibinfo{volume}{138}},
  \bibinfo{pages}{463} (\bibinfo{year}{2001}).

\bibitem[{\citenamefont{Demtr{\"o}der}(2008)}]{Demtroeder}
\bibinfo{author}{\bibfnamefont{W.}~\bibnamefont{Demtr{\"o}der}},
  \emph{\bibinfo{title}{Laser Spectroscopy}}, vol.~\bibinfo{volume}{2}
  (\bibinfo{publisher}{Springer}, \bibinfo{address}{Berlin},
  \bibinfo{year}{2008}).

\bibitem[{\citenamefont{King}(1963)}]{King}
\bibinfo{author}{\bibfnamefont{W.~H.} \bibnamefont{King}}, \bibinfo{journal}{J.
  Opt. Soc. Am.} \textbf{\bibinfo{volume}{53}}, \bibinfo{pages}{638}
  (\bibinfo{year}{1963}).

\bibitem[{\citenamefont{Niering et~al.}(2000)\citenamefont{Niering, Holzwarth,
  Reichert, Pokasov, Udem, Weitz, H\"ansch, Lemonde, Santarelli, Abgrall
  et~al.}}]{Haensch}
\bibinfo{author}{\bibfnamefont{M.}~\bibnamefont{Niering}},
  \bibinfo{author}{\bibfnamefont{R.}~\bibnamefont{Holzwarth}},
  \bibinfo{author}{\bibfnamefont{J.}~\bibnamefont{Reichert}},
  \bibinfo{author}{\bibfnamefont{P.}~\bibnamefont{Pokasov}},
  \bibinfo{author}{\bibfnamefont{T.}~\bibnamefont{Udem}},
  \bibinfo{author}{\bibfnamefont{M.}~\bibnamefont{Weitz}},
  \bibinfo{author}{\bibfnamefont{T.~W.} \bibnamefont{H\"ansch}},
  \bibinfo{author}{\bibfnamefont{P.}~\bibnamefont{Lemonde}},
  \bibinfo{author}{\bibfnamefont{G.}~\bibnamefont{Santarelli}},
  \bibinfo{author}{\bibfnamefont{M.}~\bibnamefont{Abgrall}},
  \bibnamefont{et~al.}, \bibinfo{journal}{Phys. Rev. Lett.}
  \textbf{\bibinfo{volume}{84}}, \bibinfo{pages}{5496} (\bibinfo{year}{2000}).

\bibitem[{\citenamefont{Yan and Drake}(2000)}]{Drake1}
\bibinfo{author}{\bibfnamefont{Z.-C.} \bibnamefont{Yan}} \bibnamefont{and}
  \bibinfo{author}{\bibfnamefont{G.~W.~F.} \bibnamefont{Drake}},
  \bibinfo{journal}{Phys. Rev. A} \textbf{\bibinfo{volume}{61}},
  \bibinfo{pages}{022504:1} (\bibinfo{year}{2000}).

\bibitem[{\citenamefont{Yan and Drake}(2002)}]{Drake2}
\bibinfo{author}{\bibfnamefont{Z.-C.} \bibnamefont{Yan}} \bibnamefont{and}
  \bibinfo{author}{\bibfnamefont{G.~W.~F.} \bibnamefont{Drake}},
  \bibinfo{journal}{Phys. Rev. A} \textbf{\bibinfo{volume}{66}},
  \bibinfo{pages}{042504:1} (\bibinfo{year}{2002}).

\bibitem[{\citenamefont{Yan and Drake}(2003)}]{Drake3}
\bibinfo{author}{\bibfnamefont{Z.-C.} \bibnamefont{Yan}} \bibnamefont{and}
  \bibinfo{author}{\bibfnamefont{G.~W.~F.} \bibnamefont{Drake}},
  \bibinfo{journal}{Phys. Rev. Lett.} \textbf{\bibinfo{volume}{91}},
  \bibinfo{pages}{113004:1} (\bibinfo{year}{2003}).

\bibitem[{\citenamefont{Puchalski
  et~al.}(2006{\natexlab{a}})\citenamefont{Puchalski, Moro, and
  Pachucki}}]{Pachucki1}
\bibinfo{author}{\bibfnamefont{M.}~\bibnamefont{Puchalski}},
  \bibinfo{author}{\bibfnamefont{A.~M.} \bibnamefont{Moro}}, \bibnamefont{and}
  \bibinfo{author}{\bibfnamefont{K.}~\bibnamefont{Pachucki}},
  \bibinfo{journal}{Phys. Rev. Lett.} \textbf{\bibinfo{volume}{97}},
  \bibinfo{pages}{133001:1} (\bibinfo{year}{2006}{\natexlab{a}}).

\bibitem[{\citenamefont{Puchalski and Pachucki}(2008)}]{Pachucki2}
\bibinfo{author}{\bibfnamefont{M.}~\bibnamefont{Puchalski}} \bibnamefont{and}
  \bibinfo{author}{\bibfnamefont{K.}~\bibnamefont{Pachucki}},
  \bibinfo{journal}{Phys. Rev. A} \textbf{\bibinfo{volume}{78}},
  \bibinfo{pages}{052511:1} (\bibinfo{year}{2008}).

\bibitem[{\citenamefont{Tanihata et~al.}(1985)\citenamefont{Tanihata, Hamagaki,
  Hashimoto, Shida, Yoshikawa, Sugimoto, Yamakawa, Kobayashi, and
  Takahashi}}]{Tanihata85}
\bibinfo{author}{\bibfnamefont{I.}~\bibnamefont{Tanihata}},
  \bibinfo{author}{\bibfnamefont{H.}~\bibnamefont{Hamagaki}},
  \bibinfo{author}{\bibfnamefont{O.}~\bibnamefont{Hashimoto}},
  \bibinfo{author}{\bibfnamefont{Y.}~\bibnamefont{Shida}},
  \bibinfo{author}{\bibfnamefont{N.}~\bibnamefont{Yoshikawa}},
  \bibinfo{author}{\bibfnamefont{K.}~\bibnamefont{Sugimoto}},
  \bibinfo{author}{\bibfnamefont{O.}~\bibnamefont{Yamakawa}},
  \bibinfo{author}{\bibfnamefont{T.}~\bibnamefont{Kobayashi}},
  \bibnamefont{and}
  \bibinfo{author}{\bibfnamefont{N.}~\bibnamefont{Takahashi}},
  \bibinfo{journal}{Phys. Rev. Lett.} \textbf{\bibinfo{volume}{55}},
  \bibinfo{pages}{2676} (\bibinfo{year}{1985}).

\bibitem[{\citenamefont{Ewald et~al.}(2004)\citenamefont{Ewald,
  N{\"o}rtersh{\"a}user, Dax, G{\"o}tte, Kirchner, Kluge, K{\"u}hl, Sanchez,
  Wojtaszek, Bushaw et~al.}}]{Li8-9}
\bibinfo{author}{\bibfnamefont{G.}~\bibnamefont{Ewald}},
  \bibinfo{author}{\bibfnamefont{W.}~\bibnamefont{N{\"o}rtersh{\"a}user}},
  \bibinfo{author}{\bibfnamefont{A.}~\bibnamefont{Dax}},
  \bibinfo{author}{\bibfnamefont{S.}~\bibnamefont{G{\"o}tte}},
  \bibinfo{author}{\bibfnamefont{R.}~\bibnamefont{Kirchner}},
  \bibinfo{author}{\bibfnamefont{H.-J.} \bibnamefont{Kluge}},
  \bibinfo{author}{\bibfnamefont{T.}~\bibnamefont{K{\"u}hl}},
  \bibinfo{author}{\bibfnamefont{R.}~\bibnamefont{Sanchez}},
  \bibinfo{author}{\bibfnamefont{A.}~\bibnamefont{Wojtaszek}},
  \bibinfo{author}{\bibfnamefont{B.~A.} \bibnamefont{Bushaw}},
  \bibnamefont{et~al.}, \bibinfo{journal}{Phys. Rev. Lett.}
  \textbf{\bibinfo{volume}{93}}, \bibinfo{pages}{113002:1}
  (\bibinfo{year}{2004}).

\bibitem[{\citenamefont{S{\'a}nchez et~al.}(2006)\citenamefont{S{\'a}nchez,
  N{\"o}rtersh{\"a}user, Ewald, Albers, Behr, Bricault, Bushaw, Dax, Dilling,
  Dombsky et~al.}}]{Li11}
\bibinfo{author}{\bibfnamefont{R.}~\bibnamefont{S{\'a}nchez}},
  \bibinfo{author}{\bibfnamefont{W.}~\bibnamefont{N{\"o}rtersh{\"a}user}},
  \bibinfo{author}{\bibfnamefont{G.}~\bibnamefont{Ewald}},
  \bibinfo{author}{\bibfnamefont{D.}~\bibnamefont{Albers}},
  \bibinfo{author}{\bibfnamefont{J.}~\bibnamefont{Behr}},
  \bibinfo{author}{\bibfnamefont{P.}~\bibnamefont{Bricault}},
  \bibinfo{author}{\bibfnamefont{B.~A.} \bibnamefont{Bushaw}},
  \bibinfo{author}{\bibfnamefont{A.}~\bibnamefont{Dax}},
  \bibinfo{author}{\bibfnamefont{J.}~\bibnamefont{Dilling}},
  \bibinfo{author}{\bibfnamefont{M.}~\bibnamefont{Dombsky}},
  \bibnamefont{et~al.}, \bibinfo{journal}{Phys. Rev. Lett.}
  \textbf{\bibinfo{volume}{96}}, \bibinfo{pages}{033002:1}
  (\bibinfo{year}{2006}).

\bibitem[{\citenamefont{Mueller et~al.}(2007)\citenamefont{Mueller, Sulai,
  Villari, Alc{\'a}ntara-N{\'u}{\~n}ez, Alves-Cond{\'e}, Bailey, Drake, Dubois,
  El{\'e}on, Gaubert et~al.}}]{He8}
\bibinfo{author}{\bibfnamefont{P.}~\bibnamefont{Mueller}},
  \bibinfo{author}{\bibfnamefont{I.~A.} \bibnamefont{Sulai}},
  \bibinfo{author}{\bibfnamefont{A.~C.~C.} \bibnamefont{Villari}},
  \bibinfo{author}{\bibfnamefont{J.~A.}
  \bibnamefont{Alc{\'a}ntara-N{\'u}{\~n}ez}},
  \bibinfo{author}{\bibfnamefont{R.}~\bibnamefont{Alves-Cond{\'e}}},
  \bibinfo{author}{\bibfnamefont{K.}~\bibnamefont{Bailey}},
  \bibinfo{author}{\bibfnamefont{G.~W.~F.} \bibnamefont{Drake}},
  \bibinfo{author}{\bibfnamefont{M.}~\bibnamefont{Dubois}},
  \bibinfo{author}{\bibfnamefont{C.}~\bibnamefont{El{\'e}on}},
  \bibinfo{author}{\bibfnamefont{G.}~\bibnamefont{Gaubert}},
  \bibnamefont{et~al.}, \bibinfo{journal}{Phys. Rev. Lett.}
  \textbf{\bibinfo{volume}{99}}, \bibinfo{pages}{252501:1}
  (\bibinfo{year}{2007}).

\bibitem[{\citenamefont{N{\"o}rtersh{\"a}user
  et~al.}(2009)\citenamefont{N{\"o}rtersh{\"a}user, Tiedemann,
  {$\check{\mathrm{Z}}$}{\'a}kov{\'a}, Andjelkovic, Blaum, Bissell, Cazan,
  Drake, Geppert, Kowalska et~al.}}]{Be10}
\bibinfo{author}{\bibfnamefont{W.}~\bibnamefont{N{\"o}rtersh{\"a}user}},
  \bibinfo{author}{\bibfnamefont{D.}~\bibnamefont{Tiedemann}},
  \bibinfo{author}{\bibfnamefont{M.}~\bibnamefont{{$\check{\mathrm{Z}}$}{\'a}k%
ov{\'a}}}, \bibinfo{author}{\bibfnamefont{Z.}~\bibnamefont{Andjelkovic}},
  \bibinfo{author}{\bibfnamefont{K.}~\bibnamefont{Blaum}},
  \bibinfo{author}{\bibfnamefont{M.~L.} \bibnamefont{Bissell}},
  \bibinfo{author}{\bibfnamefont{R.}~\bibnamefont{Cazan}},
  \bibinfo{author}{\bibfnamefont{G.~W.~F.} \bibnamefont{Drake}},
  \bibinfo{author}{\bibfnamefont{C.}~\bibnamefont{Geppert}},
  \bibinfo{author}{\bibfnamefont{M.}~\bibnamefont{Kowalska}},
  \bibnamefont{et~al.}, \bibinfo{journal}{Phys. Rev. Lett.}
  \textbf{\bibinfo{volume}{102}}, \bibinfo{pages}{062503:1}
  (\bibinfo{year}{2009}).

\bibitem[{\citenamefont{Geithner et~al.}(2008)\citenamefont{Geithner, Neff,
  Audi, Blaum, Delahaye, Feldmeier, George, Gu{\'e}naut, Herfurth, Herlert
  et~al.}}]{Ne17}
\bibinfo{author}{\bibfnamefont{W.}~\bibnamefont{Geithner}},
  \bibinfo{author}{\bibfnamefont{T.}~\bibnamefont{Neff}},
  \bibinfo{author}{\bibfnamefont{G.}~\bibnamefont{Audi}},
  \bibinfo{author}{\bibfnamefont{K.}~\bibnamefont{Blaum}},
  \bibinfo{author}{\bibfnamefont{P.}~\bibnamefont{Delahaye}},
  \bibinfo{author}{\bibfnamefont{H.}~\bibnamefont{Feldmeier}},
  \bibinfo{author}{\bibfnamefont{S.}~\bibnamefont{George}},
  \bibinfo{author}{\bibfnamefont{C.}~\bibnamefont{Gu{\'e}naut}},
  \bibinfo{author}{\bibfnamefont{F.}~\bibnamefont{Herfurth}},
  \bibinfo{author}{\bibfnamefont{A.}~\bibnamefont{Herlert}},
  \bibnamefont{et~al.}, \bibinfo{journal}{Phys. Rev. Lett.}
  \textbf{\bibinfo{volume}{101}}, \bibinfo{pages}{252502:1}
  (\bibinfo{year}{2008}).

\bibitem[{\citenamefont{Brockmeier et~al.}(1965)\citenamefont{Brockmeier,
  Boehm, and Hatch}}]{KalfaU}
\bibinfo{author}{\bibfnamefont{R.~T.} \bibnamefont{Brockmeier}},
  \bibinfo{author}{\bibfnamefont{F.}~\bibnamefont{Boehm}}, \bibnamefont{and}
  \bibinfo{author}{\bibfnamefont{E.~N.} \bibnamefont{Hatch}},
  \bibinfo{journal}{Phys. Rev. Lett.} \textbf{\bibinfo{volume}{15}},
  \bibinfo{pages}{132} (\bibinfo{year}{1965}).

\bibitem[{\citenamefont{Elliott et~al.}(1996)\citenamefont{Elliott,
  Beiersdorfer, and Chen}}]{SuperEBIT-U}
\bibinfo{author}{\bibfnamefont{S.~R.} \bibnamefont{Elliott}},
  \bibinfo{author}{\bibfnamefont{P.}~\bibnamefont{Beiersdorfer}},
  \bibnamefont{and} \bibinfo{author}{\bibfnamefont{M.~H.} \bibnamefont{Chen}},
  \bibinfo{journal}{Phys. Rev. Lett.} \textbf{\bibinfo{volume}{76}},
  \bibinfo{pages}{1031} (\bibinfo{year}{1996}).

\bibitem[{\citenamefont{Brandau et~al.}(2003)\citenamefont{Brandau, Kozhuharov,
  M{\"u}ller, Shi, Schippers, Bartsch, B{\"o}hm, B{\"o}hme, Hoffknecht, Knopp
  et~al.}}]{Brandau2003}
\bibinfo{author}{\bibfnamefont{C.}~\bibnamefont{Brandau}},
  \bibinfo{author}{\bibfnamefont{C.}~\bibnamefont{Kozhuharov}},
  \bibinfo{author}{\bibfnamefont{A.}~\bibnamefont{M{\"u}ller}},
  \bibinfo{author}{\bibfnamefont{W.}~\bibnamefont{Shi}},
  \bibinfo{author}{\bibfnamefont{S.}~\bibnamefont{Schippers}},
  \bibinfo{author}{\bibfnamefont{T.}~\bibnamefont{Bartsch}},
  \bibinfo{author}{\bibfnamefont{S.}~\bibnamefont{B{\"o}hm}},
  \bibinfo{author}{\bibfnamefont{C.}~\bibnamefont{B{\"o}hme}},
  \bibinfo{author}{\bibfnamefont{A.}~\bibnamefont{Hoffknecht}},
  \bibinfo{author}{\bibfnamefont{H.}~\bibnamefont{Knopp}},
  \bibnamefont{et~al.}, \bibinfo{journal}{Phys. Rev. Lett.}
  \textbf{\bibinfo{volume}{91}}, \bibinfo{pages}{073202:1}
  (\bibinfo{year}{2003}).

\bibitem[{\citenamefont{Brandau et~al.}(2008)\citenamefont{Brandau, Kozhuharov,
  Harman, M{\"u}ller, Schippers, Kozhedub, Bernhardt, B{\"o}hm, Jacobi, Schmidt
  et~al.}}]{NdPRL}
\bibinfo{author}{\bibfnamefont{C.}~\bibnamefont{Brandau}},
  \bibinfo{author}{\bibfnamefont{C.}~\bibnamefont{Kozhuharov}},
  \bibinfo{author}{\bibfnamefont{Z.}~\bibnamefont{Harman}},
  \bibinfo{author}{\bibfnamefont{A.}~\bibnamefont{M{\"u}ller}},
  \bibinfo{author}{\bibfnamefont{S.}~\bibnamefont{Schippers}},
  \bibinfo{author}{\bibfnamefont{Y.~S.} \bibnamefont{Kozhedub}},
  \bibinfo{author}{\bibfnamefont{D.}~\bibnamefont{Bernhardt}},
  \bibinfo{author}{\bibfnamefont{S.}~\bibnamefont{B{\"o}hm}},
  \bibinfo{author}{\bibfnamefont{J.}~\bibnamefont{Jacobi}},
  \bibinfo{author}{\bibfnamefont{E.~W.} \bibnamefont{Schmidt}},
  \bibnamefont{et~al.}, \bibinfo{journal}{Phys. Rev. Lett.}
  \textbf{\bibinfo{volume}{100}}, \bibinfo{pages}{073201:1}
  (\bibinfo{year}{2008}).

\bibitem[{\citenamefont{Shabaev}(1994)}]{ShabaevJPB27}
\bibinfo{author}{\bibfnamefont{V.~M.} \bibnamefont{Shabaev}},
  \bibinfo{journal}{J. Phys. B: At. Mol. Phys.} \textbf{\bibinfo{volume}{27}},
  \bibinfo{pages}{5825} (\bibinfo{year}{1994}).

\bibitem[{\citenamefont{Rosenthal and Breit}(1932)}]{BreitRosenth}
\bibinfo{author}{\bibfnamefont{J.~E.} \bibnamefont{Rosenthal}}
  \bibnamefont{and} \bibinfo{author}{\bibfnamefont{G.}~\bibnamefont{Breit}},
  \bibinfo{journal}{Phys. Rev.} \textbf{\bibinfo{volume}{41}},
  \bibinfo{pages}{459} (\bibinfo{year}{1932}).

\bibitem[{\citenamefont{Bohr and Weisskopf}(1950)}]{BohrWeisskopf}
\bibinfo{author}{\bibfnamefont{A.}~\bibnamefont{Bohr}} \bibnamefont{and}
  \bibinfo{author}{\bibfnamefont{V.~F.} \bibnamefont{Weisskopf}},
  \bibinfo{journal}{Phys. Rev.} \textbf{\bibinfo{volume}{77}},
  \bibinfo{pages}{94} (\bibinfo{year}{1950}).

\bibitem[{\citenamefont{Kusch et~al.}(1940)\citenamefont{Kusch, Millman, and
  Rabi}}]{Rabi1}
\bibinfo{author}{\bibfnamefont{P.}~\bibnamefont{Kusch}},
  \bibinfo{author}{\bibfnamefont{S.}~\bibnamefont{Millman}}, \bibnamefont{and}
  \bibinfo{author}{\bibfnamefont{I.~I.} \bibnamefont{Rabi}},
  \bibinfo{journal}{Phys. Rev.} \textbf{\bibinfo{volume}{57}},
  \bibinfo{pages}{765} (\bibinfo{year}{1940}).

\bibitem[{\citenamefont{Millman and Kusch}(1940)}]{Rabi2}
\bibinfo{author}{\bibfnamefont{S.}~\bibnamefont{Millman}} \bibnamefont{and}
  \bibinfo{author}{\bibfnamefont{P.}~\bibnamefont{Kusch}},
  \bibinfo{journal}{Phys. Rev.} \textbf{\bibinfo{volume}{58}},
  \bibinfo{pages}{438} (\bibinfo{year}{1940}).

\bibitem[{\citenamefont{Ramsey}(1950)}]{Ramsey}
\bibinfo{author}{\bibfnamefont{N.}~\bibnamefont{Ramsey}},
  \bibinfo{journal}{Phys. Rev.} \textbf{\bibinfo{volume}{78}},
  \bibinfo{pages}{695} (\bibinfo{year}{1950}).

\bibitem[{\citenamefont{Shirley et~al.}(2001)\citenamefont{Shirley, Lee, and
  Drullinger}}]{Metrologia}
\bibinfo{author}{\bibfnamefont{J.}~\bibnamefont{Shirley}},
  \bibinfo{author}{\bibfnamefont{W.}~\bibnamefont{Lee}}, \bibnamefont{and}
  \bibinfo{author}{\bibfnamefont{R.}~\bibnamefont{Drullinger}},
  \bibinfo{journal}{Metrologia} \textbf{\bibinfo{volume}{38}},
  \bibinfo{pages}{427} (\bibinfo{year}{2001}).

\bibitem[{\citenamefont{Essen et~al.}(1971)\citenamefont{Essen, Donaldson,
  Bangham, and Hope}}]{Hmaser}
\bibinfo{author}{\bibfnamefont{L.}~\bibnamefont{Essen}},
  \bibinfo{author}{\bibfnamefont{R.~W.} \bibnamefont{Donaldson}},
  \bibinfo{author}{\bibfnamefont{M.~J.} \bibnamefont{Bangham}},
  \bibnamefont{and} \bibinfo{author}{\bibfnamefont{E.~G.} \bibnamefont{Hope}},
  \bibinfo{journal}{Nature} \textbf{\bibinfo{volume}{229}},
  \bibinfo{pages}{110} (\bibinfo{year}{1971}).

\bibitem[{\citenamefont{Schuessler et~al.}(1969)\citenamefont{Schuessler,
  Fortson, and Dehmelt}}]{HeHFS}
\bibinfo{author}{\bibfnamefont{H.~A.} \bibnamefont{Schuessler}},
  \bibinfo{author}{\bibfnamefont{E.~N.} \bibnamefont{Fortson}},
  \bibnamefont{and} \bibinfo{author}{\bibfnamefont{H.~G.}
  \bibnamefont{Dehmelt}}, \bibinfo{journal}{Phys. Rev.}
  \textbf{\bibinfo{volume}{187}}, \bibinfo{pages}{5} (\bibinfo{year}{1969}).

\bibitem[{\citenamefont{Borchert et~al.}(1977)\citenamefont{Borchert, Hansen,
  Jonson, Schult, and Tidemand-Peterson}}]{XeXray}
\bibinfo{author}{\bibfnamefont{G.~L.} \bibnamefont{Borchert}},
  \bibinfo{author}{\bibfnamefont{P.~G.} \bibnamefont{Hansen}},
  \bibinfo{author}{\bibfnamefont{B.}~\bibnamefont{Jonson}},
  \bibinfo{author}{\bibfnamefont{O.~W.} \bibnamefont{Schult}},
  \bibnamefont{and}
  \bibinfo{author}{\bibfnamefont{P.}~\bibnamefont{Tidemand-Peterson}},
  \bibinfo{journal}{Phys. Lett.} \textbf{\bibinfo{volume}{63A}},
  \bibinfo{pages}{15} (\bibinfo{year}{1977}).

\bibitem[{\citenamefont{B{\"u}tggenbach}(1984)}]{Buettgenbach}
\bibinfo{author}{\bibfnamefont{S.}~\bibnamefont{B{\"u}tggenbach}},
  \bibinfo{journal}{Hyp. Interact.} \textbf{\bibinfo{volume}{20}},
  \bibinfo{pages}{1} (\bibinfo{year}{1984}).

\bibitem[{\citenamefont{R{\"u}etschi et~al.}(1984)\citenamefont{R{\"u}etschi,
  Schellenberg, Phan, Piller, Schaller, and Schneuwly}}]{Bi209M}
\bibinfo{author}{\bibfnamefont{A.}~\bibnamefont{R{\"u}etschi}},
  \bibinfo{author}{\bibfnamefont{L.}~\bibnamefont{Schellenberg}},
  \bibinfo{author}{\bibfnamefont{T.~Q.} \bibnamefont{Phan}},
  \bibinfo{author}{\bibfnamefont{G.}~\bibnamefont{Piller}},
  \bibinfo{author}{\bibfnamefont{L.~A.} \bibnamefont{Schaller}},
  \bibnamefont{and}
  \bibinfo{author}{\bibfnamefont{H.}~\bibnamefont{Schneuwly}},
  \bibinfo{journal}{Nucl. Phys.} \textbf{\bibinfo{volume}{A 422}},
  \bibinfo{pages}{461} (\bibinfo{year}{1984}).

\bibitem[{\citenamefont{Backe et~al.}(1972)\citenamefont{Backe, Engfer, Jahnke,
  Kankeleit, Pearce, Petitjean, Schellenberg, Schneuwly, Schr{\"o}der, Walter
  et~al.}}]{TlMuonic}
\bibinfo{author}{\bibfnamefont{H.}~\bibnamefont{Backe}},
  \bibinfo{author}{\bibfnamefont{R.}~\bibnamefont{Engfer}},
  \bibinfo{author}{\bibfnamefont{U.}~\bibnamefont{Jahnke}},
  \bibinfo{author}{\bibfnamefont{E.}~\bibnamefont{Kankeleit}},
  \bibinfo{author}{\bibfnamefont{R.~M.} \bibnamefont{Pearce}},
  \bibinfo{author}{\bibfnamefont{C.}~\bibnamefont{Petitjean}},
  \bibinfo{author}{\bibfnamefont{L.}~\bibnamefont{Schellenberg}},
  \bibinfo{author}{\bibfnamefont{H.}~\bibnamefont{Schneuwly}},
  \bibinfo{author}{\bibfnamefont{W.~U.} \bibnamefont{Schr{\"o}der}},
  \bibinfo{author}{\bibfnamefont{H.~K.} \bibnamefont{Walter}},
  \bibnamefont{et~al.}, \bibinfo{journal}{Nucl. Phys. A}
  \textbf{\bibinfo{volume}{189}}, \bibinfo{pages}{472} (\bibinfo{year}{1972}).

\bibitem[{\citenamefont{Crespo~L{\'o}pez-Urrutia
  et~al.}(1996)\citenamefont{Crespo~L{\'o}pez-Urrutia, Beiersdorfer, Savin, and
  Widmann}}]{Ho165}
\bibinfo{author}{\bibfnamefont{J.~R.} \bibnamefont{Crespo~L{\'o}pez-Urrutia}},
  \bibinfo{author}{\bibfnamefont{P.}~\bibnamefont{Beiersdorfer}},
  \bibinfo{author}{\bibfnamefont{D.~W.} \bibnamefont{Savin}}, \bibnamefont{and}
  \bibinfo{author}{\bibfnamefont{K.}~\bibnamefont{Widmann}},
  \bibinfo{journal}{Phys. Rev. Lett.} \textbf{\bibinfo{volume}{77}},
  \bibinfo{pages}{826} (\bibinfo{year}{1996}).

\bibitem[{\citenamefont{Crespo~L{\'o}pez-Urrutia
  et~al.}(1998)\citenamefont{Crespo~L{\'o}pez-Urrutia, Beiersdorfer, Widmann,
  Birkett, M{\aa}rtensson-Pendrill, and Gustavsson}}]{Re185187}
\bibinfo{author}{\bibfnamefont{J.~R.} \bibnamefont{Crespo~L{\'o}pez-Urrutia}},
  \bibinfo{author}{\bibfnamefont{P.}~\bibnamefont{Beiersdorfer}},
  \bibinfo{author}{\bibfnamefont{K.}~\bibnamefont{Widmann}},
  \bibinfo{author}{\bibfnamefont{B.~B.} \bibnamefont{Birkett}},
  \bibinfo{author}{\bibfnamefont{A.-M.} \bibnamefont{M{\aa}rtensson-Pendrill}},
  \bibnamefont{and} \bibinfo{author}{\bibfnamefont{M.~G.~H.}
  \bibnamefont{Gustavsson}}, \bibinfo{journal}{Phys. Rev. A}
  \textbf{\bibinfo{volume}{57}}, \bibinfo{pages}{879} (\bibinfo{year}{1998}).

\bibitem[{\citenamefont{Beiersdorfer et~al.}(2001)\citenamefont{Beiersdorfer,
  Utter, Wong, Crespo~L{\'o}pez-Urrutia, Britten, Chen, Harris, Thoe, Thorn,
  and Tr{\"a}bert}}]{Tl203205}
\bibinfo{author}{\bibfnamefont{P.}~\bibnamefont{Beiersdorfer}},
  \bibinfo{author}{\bibfnamefont{S.~B.} \bibnamefont{Utter}},
  \bibinfo{author}{\bibfnamefont{K.~L.} \bibnamefont{Wong}},
  \bibinfo{author}{\bibfnamefont{J.~R.}
  \bibnamefont{Crespo~L{\'o}pez-Urrutia}},
  \bibinfo{author}{\bibfnamefont{J.~A.} \bibnamefont{Britten}},
  \bibinfo{author}{\bibfnamefont{H.}~\bibnamefont{Chen}},
  \bibinfo{author}{\bibfnamefont{C.~L.} \bibnamefont{Harris}},
  \bibinfo{author}{\bibfnamefont{R.~S.} \bibnamefont{Thoe}},
  \bibinfo{author}{\bibfnamefont{D.~B.} \bibnamefont{Thorn}}, \bibnamefont{and}
  \bibinfo{author}{\bibfnamefont{E.}~\bibnamefont{Tr{\"a}bert}},
  \bibinfo{journal}{Phys. Rev. A} \textbf{\bibinfo{volume}{64}},
  \bibinfo{pages}{032506:1} (\bibinfo{year}{2001}).

\bibitem[{\citenamefont{Seelig et~al.}(1998)\citenamefont{Seelig, Borneis, Dax,
  Engel, Faber, Gerlach, Holbrow, Huber, K{\"u}hl, Marx et~al.}}]{Pb207}
\bibinfo{author}{\bibfnamefont{P.}~\bibnamefont{Seelig}},
  \bibinfo{author}{\bibfnamefont{S.}~\bibnamefont{Borneis}},
  \bibinfo{author}{\bibfnamefont{A.}~\bibnamefont{Dax}},
  \bibinfo{author}{\bibfnamefont{T.}~\bibnamefont{Engel}},
  \bibinfo{author}{\bibfnamefont{S.}~\bibnamefont{Faber}},
  \bibinfo{author}{\bibfnamefont{M.}~\bibnamefont{Gerlach}},
  \bibinfo{author}{\bibfnamefont{C.}~\bibnamefont{Holbrow}},
  \bibinfo{author}{\bibfnamefont{G.}~\bibnamefont{Huber}},
  \bibinfo{author}{\bibfnamefont{T.}~\bibnamefont{K{\"u}hl}},
  \bibinfo{author}{\bibfnamefont{D.}~\bibnamefont{Marx}}, \bibnamefont{et~al.},
  \bibinfo{journal}{Phys. Rev. Lett.} \textbf{\bibinfo{volume}{81}},
  \bibinfo{pages}{4824} (\bibinfo{year}{1998}).

\bibitem[{\citenamefont{Klaft et~al.}(1994)\citenamefont{Klaft, Borneis, Engel,
  Fricke, Grieser, Huber, K{\"u}hl, Marx, Neumann, Schr{\"o}der
  et~al.}}]{Bi209}
\bibinfo{author}{\bibfnamefont{I.}~\bibnamefont{Klaft}},
  \bibinfo{author}{\bibfnamefont{S.}~\bibnamefont{Borneis}},
  \bibinfo{author}{\bibfnamefont{T.}~\bibnamefont{Engel}},
  \bibinfo{author}{\bibfnamefont{B.}~\bibnamefont{Fricke}},
  \bibinfo{author}{\bibfnamefont{R.}~\bibnamefont{Grieser}},
  \bibinfo{author}{\bibfnamefont{G.}~\bibnamefont{Huber}},
  \bibinfo{author}{\bibfnamefont{T.}~\bibnamefont{K{\"u}hl}},
  \bibinfo{author}{\bibfnamefont{D.}~\bibnamefont{Marx}},
  \bibinfo{author}{\bibfnamefont{R.}~\bibnamefont{Neumann}},
  \bibinfo{author}{\bibfnamefont{S.}~\bibnamefont{Schr{\"o}der}},
  \bibnamefont{et~al.}, \bibinfo{journal}{Phys. Rev. Lett.}
  \textbf{\bibinfo{volume}{73}}, \bibinfo{pages}{2425} (\bibinfo{year}{1994}).

\bibitem[{\citenamefont{Schuch et~al.}(2005)\citenamefont{Schuch, Lindroth,
  Madzunkov, Fogle, Mohamed, and Indelicato}}]{EvaPRL95}
\bibinfo{author}{\bibfnamefont{R.}~\bibnamefont{Schuch}},
  \bibinfo{author}{\bibfnamefont{E.}~\bibnamefont{Lindroth}},
  \bibinfo{author}{\bibfnamefont{S.}~\bibnamefont{Madzunkov}},
  \bibinfo{author}{\bibfnamefont{M.}~\bibnamefont{Fogle}},
  \bibinfo{author}{\bibfnamefont{T.}~\bibnamefont{Mohamed}}, \bibnamefont{and}
  \bibinfo{author}{\bibfnamefont{P.}~\bibnamefont{Indelicato}},
  \bibinfo{journal}{Phys. Rev. Lett.} \textbf{\bibinfo{volume}{95}},
  \bibinfo{pages}{183003:1} (\bibinfo{year}{2005}).

\bibitem[{\citenamefont{Lestinsky et~al.}(2008)\citenamefont{Lestinsky,
  Lindroth, Orlov, Schmidt, Schippers, B{\"o}hm, Brandau, Sprenger, Terekhov,
  M{\"u}ller et~al.}}]{Lestinsky}
\bibinfo{author}{\bibfnamefont{M.}~\bibnamefont{Lestinsky}},
  \bibinfo{author}{\bibfnamefont{E.}~\bibnamefont{Lindroth}},
  \bibinfo{author}{\bibfnamefont{D.~A.} \bibnamefont{Orlov}},
  \bibinfo{author}{\bibfnamefont{E.~W.} \bibnamefont{Schmidt}},
  \bibinfo{author}{\bibfnamefont{S.}~\bibnamefont{Schippers}},
  \bibinfo{author}{\bibfnamefont{S.}~\bibnamefont{B{\"o}hm}},
  \bibinfo{author}{\bibfnamefont{C.}~\bibnamefont{Brandau}},
  \bibinfo{author}{\bibfnamefont{F.}~\bibnamefont{Sprenger}},
  \bibinfo{author}{\bibfnamefont{A.~S.} \bibnamefont{Terekhov}},
  \bibinfo{author}{\bibfnamefont{A.}~\bibnamefont{M{\"u}ller}},
  \bibnamefont{et~al.}, \bibinfo{journal}{Phys. Rev. Lett.}
  \textbf{\bibinfo{volume}{100}}, \bibinfo{pages}{033001:1}
  (\bibinfo{year}{2008}).

\bibitem[{\citenamefont{Schippers et~al.}(2007)\citenamefont{Schippers,
  Schmidt, Bernhardt, Yu, M{\"u}ller, Lestinsky, Orlov, Grieser, and
  Wolf}}]{StefanSchippers}
\bibinfo{author}{\bibfnamefont{S.}~\bibnamefont{Schippers}},
  \bibinfo{author}{\bibfnamefont{E.~W.} \bibnamefont{Schmidt}},
  \bibinfo{author}{\bibfnamefont{D.}~\bibnamefont{Bernhardt}},
  \bibinfo{author}{\bibfnamefont{D.}~\bibnamefont{Yu}},
  \bibinfo{author}{\bibfnamefont{A.}~\bibnamefont{M{\"u}ller}},
  \bibinfo{author}{\bibfnamefont{M.}~\bibnamefont{Lestinsky}},
  \bibinfo{author}{\bibfnamefont{D.~A.} \bibnamefont{Orlov}},
  \bibinfo{author}{\bibfnamefont{M.}~\bibnamefont{Grieser}}, \bibnamefont{and}
  \bibinfo{author}{\bibfnamefont{A.}~\bibnamefont{Wolf}},
  \bibinfo{journal}{Phys. Rev. Lett.} \textbf{\bibinfo{volume}{98}},
  \bibinfo{pages}{033001:1} (\bibinfo{year}{2007}).

\bibitem[{\citenamefont{Schmieder}(1973)}]{Ti_metastable}
\bibinfo{author}{\bibfnamefont{R.~W.} \bibnamefont{Schmieder}},
  \bibinfo{journal}{Phys. Rev. A} \textbf{\bibinfo{volume}{7}},
  \bibinfo{pages}{1458} (\bibinfo{year}{1973}).

\bibitem[{\citenamefont{Brage et~al.}(2002)\citenamefont{Brage, Judge, and
  Proffitt}}]{Brage}
\bibinfo{author}{\bibfnamefont{T.}~\bibnamefont{Brage}},
  \bibinfo{author}{\bibfnamefont{P.~G.} \bibnamefont{Judge}}, \bibnamefont{and}
  \bibinfo{author}{\bibfnamefont{C.~R.} \bibnamefont{Proffitt}},
  \bibinfo{journal}{Phys. Rev. Lett.} \textbf{\bibinfo{volume}{89}},
  \bibinfo{pages}{281101:1} (\bibinfo{year}{2002}).

\bibitem[{\citenamefont{Labzowsky et~al.}(2000)\citenamefont{Labzowsky,
  Nefiodov, Plunien, Soff, and Liesen}}]{Labzowsky}
\bibinfo{author}{\bibfnamefont{L.~N.} \bibnamefont{Labzowsky}},
  \bibinfo{author}{\bibfnamefont{A.~V.} \bibnamefont{Nefiodov}},
  \bibinfo{author}{\bibfnamefont{G.}~\bibnamefont{Plunien}},
  \bibinfo{author}{\bibfnamefont{G.}~\bibnamefont{Soff}}, \bibnamefont{and}
  \bibinfo{author}{\bibfnamefont{D.}~\bibnamefont{Liesen}},
  \bibinfo{journal}{Phys. Rev. Lett.} \textbf{\bibinfo{volume}{84}},
  \bibinfo{pages}{851} (\bibinfo{year}{2000}).

\bibitem[{\citenamefont{Lee and Yang}(1956)}]{LeeYang}
\bibinfo{author}{\bibfnamefont{T.~D.} \bibnamefont{Lee}} \bibnamefont{and}
  \bibinfo{author}{\bibfnamefont{C.~N.} \bibnamefont{Yang}},
  \bibinfo{journal}{Phys. Rev.} \textbf{\bibinfo{volume}{104}},
  \bibinfo{pages}{254} (\bibinfo{year}{1956}).

\bibitem[{\citenamefont{Wu et~al.}(1957)\citenamefont{Wu, Ambler, Hayward,
  Hoppes, and Hudson}}]{Wu}
\bibinfo{author}{\bibfnamefont{C.~S.} \bibnamefont{Wu}},
  \bibinfo{author}{\bibfnamefont{E.}~\bibnamefont{Ambler}},
  \bibinfo{author}{\bibfnamefont{R.~W.} \bibnamefont{Hayward}},
  \bibinfo{author}{\bibfnamefont{D.~D.} \bibnamefont{Hoppes}},
  \bibnamefont{and} \bibinfo{author}{\bibfnamefont{R.~P.}
  \bibnamefont{Hudson}}, \bibinfo{journal}{Phys. Rev.}
  \textbf{\bibinfo{volume}{105}}, \bibinfo{pages}{1413} (\bibinfo{year}{1957}).

\bibitem[{\citenamefont{Zel'dovich}(1959)}]{Zeldovich}
\bibinfo{author}{\bibfnamefont{Y.~B.} \bibnamefont{Zel'dovich}},
  \bibinfo{journal}{Sov. Phys. JETP} \textbf{\bibinfo{volume}{9}},
  \bibinfo{pages}{681} (\bibinfo{year}{1959}).

\bibitem[{\citenamefont{Glashow}(1961)}]{Glashow}
\bibinfo{author}{\bibfnamefont{S.~L.} \bibnamefont{Glashow}},
  \bibinfo{journal}{Nucl. Phys.} \textbf{\bibinfo{volume}{22}},
  \bibinfo{pages}{579} (\bibinfo{year}{1961}).

\bibitem[{\citenamefont{Weinberg}(1967)}]{Weinberg}
\bibinfo{author}{\bibfnamefont{S.}~\bibnamefont{Weinberg}},
  \bibinfo{journal}{Phys. Rev. Lett.} \textbf{\bibinfo{volume}{19}},
  \bibinfo{pages}{1264} (\bibinfo{year}{1967}).

\bibitem[{\citenamefont{Salam and Ward}(1964)}]{SalamWard}
\bibinfo{author}{\bibfnamefont{A.}~\bibnamefont{Salam}} \bibnamefont{and}
  \bibinfo{author}{\bibfnamefont{J.}~\bibnamefont{Ward}},
  \bibinfo{journal}{Physics Letters} \textbf{\bibinfo{volume}{13}},
  \bibinfo{pages}{168} (\bibinfo{year}{1964}).

\bibitem[{\citenamefont{Bouchiat and Bouchiat}(1997)}]{Bouchiats_review}
\bibinfo{author}{\bibfnamefont{M.-A.} \bibnamefont{Bouchiat}} \bibnamefont{and}
  \bibinfo{author}{\bibfnamefont{C.}~\bibnamefont{Bouchiat}},
  \bibinfo{journal}{Rep. Prog. Phys.} \textbf{\bibinfo{volume}{60}},
  \bibinfo{pages}{1351} (\bibinfo{year}{1997}).

\bibitem[{\citenamefont{Bouchiat and Bouchiat}(1974)}]{Bouchiats1974}
\bibinfo{author}{\bibfnamefont{M.-A.} \bibnamefont{Bouchiat}} \bibnamefont{and}
  \bibinfo{author}{\bibfnamefont{C.}~\bibnamefont{Bouchiat}},
  \bibinfo{journal}{Phys. Lett.} \textbf{\bibinfo{volume}{48B}},
  \bibinfo{pages}{111} (\bibinfo{year}{1974}).

\bibitem[{\citenamefont{Bouchiat and Bouchiat}(1975)}]{Bouchiats1975}
\bibinfo{author}{\bibfnamefont{M.-A.} \bibnamefont{Bouchiat}} \bibnamefont{and}
  \bibinfo{author}{\bibfnamefont{C.}~\bibnamefont{Bouchiat}},
  \bibinfo{journal}{J. Physique} \textbf{\bibinfo{volume}{36}},
  \bibinfo{pages}{493} (\bibinfo{year}{1975}).

\bibitem[{\citenamefont{Wood et~al.}(1997)\citenamefont{Wood, Bennett, Cho,
  Masterson, Roberto, Tanner, and Wieman}}]{Wood_Science}
\bibinfo{author}{\bibfnamefont{C.~S.} \bibnamefont{Wood}},
  \bibinfo{author}{\bibfnamefont{S.~C.} \bibnamefont{Bennett}},
  \bibinfo{author}{\bibfnamefont{D.}~\bibnamefont{Cho}},
  \bibinfo{author}{\bibfnamefont{B.~P.} \bibnamefont{Masterson}},
  \bibinfo{author}{\bibfnamefont{J.~L.} \bibnamefont{Roberto}},
  \bibinfo{author}{\bibfnamefont{C.~E.} \bibnamefont{Tanner}},
  \bibnamefont{and} \bibinfo{author}{\bibfnamefont{C.~E.}
  \bibnamefont{Wieman}}, \bibinfo{journal}{Science}
  \textbf{\bibinfo{volume}{275}}, \bibinfo{pages}{1759} (\bibinfo{year}{1997}).

\bibitem[{\citenamefont{Bennett and Wieman}(1999)}]{BennettPRL}
\bibinfo{author}{\bibfnamefont{S.~C.} \bibnamefont{Bennett}} \bibnamefont{and}
  \bibinfo{author}{\bibfnamefont{C.~E.} \bibnamefont{Wieman}},
  \bibinfo{journal}{Phys. Rev. Lett.} \textbf{\bibinfo{volume}{82}},
  \bibinfo{pages}{2484} (\bibinfo{year}{1999}).

\bibitem[{\citenamefont{Porsev et~al.}(2009)\citenamefont{Porsev, Beloy, and
  Derevianko}}]{Porsev}
\bibinfo{author}{\bibfnamefont{S.~G.} \bibnamefont{Porsev}},
  \bibinfo{author}{\bibfnamefont{K.}~\bibnamefont{Beloy}}, \bibnamefont{and}
  \bibinfo{author}{\bibfnamefont{A.}~\bibnamefont{Derevianko}},
  \bibinfo{journal}{Phys. Rev. Lett.} \textbf{\bibinfo{volume}{102}},
  \bibinfo{pages}{181601:1} (\bibinfo{year}{2009}).

\bibitem[{\citenamefont{Langacker et~al.}(1992)\citenamefont{Langacker, Luo,
  and Mann}}]{Langacker}
\bibinfo{author}{\bibfnamefont{P.}~\bibnamefont{Langacker}},
  \bibinfo{author}{\bibfnamefont{M.}~\bibnamefont{Luo}}, \bibnamefont{and}
  \bibinfo{author}{\bibfnamefont{A.~K.} \bibnamefont{Mann}},
  \bibinfo{journal}{Rev. Mod. Phys.} \textbf{\bibinfo{volume}{64}},
  \bibinfo{pages}{87} (\bibinfo{year}{1992}).

\bibitem[{\citenamefont{Ramsey-Musolf}(1999)}]{Musolf}
\bibinfo{author}{\bibfnamefont{M.~J.} \bibnamefont{Ramsey-Musolf}},
  \bibinfo{journal}{Phys. Rev. C} \textbf{\bibinfo{volume}{60}},
  \bibinfo{pages}{015501:1} (\bibinfo{year}{1999}).

\bibitem[{\citenamefont{Plunien and Soff}(1995)}]{PlunienPRA43}
\bibinfo{author}{\bibfnamefont{G.}~\bibnamefont{Plunien}} \bibnamefont{and}
  \bibinfo{author}{\bibfnamefont{G.}~\bibnamefont{Soff}},
  \bibinfo{journal}{Phys. Rev. A} \textbf{\bibinfo{volume}{51}},
  \bibinfo{pages}{1119} (\bibinfo{year}{1995}).

\bibitem[{\citenamefont{Plunien et~al.}(1991)\citenamefont{Plunien, M{\"u}ller,
  Greiner, and Soff}}]{PlunienPRA51}
\bibinfo{author}{\bibfnamefont{G.}~\bibnamefont{Plunien}},
  \bibinfo{author}{\bibfnamefont{B.}~\bibnamefont{M{\"u}ller}},
  \bibinfo{author}{\bibfnamefont{W.}~\bibnamefont{Greiner}}, \bibnamefont{and}
  \bibinfo{author}{\bibfnamefont{G.}~\bibnamefont{Soff}},
  \bibinfo{journal}{Phys. Rev. A} \textbf{\bibinfo{volume}{43}},
  \bibinfo{pages}{5853} (\bibinfo{year}{1991}).

\bibitem[{\citenamefont{Mohr et~al.}(1998)\citenamefont{Mohr, Plunien, and
  Soff}}]{MPSoff}
\bibinfo{author}{\bibfnamefont{P.~J.} \bibnamefont{Mohr}},
  \bibinfo{author}{\bibfnamefont{G.}~\bibnamefont{Plunien}}, \bibnamefont{and}
  \bibinfo{author}{\bibfnamefont{G.}~\bibnamefont{Soff}},
  \bibinfo{journal}{Phys. Rep.} \textbf{\bibinfo{volume}{293}},
  \bibinfo{pages}{227} (\bibinfo{year}{1998}).

\bibitem[{\citenamefont{Pachucki et~al.}(1993)\citenamefont{Pachucki,
  Leibfried, and H{\"a}nsch}}]{PachuckiPRA48}
\bibinfo{author}{\bibfnamefont{K.}~\bibnamefont{Pachucki}},
  \bibinfo{author}{\bibfnamefont{D.}~\bibnamefont{Leibfried}},
  \bibnamefont{and} \bibinfo{author}{\bibfnamefont{T.~W.}
  \bibnamefont{H{\"a}nsch}}, \bibinfo{journal}{Phys. Rev. A}
  \textbf{\bibinfo{volume}{48}}, \bibinfo{pages}{R1} (\bibinfo{year}{1993}).

\bibitem[{\citenamefont{Pachucki et~al.}(1994)\citenamefont{Pachucki, Weitz,
  and H{\"a}nsch}}]{PachuckiPRA49}
\bibinfo{author}{\bibfnamefont{K.}~\bibnamefont{Pachucki}},
  \bibinfo{author}{\bibfnamefont{M.}~\bibnamefont{Weitz}}, \bibnamefont{and}
  \bibinfo{author}{\bibfnamefont{T.~W.} \bibnamefont{H{\"a}nsch}},
  \bibinfo{journal}{Phys. Rev. A} \textbf{\bibinfo{volume}{49}},
  \bibinfo{pages}{2255} (\bibinfo{year}{1994}).

\bibitem[{\citenamefont{Yamanaka et~al.}(2001)\citenamefont{Yamanaka, Haga,
  Horikawa, and Ichimura}}]{Yamanaka}
\bibinfo{author}{\bibfnamefont{N.}~\bibnamefont{Yamanaka}},
  \bibinfo{author}{\bibfnamefont{A.}~\bibnamefont{Haga}},
  \bibinfo{author}{\bibfnamefont{Y.}~\bibnamefont{Horikawa}}, \bibnamefont{and}
  \bibinfo{author}{\bibfnamefont{A.}~\bibnamefont{Ichimura}},
  \bibinfo{journal}{Phys. Rev. A} \textbf{\bibinfo{volume}{63}},
  \bibinfo{pages}{062502:1} (\bibinfo{year}{2001}).

\bibitem[{\citenamefont{Puchalski
  et~al.}(2006{\natexlab{b}})\citenamefont{Puchalski, Moro, and
  Pachucki}}]{NPLi11}
\bibinfo{author}{\bibfnamefont{M.}~\bibnamefont{Puchalski}},
  \bibinfo{author}{\bibfnamefont{A.~M.} \bibnamefont{Moro}}, \bibnamefont{and}
  \bibinfo{author}{\bibfnamefont{K.}~\bibnamefont{Pachucki}},
  \bibinfo{journal}{Phys. Rev. Lett.} \textbf{\bibinfo{volume}{97}},
  \bibinfo{pages}{133001:1} (\bibinfo{year}{2006}{\natexlab{b}}).

\bibitem[{\citenamefont{Pachucki and Moro}(2007)}]{NPHe}
\bibinfo{author}{\bibfnamefont{K.}~\bibnamefont{Pachucki}} \bibnamefont{and}
  \bibinfo{author}{\bibfnamefont{A.~M.} \bibnamefont{Moro}},
  \bibinfo{journal}{Phys. Rev. A} \textbf{\bibinfo{volume}{75}},
  \bibinfo{pages}{032521:1} (\bibinfo{year}{2007}).

\bibitem[{\citenamefont{Nefiodov et~al.}(2002)\citenamefont{Nefiodov, Plunien,
  and Soff}}]{PlunienGFactor}
\bibinfo{author}{\bibfnamefont{A.~V.} \bibnamefont{Nefiodov}},
  \bibinfo{author}{\bibfnamefont{G.}~\bibnamefont{Plunien}}, \bibnamefont{and}
  \bibinfo{author}{\bibfnamefont{G.}~\bibnamefont{Soff}},
  \bibinfo{journal}{Phys. Rev. Lett.} \textbf{\bibinfo{volume}{89}},
  \bibinfo{pages}{081802:1} (\bibinfo{year}{2002}).

\bibitem[{\citenamefont{Beiersdorfer et~al.}(1998)\citenamefont{Beiersdorfer,
  Osterheld, Scofield, Crespo~L{\'o}pez-Urrutia, and Widmann}}]{BiXray}
\bibinfo{author}{\bibfnamefont{P.}~\bibnamefont{Beiersdorfer}},
  \bibinfo{author}{\bibfnamefont{A.~L.} \bibnamefont{Osterheld}},
  \bibinfo{author}{\bibfnamefont{J.~H.} \bibnamefont{Scofield}},
  \bibinfo{author}{\bibfnamefont{J.~R.}
  \bibnamefont{Crespo~L{\'o}pez-Urrutia}}, \bibnamefont{and}
  \bibinfo{author}{\bibfnamefont{K.}~\bibnamefont{Widmann}},
  \bibinfo{journal}{Phys. Rev. Lett.} \textbf{\bibinfo{volume}{80}},
  \bibinfo{pages}{3022} (\bibinfo{year}{1998}).

\bibitem[{\citenamefont{Beiersdorfer et~al.}(2005)\citenamefont{Beiersdorfer,
  Chen, Thorn, and Tr{\"a}bert}}]{QEDNP}
\bibinfo{author}{\bibfnamefont{P.}~\bibnamefont{Beiersdorfer}},
  \bibinfo{author}{\bibfnamefont{H.}~\bibnamefont{Chen}},
  \bibinfo{author}{\bibfnamefont{D.~B.} \bibnamefont{Thorn}}, \bibnamefont{and}
  \bibinfo{author}{\bibfnamefont{E.}~\bibnamefont{Tr{\"a}bert}},
  \bibinfo{journal}{Phys. Rev. Lett.} \textbf{\bibinfo{volume}{95}},
  \bibinfo{pages}{233003:1} (\bibinfo{year}{2005}).

\bibitem[{\citenamefont{Sakabe et~al.}(2005)\citenamefont{Sakabe, Takahashi,
  Hashida, Shimizu, and Iida}}]{NEET_list}
\bibinfo{author}{\bibfnamefont{S.}~\bibnamefont{Sakabe}},
  \bibinfo{author}{\bibfnamefont{K.}~\bibnamefont{Takahashi}},
  \bibinfo{author}{\bibfnamefont{M.}~\bibnamefont{Hashida}},
  \bibinfo{author}{\bibfnamefont{S.}~\bibnamefont{Shimizu}}, \bibnamefont{and}
  \bibinfo{author}{\bibfnamefont{T.}~\bibnamefont{Iida}},
  \bibinfo{journal}{At.~Dat.~Nucl.~Dat.~Tabl.} \textbf{\bibinfo{volume}{91}},
  \bibinfo{pages}{1} (\bibinfo{year}{2005}).

\bibitem[{\citenamefont{Carreyre et~al.}(2000)\citenamefont{Carreyre, Harston,
  Aiche, Bourgine, Chemin, Claverie, Goudour, Scheurer, Attallah, Bogaert
  et~al.}}]{Carreyre}
\bibinfo{author}{\bibfnamefont{T.}~\bibnamefont{Carreyre}},
  \bibinfo{author}{\bibfnamefont{M.~R.} \bibnamefont{Harston}},
  \bibinfo{author}{\bibfnamefont{M.}~\bibnamefont{Aiche}},
  \bibinfo{author}{\bibfnamefont{F.}~\bibnamefont{Bourgine}},
  \bibinfo{author}{\bibfnamefont{J.~F.} \bibnamefont{Chemin}},
  \bibinfo{author}{\bibfnamefont{G.}~\bibnamefont{Claverie}},
  \bibinfo{author}{\bibfnamefont{J.~P.} \bibnamefont{Goudour}},
  \bibinfo{author}{\bibfnamefont{J.~N.} \bibnamefont{Scheurer}},
  \bibinfo{author}{\bibfnamefont{F.}~\bibnamefont{Attallah}},
  \bibinfo{author}{\bibfnamefont{G.}~\bibnamefont{Bogaert}},
  \bibnamefont{et~al.}, \bibinfo{journal}{Phys.\ Rev.\ C}
  \textbf{\bibinfo{volume}{62}}, \bibinfo{pages}{024311:1}
  (\bibinfo{year}{2000}).

\bibitem[{\citenamefont{Kishimoto et~al.}(2000)\citenamefont{Kishimoto, Yoda,
  Seto, Kobayashi, Kitao, Haruki, Kawauchi, Fukutani, and Okano}}]{Kishimoto}
\bibinfo{author}{\bibfnamefont{S.}~\bibnamefont{Kishimoto}},
  \bibinfo{author}{\bibfnamefont{Y.}~\bibnamefont{Yoda}},
  \bibinfo{author}{\bibfnamefont{M.}~\bibnamefont{Seto}},
  \bibinfo{author}{\bibfnamefont{Y.}~\bibnamefont{Kobayashi}},
  \bibinfo{author}{\bibfnamefont{S.}~\bibnamefont{Kitao}},
  \bibinfo{author}{\bibfnamefont{R.}~\bibnamefont{Haruki}},
  \bibinfo{author}{\bibfnamefont{T.}~\bibnamefont{Kawauchi}},
  \bibinfo{author}{\bibfnamefont{K.}~\bibnamefont{Fukutani}}, \bibnamefont{and}
  \bibinfo{author}{\bibfnamefont{T.}~\bibnamefont{Okano}},
  \bibinfo{journal}{Phys.~Rev.~Lett.} \textbf{\bibinfo{volume}{85}},
  \bibinfo{pages}{1831} (\bibinfo{year}{2000}).

\bibitem[{\citenamefont{Takahashi and Yokoi}(1983)}]{Takahashi1}
\bibinfo{author}{\bibfnamefont{K.}~\bibnamefont{Takahashi}} \bibnamefont{and}
  \bibinfo{author}{\bibfnamefont{K.}~\bibnamefont{Yokoi}},
  \bibinfo{journal}{Nucl.~Phys.~A} \textbf{\bibinfo{volume}{404}},
  \bibinfo{pages}{578} (\bibinfo{year}{1983}).

\bibitem[{\citenamefont{Takahashi et~al.}(1987)\citenamefont{Takahashi, Boyd,
  Mathiews, and Yokoi}}]{Takahashi2}
\bibinfo{author}{\bibfnamefont{K.}~\bibnamefont{Takahashi}},
  \bibinfo{author}{\bibfnamefont{R.~N.} \bibnamefont{Boyd}},
  \bibinfo{author}{\bibfnamefont{G.~J.} \bibnamefont{Mathiews}},
  \bibnamefont{and} \bibinfo{author}{\bibfnamefont{K.}~\bibnamefont{Yokoi}},
  \bibinfo{journal}{Phys.~Rev.~C} \textbf{\bibinfo{volume}{36}},
  \bibinfo{pages}{1522} (\bibinfo{year}{1987}).

\bibitem[{\citenamefont{Ohtsubo et~al.}(2005)\citenamefont{Ohtsubo, Bosch,
  Geissel, Maier, Scheidenberger, Attallah, Beckert, Beller, Boutin,
  Faestermann et~al.}}]{Bosch2}
\bibinfo{author}{\bibfnamefont{T.}~\bibnamefont{Ohtsubo}},
  \bibinfo{author}{\bibfnamefont{F.}~\bibnamefont{Bosch}},
  \bibinfo{author}{\bibfnamefont{H.}~\bibnamefont{Geissel}},
  \bibinfo{author}{\bibfnamefont{L.}~\bibnamefont{Maier}},
  \bibinfo{author}{\bibfnamefont{C.}~\bibnamefont{Scheidenberger}},
  \bibinfo{author}{\bibfnamefont{F.}~\bibnamefont{Attallah}},
  \bibinfo{author}{\bibfnamefont{K.}~\bibnamefont{Beckert}},
  \bibinfo{author}{\bibfnamefont{P.}~\bibnamefont{Beller}},
  \bibinfo{author}{\bibfnamefont{D.}~\bibnamefont{Boutin}},
  \bibinfo{author}{\bibfnamefont{T.}~\bibnamefont{Faestermann}},
  \bibnamefont{et~al.}, \bibinfo{journal}{Phys. Rev. Lett.}
  \textbf{\bibinfo{volume}{95}}, \bibinfo{pages}{052501:1}
  (\bibinfo{year}{2005}).

\bibitem[{\citenamefont{Bosch et~al.}(1996)\citenamefont{Bosch, Faestermann,
  Friese, Heine, Kienle, Wefers, Zeitelhack, Beckert, Franzke, Klepper
  et~al.}}]{Bosch1}
\bibinfo{author}{\bibfnamefont{F.}~\bibnamefont{Bosch}},
  \bibinfo{author}{\bibfnamefont{T.}~\bibnamefont{Faestermann}},
  \bibinfo{author}{\bibfnamefont{J.}~\bibnamefont{Friese}},
  \bibinfo{author}{\bibfnamefont{F.}~\bibnamefont{Heine}},
  \bibinfo{author}{\bibfnamefont{P.}~\bibnamefont{Kienle}},
  \bibinfo{author}{\bibfnamefont{E.}~\bibnamefont{Wefers}},
  \bibinfo{author}{\bibfnamefont{K.}~\bibnamefont{Zeitelhack}},
  \bibinfo{author}{\bibfnamefont{K.}~\bibnamefont{Beckert}},
  \bibinfo{author}{\bibfnamefont{B.}~\bibnamefont{Franzke}},
  \bibinfo{author}{\bibfnamefont{O.}~\bibnamefont{Klepper}},
  \bibnamefont{et~al.}, \bibinfo{journal}{Phys. Rev. Lett.}
  \textbf{\bibinfo{volume}{77}}, \bibinfo{pages}{5190} (\bibinfo{year}{1996}).

\bibitem[{\citenamefont{Goldanskii and Namiot}(1976)}]{Goldanskii}
\bibinfo{author}{\bibfnamefont{V.}~\bibnamefont{Goldanskii}} \bibnamefont{and}
  \bibinfo{author}{\bibfnamefont{V.~A.} \bibnamefont{Namiot}},
  \bibinfo{journal}{Phys.\ Lett.} \textbf{\bibinfo{volume}{62B}},
  \bibinfo{pages}{393} (\bibinfo{year}{1976}).

\bibitem[{\citenamefont{Harston and Chemin}(1999)}]{Harston}
\bibinfo{author}{\bibfnamefont{M.}~\bibnamefont{Harston}} \bibnamefont{and}
  \bibinfo{author}{\bibfnamefont{J.}~\bibnamefont{Chemin}},
  \bibinfo{journal}{Phys.\ Rev.\ C} \textbf{\bibinfo{volume}{59}},
  \bibinfo{pages}{2462} (\bibinfo{year}{1999}).

\bibitem[{\citenamefont{Cue et~al.}(1989)\citenamefont{Cue, Poizat, and
  Remillieux}}]{Cue}
\bibinfo{author}{\bibfnamefont{N.}~\bibnamefont{Cue}},
  \bibinfo{author}{\bibfnamefont{J.-C.} \bibnamefont{Poizat}},
  \bibnamefont{and}
  \bibinfo{author}{\bibfnamefont{J.}~\bibnamefont{Remillieux}},
  \bibinfo{journal}{Europhys.~Lett.} \textbf{\bibinfo{volume}{8}},
  \bibinfo{pages}{19} (\bibinfo{year}{1989}).

\bibitem[{\citenamefont{Kimball et~al.}(1991)\citenamefont{Kimball, Bittle, and
  Cue}}]{Kimball1}
\bibinfo{author}{\bibfnamefont{J.}~\bibnamefont{Kimball}},
  \bibinfo{author}{\bibfnamefont{D.}~\bibnamefont{Bittle}}, \bibnamefont{and}
  \bibinfo{author}{\bibfnamefont{N.}~\bibnamefont{Cue}},
  \bibinfo{journal}{Phys.~Lett.~A} \textbf{\bibinfo{volume}{152}},
  \bibinfo{pages}{367} (\bibinfo{year}{1991}).

\bibitem[{\citenamefont{Yuan and Kimball}(1993)}]{Kimball2}
\bibinfo{author}{\bibfnamefont{Z.-S.} \bibnamefont{Yuan}} \bibnamefont{and}
  \bibinfo{author}{\bibfnamefont{J.}~\bibnamefont{Kimball}},
  \bibinfo{journal}{Phys.\ Rev.\ C} \textbf{\bibinfo{volume}{47}},
  \bibinfo{pages}{323} (\bibinfo{year}{1993}).

\bibitem[{\citenamefont{P{\'a}lffy et~al.}(2006)\citenamefont{P{\'a}lffy,
  Scheid, and Harman}}]{PalffyPRA73}
\bibinfo{author}{\bibfnamefont{A.}~\bibnamefont{P{\'a}lffy}},
  \bibinfo{author}{\bibfnamefont{W.}~\bibnamefont{Scheid}}, \bibnamefont{and}
  \bibinfo{author}{\bibfnamefont{Z.}~\bibnamefont{Harman}},
  \bibinfo{journal}{Phys.~Rev.~A} \textbf{\bibinfo{volume}{73}},
  \bibinfo{pages}{012715:1} (\bibinfo{year}{2006}).

\bibitem[{\citenamefont{P{\'a}lffy
  et~al.}(2007{\natexlab{a}})\citenamefont{P{\'a}lffy, Harman, and
  Scheid}}]{interference}
\bibinfo{author}{\bibfnamefont{A.}~\bibnamefont{P{\'a}lffy}},
  \bibinfo{author}{\bibfnamefont{Z.}~\bibnamefont{Harman}}, \bibnamefont{and}
  \bibinfo{author}{\bibfnamefont{W.}~\bibnamefont{Scheid}},
  \bibinfo{journal}{Phys.~Rev.~A} \textbf{\bibinfo{volume}{75}},
  \bibinfo{pages}{012709:1} (\bibinfo{year}{2007}{\natexlab{a}}).

\bibitem[{\citenamefont{P{\'a}lffy et~al.}(2008)\citenamefont{P{\'a}lffy,
  Harman, Kozhuharov, Brandau, Keitel, Scheid, and St{\"o}hlker}}]{PalffyPLB}
\bibinfo{author}{\bibfnamefont{A.}~\bibnamefont{P{\'a}lffy}},
  \bibinfo{author}{\bibfnamefont{Z.}~\bibnamefont{Harman}},
  \bibinfo{author}{\bibfnamefont{C.}~\bibnamefont{Kozhuharov}},
  \bibinfo{author}{\bibfnamefont{C.}~\bibnamefont{Brandau}},
  \bibinfo{author}{\bibfnamefont{C.~H.} \bibnamefont{Keitel}},
  \bibinfo{author}{\bibfnamefont{W.}~\bibnamefont{Scheid}}, \bibnamefont{and}
  \bibinfo{author}{\bibfnamefont{T.}~\bibnamefont{St{\"o}hlker}},
  \bibinfo{journal}{Phys. Lett. B} \textbf{\bibinfo{volume}{661}},
  \bibinfo{pages}{330} (\bibinfo{year}{2008}).

\bibitem[{\citenamefont{Walker and Dracoulis}(1999)}]{walker}
\bibinfo{author}{\bibfnamefont{P.~M.} \bibnamefont{Walker}} \bibnamefont{and}
  \bibinfo{author}{\bibfnamefont{G.~D.} \bibnamefont{Dracoulis}},
  \bibinfo{journal}{Nature} \textbf{\bibinfo{volume}{399}}, \bibinfo{pages}{35}
  (\bibinfo{year}{1999}).

\bibitem[{\citenamefont{Aprahamian and Sun}(2005)}]{nat_phys}
\bibinfo{author}{\bibfnamefont{A.}~\bibnamefont{Aprahamian}} \bibnamefont{and}
  \bibinfo{author}{\bibfnamefont{Y.}~\bibnamefont{Sun}},
  \bibinfo{journal}{Nature~Phys.} \textbf{\bibinfo{volume}{1}},
  \bibinfo{pages}{81} (\bibinfo{year}{2005}).

\bibitem[{\citenamefont{P{\'a}lffy
  et~al.}(2007{\natexlab{b}})\citenamefont{P{\'a}lffy, Evers, and
  Keitel}}]{PalffyPRL}
\bibinfo{author}{\bibfnamefont{A.}~\bibnamefont{P{\'a}lffy}},
  \bibinfo{author}{\bibfnamefont{J.}~\bibnamefont{Evers}}, \bibnamefont{and}
  \bibinfo{author}{\bibfnamefont{C.~H.} \bibnamefont{Keitel}},
  \bibinfo{journal}{Phys. Rev. Lett.} \textbf{\bibinfo{volume}{99}},
  \bibinfo{pages}{172502:1} (\bibinfo{year}{2007}{\natexlab{b}}).

\bibitem[{\citenamefont{Xia et~al.}(2002)\citenamefont{Xia, Zhan, Wei, Yuan,
  Song, Zhang, Yang, Yuan, Gao, Zhao et~al.}}]{HIRFL}
\bibinfo{author}{\bibfnamefont{J.~W.} \bibnamefont{Xia}},
  \bibinfo{author}{\bibfnamefont{W.~L.} \bibnamefont{Zhan}},
  \bibinfo{author}{\bibfnamefont{B.~W.} \bibnamefont{Wei}},
  \bibinfo{author}{\bibfnamefont{Y.~J.} \bibnamefont{Yuan}},
  \bibinfo{author}{\bibfnamefont{M.~T.} \bibnamefont{Song}},
  \bibinfo{author}{\bibfnamefont{W.~Z.} \bibnamefont{Zhang}},
  \bibinfo{author}{\bibfnamefont{X.~D.} \bibnamefont{Yang}},
  \bibinfo{author}{\bibfnamefont{P.}~\bibnamefont{Yuan}},
  \bibinfo{author}{\bibfnamefont{D.~Q.} \bibnamefont{Gao}},
  \bibinfo{author}{\bibfnamefont{H.~W.} \bibnamefont{Zhao}},
  \bibnamefont{et~al.}, \bibinfo{journal}{Nucl. Instrum. Methods Phys. Res.,
  Sect. A} \textbf{\bibinfo{volume}{488}}, \bibinfo{pages}{11}
  (\bibinfo{year}{2002}).

\bibitem[{\citenamefont{Beck et~al.}(2007)\citenamefont{Beck, Becker,
  Beiersdorfer, Brown, Moody, Wilhelmy, Porter, Kilbourne, and
  Kelley}}]{Th229m}
\bibinfo{author}{\bibfnamefont{B.~R.} \bibnamefont{Beck}},
  \bibinfo{author}{\bibfnamefont{J.~A.} \bibnamefont{Becker}},
  \bibinfo{author}{\bibfnamefont{P.}~\bibnamefont{Beiersdorfer}},
  \bibinfo{author}{\bibfnamefont{G.~V.} \bibnamefont{Brown}},
  \bibinfo{author}{\bibfnamefont{K.~J.} \bibnamefont{Moody}},
  \bibinfo{author}{\bibfnamefont{J.~B.} \bibnamefont{Wilhelmy}},
  \bibinfo{author}{\bibfnamefont{F.~S.} \bibnamefont{Porter}},
  \bibinfo{author}{\bibfnamefont{C.~A.} \bibnamefont{Kilbourne}},
  \bibnamefont{and} \bibinfo{author}{\bibfnamefont{R.~L.}
  \bibnamefont{Kelley}}, \bibinfo{journal}{Phys. Rev. Lett.}
  \textbf{\bibinfo{volume}{98}}, \bibinfo{pages}{142501:1}
  (\bibinfo{year}{2007}).

\bibitem[{\citenamefont{Dragani{\'c} et~al.}(2003)\citenamefont{Dragani{\'c},
  Crespo~L{\'o}pez-Urrutia, DuBois, Fritzsche, Shabaev, Orts, Tupitsyn, Zou,
  and Ullrich}}]{Draganic}
\bibinfo{author}{\bibfnamefont{I.}~\bibnamefont{Dragani{\'c}}},
  \bibinfo{author}{\bibfnamefont{J.~R.}
  \bibnamefont{Crespo~L{\'o}pez-Urrutia}},
  \bibinfo{author}{\bibfnamefont{R.}~\bibnamefont{DuBois}},
  \bibinfo{author}{\bibfnamefont{S.}~\bibnamefont{Fritzsche}},
  \bibinfo{author}{\bibfnamefont{V.~M.} \bibnamefont{Shabaev}},
  \bibinfo{author}{\bibfnamefont{R.~S.} \bibnamefont{Orts}},
  \bibinfo{author}{\bibfnamefont{I.~I.} \bibnamefont{Tupitsyn}},
  \bibinfo{author}{\bibfnamefont{Y.}~\bibnamefont{Zou}}, \bibnamefont{and}
  \bibinfo{author}{\bibfnamefont{J.}~\bibnamefont{Ullrich}},
  \bibinfo{journal}{Phys. Rev. Lett.} \textbf{\bibinfo{volume}{91}},
  \bibinfo{pages}{183001:1} (\bibinfo{year}{2003}).

\bibitem[{\citenamefont{P{\'a}lffy}(2008)}]{PalffyJMO}
\bibinfo{author}{\bibfnamefont{A.}~\bibnamefont{P{\'a}lffy}},
  \bibinfo{journal}{J. Mod. Opt.} \textbf{\bibinfo{volume}{55}},
  \bibinfo{pages}{2603} (\bibinfo{year}{2008}).

\end{thebibliography}
\newpage

\begin{table}
  \caption{Nuclear effects in atomic transitions}
{\begin{tabular}{llc}\toprule
 Nuclear property & Effect on atomic structure \\
\colrule 
Charge $+Ze$  & Binding energy     \\
  Size -- radius $r_{RMS}$ & Field shift  \\
   Mass $M<\infty$, nuclear recoil & Mass shift: NMS, SMS  \\
\colrule 
Spin and magnetic moment & Magnetic HFS \\
Quadrupole  moment & Quadrupole HFS \\
\colrule
Weak interaction & Parity mixing of atomic states \\
Polarizability -- virtual nuclear excitations & Nuclear polarization   \\
Nuclear transitions & IC/NEEC, BIC/NEET  \\
   \botrule
  \end{tabular}}

\label{summary}
\end{table}


\newpage


%
%


%
%


%
%

%
%


%
%


%
\vspace{5cm}


%
%

%
%


%
%


%
%


%
%


%
%

%
%

%

\label{lastpage}

\end{document}